# Benefits of Applying Software Design Patterns to Backend Rust Applications

*Leon Heuer*

leon.heuer.a22b@nordakademie.org
NORDAKADEMIE Hochschule der Wirtschaft
Elmshorn, Germany

Submitted on July 02, 2026

---

Bachelor's Thesis
at NORDAKADEMIE Hochschule der Wirtschaft
in cooperation with Otto GmbH & Co. KGaA

---

Supervising Examiner: **Prof. Dr. habil. Jan Haase**
Secondary Examiner: **Prof. Dr. rer. nat. Joachim Sauer**
Advisor: **B.Sc. Falk Woldmann Lu**

**TITLE OF THESIS**

Benefits of Applying Software Design Patterns to Backend Rust Applications

**ABSTRACT**

Software design patterns' effects on code quality have mostly been studied in the context of object-oriented languages. In the programming language Rust, which comes with novel language concepts, compile-time safety guarantees and a distinct type system, there has been little research on design patterns. This work investigates how patterns affect software quality and compile-time enforcement of invariants through a case study on three representative components of production backend applications. An evaluation framework based on criteria derived from the SQuaRE quality model, incorporating benchmarking, static code analysis and expert interviews, is developed to assess the refactored code. The patterns typestate and newtype are applied to address existing code smells in the selected use cases. While the typestate pattern improves faultlessness and testability significantly, it comes at the cost of more structural code that can degrade readability. Code with extensive branching logic and a high number of invariants is likely to benefit most from the pattern. The newtype pattern combined with the "Parse, don't validate" principle offers high returns in software quality at a low cost and prevents invalid states during runtime. Overall, this work provides an initial empirical assessment of design patterns in Rust and establishes a foundation for further studies involving additional use cases and patterns.

**KEYWORDS**

Rust, Design Patterns, Refactoring, Code Quality, Typestate Pattern, Newtype Pattern, Static Code Analysis, Type System

**TITEL DER ARBEIT**

Vorteile des Anwendens von Software Design Patterns in Backend Rust Anwendungen

**KURZZUSAMMENFASSUNG**

Die Auswirkungen von Software Design Patterns auf die Codequalität wurden bislang überwiegend im Kontext objektorientierter Programmiersprachen untersucht. Für die Programmiersprache Rust, die neuartige Sprachkonzepte, Sicherheitsgarantien zur Compilezeit und ein eigenständiges Typsystem mitbringt, existiert bisher nur wenig Forschung zu Design Patterns. Diese Arbeit untersucht anhand einer Fallstudie mit drei repräsentativen Komponenten aus laufenden Backend-Anwendungen, wie sich Design Patterns auf die Softwarequalität und die Sicherstellung von Invarianten zur Compilezeit auswirken. Zur Bewertung des modifizierten Codes wird eine Methodik entwickelt, die auf abgeleiteten Kriterien aus dem SQuaRE-Qualitätsmodell basiert und Benchmarking, statische Codeanalyse sowie Experteninterviews kombiniert. Zur Behebung bestehender Probleme im Code der ausgewählten Use Cases werden die Patterns Typestate und Newtype eingesetzt. Während das Typestate-Pattern die Fehlerfreiheit und Testbarkeit deutlich verbessert, geht dies mit einem erhöhten Volumen von strukturellem Code einher, der die Lesbarkeit beeinträchtigen kann. Insbesondere Code mit umfangreicher Verzweigungslogik und einer hohen Anzahl an Invarianten profitiert am meisten von diesem Pattern. Das Newtype-Pattern in Kombination mit dem Prinzip „Parse, don't validate“ erzielt hingegen bei geringem Aufwand deutliche Verbesserungen der Softwarequalität und verhindert invalide Zustände zur Laufzeit. Insgesamt liefert diese Arbeit eine erste empirische Bewertung von Design Patterns in Rust und schafft eine Grundlage für weiterführende Untersuchungen mit zusätzlichen Anwendungsfällen und Patterns.

# Contents

# List of Figures

# List of Tables

# List of Listings

# Abbreviations

***ADT* – Abstract Data Type**

***BPMN* – Business Process Model and Notation**

***CFG* – Control Flow Graph**

***CoC* – Cognitive Complexity**

***CyC* – Cyclomatic Complexity**

***FC-IS* – Functional Core - Imperative Shell**

***FP* – Functional Programming**

***GoF* – Gang of Four**

***HDiff* – Halstead's Difficulty**

***I/O* – Input / Output**

***LoC* – Lines of Code**

***MI* – Maintainability Index**

***OOP* – Object Oriented Programming**

***PDF* – Probability Density Function**

***QMOOD* – Quality Model for Object Oriented Design**

***SQuaRE* – Systems and Software Quality Requirements and Evaluation**

***SRP* – Single Responsibility Principle**

***UML* – Unified Modeling Language**

# 1 Introduction

The programming language Rust, consistently ranking as “most loved” among developers surveyed in the StackOverflow developer survey [1], brings programming concepts like Abstract Data Types (ADTs) and functional programming to the mainstream, while offering performance levels comparable to low-level programming languages like C. The unique ownership concept makes it memory safe without the cost of a runtime garbage collector [MK14]. Combining the ownership concept and its strong type system, Rust detects many errors at compile time, such as data races and access to deallocated memory [MK14], that are often only noticed during runtime when using other languages. Additional features like developer-friendly error messages and the `cargo` toolchain make it a practicable general purpose language.

Rust’s increasing adoption in critical systems that require memory safety and low-level control [PCO19], efficient implementation of security and reliability mechanisms [Bal+17], tools and software in academics [Per20] and in the Android platform [2] is evident of its influence on modern-day programming. According to the StackOverflow developer survey, Rust has seen a steady increase in popularity from 3.2% to 14.8% since it was first included in the ranking in 2019 [1]. The US government has commited to using memory safe languages like Rust as a countermeasure to increasing cyberattacks [3]. For the Android platform, it was reported that by switching to Rust, memory safety vulnerabilities were reduced by 1000x in comparison to C and C++, the rollback rate of code changes was 4x lower and code review took 25% less time [4].

While Rust is often mentioned as a low-level systems programming language, it has particularly become a popular choice for developing web applications. According to the 2024 State of Rust Survey, 53.4% of developers use Rust for backend applications [5], the highest among all categories. At the German online retailer OTTO, developers have begun migrating various services to Rust, with one team running fully on Rust since 2025. Developers are investigating how services can be designed to be more cost-efficient, performant, scalable and fail-safe through compile-time guarantees [6], [7], and more productive through the `cargo` toolchain .

Although Rust’s language design enforces memory safety and compile-time guarantees through its type system, it doesn’t inherently guarantee maintainable software or good architecture. As Rust adoption and code complexity grow, developers need structural guidance and must look beyond pure syntax. A common way to transfer knowledge about software design and architecture are design patterns. Design patterns are an abstraction of solutions to common problems that have already been solved in existing systems, thereby allowing the transfer of this knowledge to new systems [Gam+95, pp. 2-3]. For example, the book by the Gang of Four (GoF) introduced many design patterns still popular today, like the Builder [Gam+95, p.97] or Factory Method [Gam+95, p.107] pattern. Their patterns have been among the most widely studied.

Most of the design patterns known as of now originate in Object Oriented Programming (OOP) [Gam+95, p. 4]. However, Baumgartner et al. state that a design pattern can become obsolete when transferred to another programming language, if the problem it solves can already be expressed through an existing language construct in the target language [BLR96, p.2]. Some patterns often make up for missing language constructs in certain languages and are called “paradigmatic idioms” [BLR96, p.3]. Moreover, a design pattern might not be supported by a language when the language lacks the constructs or features required to implement the pattern.

Rust is a multi-paradigm language and cannot be categorized as strictly object-oriented or functional. According to the official book “The Rust Programming Language”, the language incorporates both functional [KN23, pp. 273-294] and object-oriented features [KN23, pp. 375-396]. In combination with the restrictions mentioned earlier, and due to many of the well studied design patterns being grounded in OOP, this means that many of them are not directly transferrable to Rust. According to senior software developers interviewed at OTTO, team members often have experience in designing object-oriented applications, but this doesn’t directly translate to Rust’s language features [6], [7], creating a need for principles and patterns that make use of Rust’s type system and enforce idiomatic language use.

Therefore, this thesis will perform a study on selected use cases from teams at OTTO. Three use cases will be analyzed and refactored using well-suited design patterns. The goal is to identify design patterns that improve Rust code quality while leveraging its type system, a field that currently lacks research. To evaluate whether a pattern influences code quality positively, the criteria listed in Table 1 will be used. Their selection process and evaluation methods will be explained in detail in Chapter 3. As a result of this study, a set of recommendations regarding when and how to use the selected design patterns in Rust will be formulated.

| ID | Name |
|---|---|
| C1 | Time Behaviour |
| C2 | Modularity |
| C3 | Reusability |
| C4 | Analysability |
| C5 | Modifiability |
| C6 | Testability |
| C7 | Faultlessness |

**Table 1:** Criteria the refactored code needs to meet. These criteria will be referenced throughout the study, and will be used to evaluate the refactored code and the effect of the applied design pattern.

# 2 Related Work

This chapter will explore the existing knowledge and understanding on the effectiveness of design patterns, investigate guidelines and processes used to decide when to use which design pattern, as well as design patterns specifically for Rust. Finally, the research gap that this thesis aims to address will be identified.

## 2.1 State of research on the effectiveness of design patterns

A lack of consensus exists on the point of the effectiveness of design patterns. According to a literature survey from 2013, previous studies on the impact of design patterns yielded controversial results, with some studies reporting positive and others reporting negative outcomes [AE13]. Moreover, the literature survey found that previous studies don't cover design patterns extensively enough, with not all of the GoF design patterns being covered, for example. A potential reason for this lack of consensus and controversial results could be the fact that design patterns come with tradeoffs [ACS13]. While a design pattern might improve one aspect of quality, it might degrade another. Furthermore, there was a lack of objective data on design patterns as of 2012, and no "firm support for any of the claims" about design patterns, like improving reusability or reducing complexity universally [ZB12]. In 2020, another systematic literature review found that there was no consensus about how design patterns impact maintainability [WA20]. While maintainability was found to be the most frequently assessed quality attribute, other attributes like change- and fault-proneness yielded contradictory results. In alignment with the previous studies, it points out that design pattern's effectiveness depends on the context it is used in. This shows that there is a lack of context-dependent assessment of design patterns' effects on software quality.

When it comes to measuring source code quality, a mapping study from 2017 that summarizes and categorizes research on metrics found that most research on code metrics has been conducted in OOP languages [Nuñ+17]. Specifically, most research was conducted in Java at 154 occurrences, followed by AspectJ and C++ at 30 and 24 occurrences, respectively. The study identifies fail safety, quality and complexity as most studied topics in object-oriented code. For procedural languages, the most studied metrics were Cyclomatic Complexity (CyC) and Lines of Code (LoC). The study concludes that for new metrics, more secondary studies that evaluate them are necessary so practitioners and researchers can rely on them. This highlights a lack of research in non-object-oriented languages, particularly in languages other than Java.

Primary sources that investigate the effects of design patterns show mixed evidence. As for positive evidence, a study from 2012 evaluates design patterns in the Java library JHotDraw by analyzing more than 300 revisions of it, using a probabilistic quality model [Heg+12]. The design pattern instances were retrieved from documentation in the code. It was found that all of the assessed quality characteristics improved with each introduced design pattern instance. However, the analyzed library JHotDraw can be seen as an exception, as it was co-developed by Erich Gamma for educational purposes [Chr04], who is also one of the authors of the GoF patterns [Gam+95], and therefore influenced the design of the software. An earlier study focuses on three selected GoF design patterns and demonstrated that they are effective at reducing high metric scores, thus causing less violations of alarm thresholds [Hus01]. According to a more recent study from 2021, the use of design patterns significantly reduces complexity and improves maintainability [JR21]. The results are gathered through evaluating metrics such as CyC and Maintainability Index (MI), making use of the static code analysis tool SonarQube. Another study from 2017 found a positive effect of investigated design patterns on a system level [HKK17] by using the Quality Model for Object Oriented Design (QMOOD) [BD02]. Because the latter two studies each only analyze three different design patterns, their findings are contextual and it is unclear whether other patterns yield similar improvements.

Other researchers developed a methodology to assess the impact of design patterns, and their results show that design patterns are not universally good or bad [AFS12]. Still, the study found that patterns generally produce more extendible code, and that they should be preferably used for systems with high reusability and maintainability needs. Another study employed a qualitative evaluation method by questioning software engineers, and found that design patterns do not necessarily influence all quality attributes positively [KG08]. Based on this result, the researchers suggest that design patterns need to be used with caution, aligning with the findings of several other studies [ACS13; AE13; AFS12].

When design patterns are used, it was found that documenting them with Unified Modeling Language (UML) diagrams or comments in the code improves correctness and ease of understanding [Sca+15]. The researchers conducted the experiment with participants that had varying experience levels. Especially for developers with higher experience levels, documenting design patterns leads to better correctness of understanding the code.

## 2.2 Guidelines for choosing the right design pattern

Numerous studies developed guidelines for when to use which design pattern, since it is a challenging task to choose a design pattern from the wide range available. In 2003, Kung et al. developed an expert system that suggests an appropriate design pattern based on the user's requirements [Kun+03, p.291-292]. A rule-based knowledge base is used to perform a question-answer process with the user. Another approach based on a knowledge base was proposed by Palma et al. [Pal+12] and also requires the user to answer questions about the design problem. Each question has an assigned weight and is mapped to appropriate patterns, so that a final ranking can be created. Another approach is based on defining the problem domain of a design pattern first [Kim08]. The study states that it is a difficult task to generalize the problem a pattern solves, and uses a methodology developed in previous work for formalizing pattern problem domains. As a result, the study presented an approach that enables developers to search a pattern that is applicable to a given problem model, including a prototype tool that transforms the original class diagram into a solution diagram that conforms the transformation specification. Yet it should be questioned whether this approach is practical, since it requires initial effort for formalizing the problem, and the problem domain for each design pattern.

Nevertheless, the underlying evidence for design pattern guidelines is mixed. A mapping study from 2012 concluded that there wasn't enough evidence and knowledge for creating a guideline on when to use which pattern [ZB12]. At the time, the study identified a lack of empirical evidence for proving design patterns' effectiveness. In addition, the study concludes that design patterns don't help developers with low experience levels at making better design choices, and that applying patterns "blindly" is counterproductive.

## 2.3 Existing design patterns for Rust

As of now, there is little data on the effectiveness of design patterns in Rust. Researchers recently developed a tool capable of analysis and extraction of bug fix patterns in Rust code, and used it to obtain the most common bug fix patterns from popular Rust projects [RL24]. The study however doesn't describe design patterns in the classical sense on an architectural level, but rather single changes like modifying struct fields or `clone` calls.

Unofficial guidelines with design patterns applied to or specifically designed for Rust exist in online documentation. One such documentation is Refactoring Guru, which is a guide on GoF patterns with code examples for many implementation languages, including Rust [8]. Yet, it lacks verification of their effects on code quality. Furthermore, it excludes design patterns apart from GoF patterns that are Rust-specific. Another online documentation is maintained by the Rust open source community and lists patterns that are commonly used in Rust, including some of the GoF patterns and Rust-specific patterns based on its language features [9]. It is a practical guideline with examples, advantages and disadvantages for each pattern, but it doesn't study or verify the quality impacts of each pattern in detail.

There have been recent advances in investigating design patterns in Rust. Researchers identified that there is no equivalent catalogue like the object-oriented GoF patterns for Functional Programming (FP) [Cri23]. The study extracts four recurring patterns from Rust code and verifies their positive impact on code correctness through case studies. While these are novel patterns that should be investigated more in future research, Rust as a multi-paradigm language might also support other design patterns that need to be investigated again in a Rust context.

## 2.4 Research gap

In summary, it can be seen that research has been heavily focused on GoF patterns applied to Java code, and there is not much evidence on design patterns in the context of Rust, despite its growing relevance as shown in Chapter 1. It was pointed out that several literature surveys and mapping studies see the effectiveness of design patterns as controversial and context dependent, thereby further investigation is required.

# 3 Method

In this chapter, the requirements for the refactored code will be derived through a requirement analysis. Then, the methods used for evaluating the refactored code in regards to the requirements will be explained. Quantitative methods will be assigned a threshold, if possible, that indicates whether the criterion is passed or not.

## 3.1 Requirement analysis

For investigating the benefit design patterns have on Rust in the context of backend applications, this thesis refactored use cases from real-world applications by applying design patterns. According to the definition from Martin Fowler, a refactoring is the modification of existing code that improves its design and internal structure without altering external behaviour [Fow+12]. This implies that refactoring is concerned with reaching a certain quality of code design.

To determine whether the design patterns involved in the refactoring are beneficial, quality requirements were derived from the Systems and Software Quality Requirements and Evaluation (SQuaRE) model, defined in the ISO/IEC 25010 standard [ISO23]. The standard defines 9 main criteria and 40 subcriteria, listed in Table 6 in Appendix A.2 for reference.

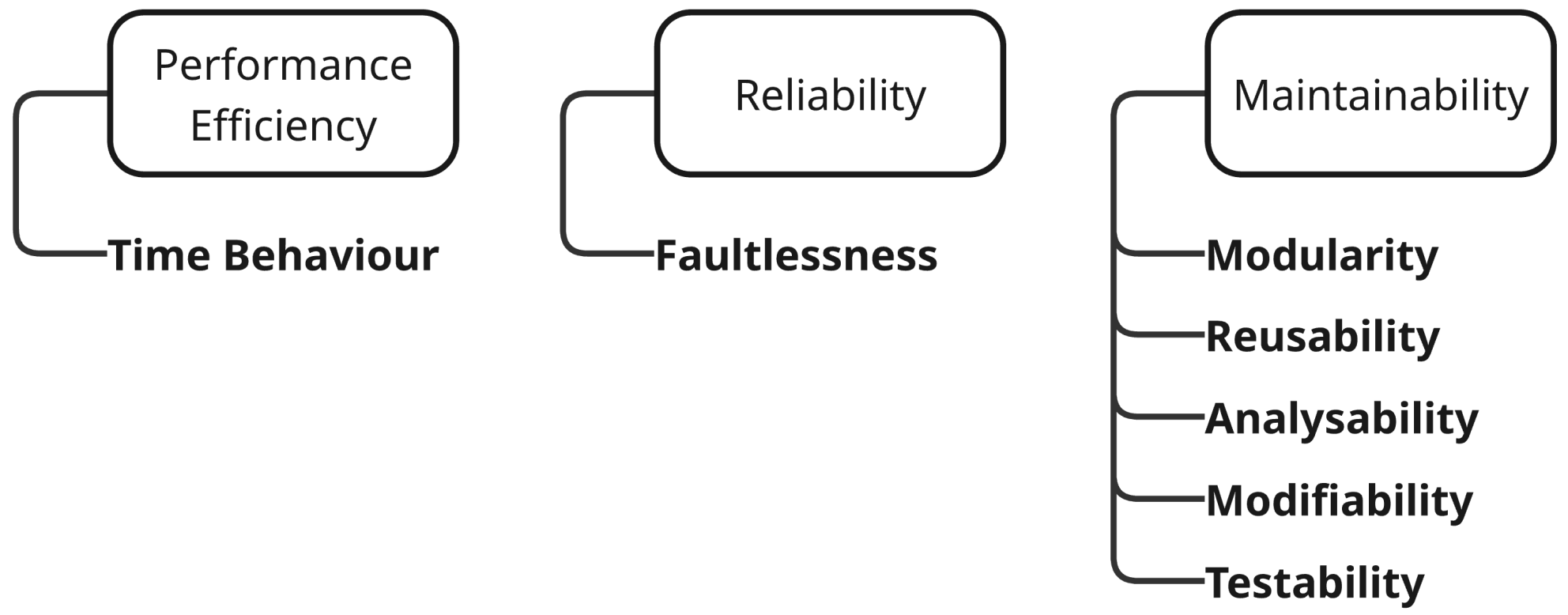

**Figure 1:** Selected criteria from the SQuaRE model. For Performance Efficiency and Reliability, only one subcriterion has been selected. Maintainability will be evaluated as a whole, thus all subcriteria were selected. This picture omits other criteria that were not selected.

However, not all criteria are equally relevant or addressed by design patterns. Consequently, relevant criteria were narrowed down to *Time Behaviour*, *Faultlessness*, *Modularity*, *Reusability*, *Analysability*, *Modifiability* and *Testability*, which are marked in **bold** in Figure 1. The definition and reasoning for eliminating or including specific criteria are explained in the following.

- *Functional Suitability*: Whether the software provides functions that fulfill its specified requirements, when used according to specified conditions. Refactorings shouldn't alter this external behaviour [Fow+12]. Therefore, evaluating this criterion would not yield meaningful results, as functional requirements stay consistent before and after. Passing the existing unit and integration tests and compiling successfully are preconditions to make sure the code's external behaviour wasn't altered. However, verifying the subcriterion *Functional Correctness* would require a formal proof of correctness, which is out of scope for this thesis. Overall, the criterion *Functional Suitability* is excluded.

- *Performance Efficiency*: Whether the software performs within given time and throughput thresholds, and uses hardware resources like CPU and memory efficiently. Design patterns most commonly focus on improving maintainability and readability of code, not its performance. However, it is a critical aspect relevant to business. For instance, if a critical component is refactored and much more readable as a result, but consumes significantly more resources, the refactored version might not be a viable choice for the business. Thereby, performance aspects should be measured to judge whether the design pattern is a viable choice. *Time Behaviour* can be assessed in the scope of a function and is therefore included in the criteria. *Resource Utilization*, however, is usually measured for the running system as a whole under realistic conditions or load test simulations. In this study, the use cases are a minor part of a bigger system, thus the performance impact is assumed to be minimal, and therefore this subcriterion is excluded. The same argument applies for *Capacity*, which describes whether a system's maximum limits are high enough so that other requirements can be met, and is excluded as well.
- *Compatibility*: Whether a software can communicate with other software appropriately, and run correctly when sharing the same environment and resources. The interface of the refactored software and its running environment are not discussed in this thesis, because a refactoring, as defined previously, doesn't alter external behaviour. Thereby, this criterion is excluded.
- *Interaction Capability*: Whether intended users of the software are able to exchange information with the user interface and complete specific tasks. Because this criterion focuses on the user interface, while this study only refactored source code of backend applications, this criterion is excluded.
- *Reliability*: Describes the ability of software to run as intended under the specified conditions for a specific period of time. The assessment of software's maintainability and readability will only be from a source code and developer perspective, and runtime behaviour will not be analyzed. Moreover, it is out of scope of this thesis to assess *Reliability* under realistic conditions for an extended period of time due to time and resource constraints. However, specific design patterns can improve reliability of the code by eliminating the possibility of invalid states, as in Crichton's State Machine and Witness patterns [Cri23]. Thus, *Faultlessness* can be judged through logical reasoning about the states possible in an application, and is included in the criteria.
- *Security*: The ability of software to be accessible to intended users within their authorization level, while protecting itself and its data against attacks by malicious actors. Whereas design patterns that focus specifically on security aspects exist, for instance the Proxy pattern [Gam+95, p.207], those patterns will not be part of this study, and thereby the criterion *Security* is excluded.

- *Maintainability*: How effectively and efficiently software can be modified, corrected or adapted to new requirements and environments. This criterion is the main aspect of code quality that design patterns and refactorings address, as it is related to how developers perceive and interact with the code. All subcriteria are included for evaluation.
- *Flexibility*: How adaptable software is to changes in external requirements and system environment. This criterion focuses on the interaction between the software and the hardware, like efficient scaling, adaptability to new hardware and how easy it is to install the software. As these are mostly runtime factors or factors that lie outside of the source code itself, and are not affected by design patterns, *Flexibility* is excluded.
- *Safety*: Describes whether the software avoids states that can be dangerous for human life, health or the environment. The chosen use cases will be parts of an online application that has no real-world criticality. Moreover, this is an aspect of software's behaviour during execution as well, and therefore *Safety* is excluded.

## 3.2 Time behaviour

The first criterion time behaviour identified in the requirement analysis will be evaluated by measuring execution time. As only selected parts of applications will be modified, the most straightforward way of measuring time behaviour is therefore measuring time on a function level. For this measurement, benchmarks will be created using the microbenchmarking tool "criterion-rs", available on GitHub [10]. This way, the refactored functions can be tested in an isolated manner by running the entry function, and small deviations in execution time can be identified. Concrete test setup, hardware specifications and configuration will be explained in detail in Chapter 6. The tool automatically performs the benchmark through a warmup phase, followed by a measurement phase where multiple samples are taken. For each subsequent sample, the amount of iterations is incremented by the sample size configured. Criterion then classifies outliers and reports them. This is followed by a linear regression and the calculation of statistical estimates and confidence intervals through a resampling process. Finally, the outcomes are then compared by performing another resampling and T-test.

For passing the criterion time behaviour, the code must be **within Criterion's noise threshold**, which is $\pm 2\%$ by default [11]. A benchmarking result that is statistically significant but within the noise threshold cannot be used as interpretation for either an improvement or degradation, as other background processes, CPU scheduling or other external effects of the system environment can cause irregularities.

## 3.3 Source code metrics

Static code analysis tools enable data-driven assessment and automated quality evaluations on a source code level by analyzing the code without executing it. That way, reliability and maintainability can be estimated based on structural and complexity metrics without needing to collect operational data [FB14, p.20]. Those metrics enable the creation of rules and thresholds that indicate when the code doesn't meet certain quality standards. In the following, appropriate metrics will be chosen and mapped to the criteria in Table 1.

### 3.3.1 Lines of code

LoC describes the count of lines in a specific scope, such as a function or source code file. Even though it is a trivial metric, it is still widely used due to its ease and understandability. Alpernas et al. found that while paper authors often try to back qualitative claims about source code with empirical data, LoC is no better support for those claims [AFP20]. It has limited expressiveness and is often used simply to adhere to the standard of measuring LoC. Although there are cases where LoC can be a meaningful metric, the aim of design patterns is not to write shorter, but more maintainable code. Moreover, two interviewed experts stated that they don't consider LoC a particularly good metric [12], [6]. Especially in Rust, more code often means more explicit types and data structures, which is not seen as problematic, while other features such as a large number of conditional branches are seen as more impactful [6]. Therefore, this study will not measure LoC.

### 3.3.2 Cyclomatic complexity

CyC was introduced to identify modules that will be difficult to test or maintain and is one of the most widely used source code metrics [McC76]. It is a single numeric value based on the cyclomatic number of the Control Flow Graph (CFG) of the program. In a strongly connected graph, achieved by connecting the exit node back to the entry node, it can be calculated as:

$$v(G) = e - n + 2p$$

where $G$ is the graph, $n$ the number of nodes, $e$ the number of edges, and $p$ the number of connected components. It represents the maximum number of linearly independent paths in the graph, indicating how many test cases might be necessary to test the module under observation. In practice, when a single component like a function is assessed, CyC can be simplified to:

$$v(G) = \pi + 1$$

where $\pi$ is the number of conditions in a program, for example `if`, `while` and the cases of a `switch` statement.

The metric is often used for measuring source code complexity. However, a limit of CyC is that it focuses only on control flow and not on other aspects of complexity, such as data flow or depth of nesting. The theoretical model of the metric makes it unsuitable for the general measurement of complexity, according to Fenton and Bieman [FB14, p.392]. Still, they state that it is a useful metric for identifying how difficult a program or module will be to test and maintain, following McCabe's original statement that his metric closely relates to the amount of work necessary for testing a program [McC76]. According to McCabe, if the amount of tests is less than the cyclomatic number, there are either tests missing, or the program can be reduced in complexity. As an upper limit based on empirical evidence, he suggested the number **10**, which will be used in this study to assess the criterion **C6 Testability**.

### 3.3.3 Cognitive complexity

Cognitive Complexity (CoC) is a metric developed at Sonar, a company that builds software for code quality and security. The metric was first introduced in a conference paper by Campbell [Cam18], who selected open source projects and evaluated whether their maintainers accept the metric. The overall acceptance rate of CoC was 77%, forming the basis of the metric's validity. The metric was created to address the shortcomings of CyC in predicting the mental effort required to understand a program, and is limited to the control flow as well. However, it is based on qualitative argumentation and a set of rules for specific language structures rather than a mathematical model. According to the official CoC white paper from Sonar [13], it is calculated as follows:

1. *Control flow*: Expressions that change control flow of a program increase the number by 1, with some additional rules. This includes:
   - Conditional statements, including `if`-conditions and ternary operators
   - Loop structures, including `for` and `while`
   - `catch` clauses, but any `try` or `finally` clauses are ignored
   - Simple `switch` conditions that match primitive data like integers, booleans and strings cause a single increase, but no additional increase for the cases
   - Each sequence with the same logical operator in a predicate (for example a sequence with a logical `AND`, then with a logical `OR` and then with an `AND` again has CoCo = 3)
   - Recursion
   - Jumps, such as `goto`, and `break` or `continue` statements to a specific label
2. *Nesting*: When an expression falls under the category *Control flow* and thus already incremented the number, the current level of nesting is added to it. There are structures such as lambdas that increase the level of nesting, but don't cause any increment on their own.
3. *Ignore shorthands*: There is no increase for shorthands, which are language features or constructs that express the original semantic meaning in a shorter form. For instance, extracting code to another method and invoking it, or using the null-coalescing operator (like `?` in Kotlin) instead of a null-check using an `if`-condition, doesn't increase the CoC number.

In summary, the main difference from CyC is the consideration of nesting levels, recursion and special rules when to ignore control flow structures. Due to its presence in the static code analysis tool Sonarqube and novelty in comparison to the other measures, it will be used for evaluation of **C4 Analysability**. While Campbell doesn't define a specific threshold in the current white paper for CoC [13], they answered in a post on StackOverflow that the value **15** is the recommended maximum for single functions [14].

However, the metric should be evaluated with caution, as there are exceptions for certain programming languages listed in the CoC white paper [13]. An email correspondence with Campbell, the author of the white paper and community manager at Sonar, shows that Rust hasn't been assessed yet in terms of caveats when measuring CoC. The email conversation can be found in Appendix A.4.

### 3.3.4 Halstead's measures

Halstead was the first to propose software metrics based on a mathematical model in his book *Software Science* [Hal77]. The model is based on the following fundamental measures, which can be derived directly from the source code:

- $\eta_1$: Number of unique operators
- $\eta_2$: Number of unique operands
- $\eta = \eta_1 + \eta_2$: Vocabulary of the program
- $N_1$: Total number of operators
- $N_2$: Total number of operands
- $N = N_1 + N_2$: Length of the program

On the basis of these measures, Halstead invented multiple composite metrics that can be used to judge certain aspects of the program, such as *Program Volume*, *Effort* and *Difficulty*. The latter is defined as the inverse of Halstead's *Program Level* metric [Abr10, p.157-158]:

$$D = \frac{\eta_1}{2} \times \frac{N_2}{\eta_2}$$

Halstead's Difficulty (HDiff) is proportional to the number of unique operators and usage of operands and represents the ease of reading. Therefore, it can be mapped to the criterion **C4 Analysability**. Despite being one of the earliest mathematical source code metrics and receiving criticism for being based on a poor cognitive model [Cur+84], a recent study found Halstead's Effort and Difficulty to have higher correlations with cognitive load than CyC and CoC [Hao+26], making it a relevant complementary metric. However, there is no normative prescription on what the size of HDiff means in practice, or how high or low it should be. It is unclear whether Halstead's metrics are measurement or prediction systems, and what the relationship between the mathematical model and real-world consequences is [FB14, p.345]. Therefore, HDiff will only be analyzed in terms of relative difference in this study.

### 3.3.5 Maintainability index

The initially proposed polynomial regression formula to calculate the MI is based on four metrics and was created in 1994 [OH94]. It is based on an earlier study that identified the smallest set of metrics useful for predicting software maintainability [OH92]. In their paper from 1994, Oman and Hagemeister create different regression models from a minimal set of metrics that can be easily calculated, based on test data from software systems at Hewlett-Peckard [OH94]. For evaluation, they correlate the models' results with the subjective assessment of maintainability found through surveying software engineers. Although more complex polynomials involving more metrics were found to be more accurate, they kept the final proposal simple in order to make it "quick, easy and reasonably accurate" for engineers to measure [OH94]. The model that was found most suitable was then refined and automatically assessed by Coleman et al. for 11 industrial software systems [Col+94]. According to them, the resulting data corresponds to the engineers' subjective assessment of the components. The following formula was used:

$$\text{MI} = 171 - 5.2 \times \ln(\text{aveVol}) - 0.23 \times \text{ave } V(g') - 16.2 \times \ln(\text{aveLOC})$$
$$+ \left(50 \times \sin\left(\sqrt{2.46 \times \text{perCM}}\right)\right)$$

where aveVol is the average Halstead Volume, ave $V(g')$ the average cyclomatic complexity, aveLOC the average lines of code and perCM the percentage of comments. The average is calculated within the scope that is analyzed, for example a function, a class or a whole program. Furthermore, the article proposes thresholds that indicate when a software is:

- *highly maintainable*: MI $\geq 85$
- *moderately maintainable*: $65 <$ MI $< 85$
- *difficult to maintain*: MI $\leq 65$

It was later identified that measuring comments in code is inaccurate, resulting in an updated model [WOA97]. Comments can contain commented out code, standard headers or other content that has no additional use, and therefore should only be measured if adequate for the use case. The following MI formula doesn't assess comments, and will be used for evaluation in this study:

$$\text{MI} = 171 - 5.2 \times \ln(\text{aveVol}) - 0.23 \times \text{ave } V(g') - 16.2 \times \ln(\text{aveLOC})$$

Despite early studies verifying the correlation of the MI with actual maintainability [Col+94], it should be questioned whether it is still a relevant indicator for maintainability. The programming languages used today have changed and software engineering practices have evolved since the introduction of MI. A more recent case study found that it is not a good predictor for maintainability and that simple size metrics had a higher correlation with maintainability [SAM12]. Another study that evaluated MI from an object-oriented perspective found a correlation of MI, but only due to its size component [Cou+15]. Therefore, MI will be applied only as a supporting metric for claims of qualitative analysis.

Due to its name, MI might intuitively need to be mapped to the main criterion Maintainability from the SQuaRE model. However, because Maintainability is a main criterion, and main criteria are made up of various subcriteria, MI should be mapped to a subcriterion instead. The creators of the MI metric [WOA97] based it on the definition of maintainability from the IEEE Standard Computer Dictionary, which defines it as "The ease with which a software system or component can be modified to correct faults, improve performance or other attributes, or adapt to a changed environment" [IEE91]. Therefore, MI will be mapped to the criterion **C5 Modifiability**, which has a similar definition: "capability of a product to be effectively and efficiently modified without introducing defects or degrading existing product quality" [ISO23].

### 3.3.6 Limits of static code metrics

Metrics that quantify code understandability and comprehension, such as CoC and Halstead's Difficulty, are insufficient in some regards according to numerous studies. For example, a systematic review from 2009 showed that there is not enough evidence on whether the available prediction models and metrics are effective [RMT09]. A more recent study from 2023 concludes that it is not possible to predict code's understandability based on these metrics only [LMG23]. While all metrics correlate with understandability, their prediction error was at around 30%, making them unreliable. Particularly, the CoC metric, which was introduced to mitigate some of CyC's shortcomings [Cam18], wasn't found to outperform metrics established earlier [Lav+23]. Supporting this, an earlier study from 2020 found that CoC correlates with comprehension time and subjective understandability, but less with comprehension correctness and physiological measures [MWW20].

Additionally, there are studies that measure brain activity in participants during code comprehension tasks. Peitek et al. found that CyC shows no correlation with an increased demand in any brain area, while LoC and Halstead show a medium correlation with cognitive load [Pei+21]. The CoC metric only shows a small improvement over CyC, supporting the claims of other studies mentioned in the former paragraph. While some metrics were found to be suitable to predict cognitive load in specific areas of the brain, none of them predict overall cognitive effort [Pei+21]. Another study that also measured brain activity yielded similar results [Hao+23]. For example, the study's results indicate that understanding complex algorithms requires higher mental effort, even though they consist of simple code constructs, which kept classical metrics low in the experiment. The authors further observed that participants' perceived complexity saturates at a certain point, which is not reflected in the metrics. Nested structures, which are captured by some metrics such as CoC, were not found to strictly make comprehension more difficult.

Additionally, a recent study found that structural metrics like CyC and CoC alone fail at measuring complexity [Hao+26]. The authors suggest to combine structural metrics with metrics that quantify data flow, such as Halstead's Difficulty and Effort. The authors moreover propose to create composite metrics based on a combination of metrics that assess control flow, data complexity and cognitive factors. Further supporting evidence for this recommendation can be found in a study from Scalabrino et al., who found models that combine multiple metrics to correlate slightly higher with code understandability [Sca+21]. They were however reported far from being usable in practice.

To conclude, evidence suggests that there is a lack of reliable methods for quantifying code comprehension. Therefore Crichton, the author of the conference paper "Typed Design Patterns for the Functional Era" [Cri23], has been asked via email. The questions and answers can be found in Appendix A.3 in full length. According to Crichton, there are currently no good ways of measuring code understandability, and suggests it is best to conduct a qualitative analysis through case studies. He further states in a blog post, that quantitative criteria are easy to measure, but often create the "illusion of rigor" and don't have much value for evaluating actual code understandability [15].

## 3.4 Qualitative analysis

Due to the limits of static code analysis alone that were explained in Chapter 3.3.6, this thesis will employ a mixed-method approach to supplement the quantitative with a qualitative evaluation. For the latter, expert interviews will be used to cover the criteria C2 - C7. A *semi-structured* interview as defined by Wohlin et al. will be conducted [Woh+24], which features concrete questions for evaluation, but allows room for open questions, follow up questions or discussion. As a result, the expert's opinions can be captured holistically while retrieving the relevant data for evaluation.

The guideline that will be used for expert interviews can be found in Appendix A.5, and was structured into the following sections:

1. **Introduction**: In this section, the background of the interviewee and their personal opinion about development in Rust will be determined.
2. **Current problems in the code**: This section has the goal of identifying general patterns and examples of poorly maintainable Rust code and finding out difficulties junior developers might have in understanding the code. Furthermore, this section identifies the interviewee's opinion on code metrics like the ones mentioned in Chapter 3.3, and whether they find them useful for quantifying problematic code. At the same time, the criteria used later for assessment will be introduced to the interviewee.
3. **Status quo**: The purpose of this section is to determine what interviewees dislike about the use case's former code and which criteria are affected negatively by that. The status quo will be evaluated for each use case.
4. **Refactoring**: In this section, each criterion and the refactored code's impact on them will be discussed. This will be the basis for evaluating the success of each refactoring and design pattern later. Similar to evaluating the status quo, the impact will be discussed for each use case.

Four Rust developers that currently work at OTTO will be chosen and must meet the following criteria:

- At least six years of software development experience
- At least two years of experience with Rust
- Implementation language used at their team is mainly Rust

Each interviewee will be given at least one day of preparation. They are allowed to take notes prior to the interview. Interviews will be conducted in 90-minute meetings, one-on-one. The interviewer and interviewee might ask additional questions or discuss further topics than the ones mentioned in the interview guideline, adhering to the semi-structured interview type. For capturing the interviewees' opinions unfiltered, all interviews will be conducted in their native language German. All questions asked and a summary of the corresponding answers will be attached in English in Appendix A.6. The summary will be double-checked by the interviewee to confirm its correctness. An audio recording of each interview will be attached for reference.

## 3.5 Method overview

Following the quantitative and qualitative evaluation methods developed in the previous chapters, Table 2 gives an overview of each criterion and the corresponding evaluation methods that will be used. Each criterion has been assigned at least one evaluation method. C2-C7 will be evaluated through expert interviews, and some of them through the additional quantitative metrics identified.

| **ID** | **Name** | **Method(s)** |
|---|---|---|
| C1 | Time Behaviour | Benchmarking: Time diff $\leq \pm 2\%$ |
| C2 | Modularity | Expert Interview |
| C3 | Reusability | Expert Interview |
| C4 | Analysability | Expert Interview<br>CoC $\leq 15$<br>HDiff (relative) |
| C5 | Modifiability | Expert Interview<br>MI $\geq 85$ |
| C6 | Testability | Expert Interview<br>CyC $\leq 10$ |
| C7 | Faultlessness | Expert Interview |

**Table 2:** Criteria selected through the requirement analysis. Each row lists a criterion, its ID and corresponding method(s).

# 4 Case Study

This chapter will introduce the three use cases selected for the case study. The use cases will be selected from applications maintained by teams at OTTO that use Rust. Each use case will be explained briefly, guided by a diagram following the Business Process Model and Notation (BPMN). BPMN is a standardized notation for business process flows, maintained by the Object Management Group, and can be seen as a common medium of communication between IT and business stakeholders [Sil11, p.3]. Each use case's relevance, current problems and proposed design pattern solution will be explained. Test code like unit and integration tests will not be part of the refactoring. Instead, they will be used as verification that the refactored code still works as intended. All use cases' status quo code can be found in Appendix A.1.1.

## 4.1 Use case 1

The first use case is the "Detailview" service maintained by FT9, which is a team at OTTO that uses Rust as main implementation language and is responsible for customer benefits on the store website. The Detailview service is responsible for creating view models that represent a small area on the product detail page and customer wishlist, which show a customer benefit (e.g. limited sale) that has been matched with the product and a slider for activating it. While the Detailview service itself is responsible for domain-specific logic only, requests are handled by another handler module that depends on the service. The BPMN model in Figure 2 illustrates how the service function creates the view model.

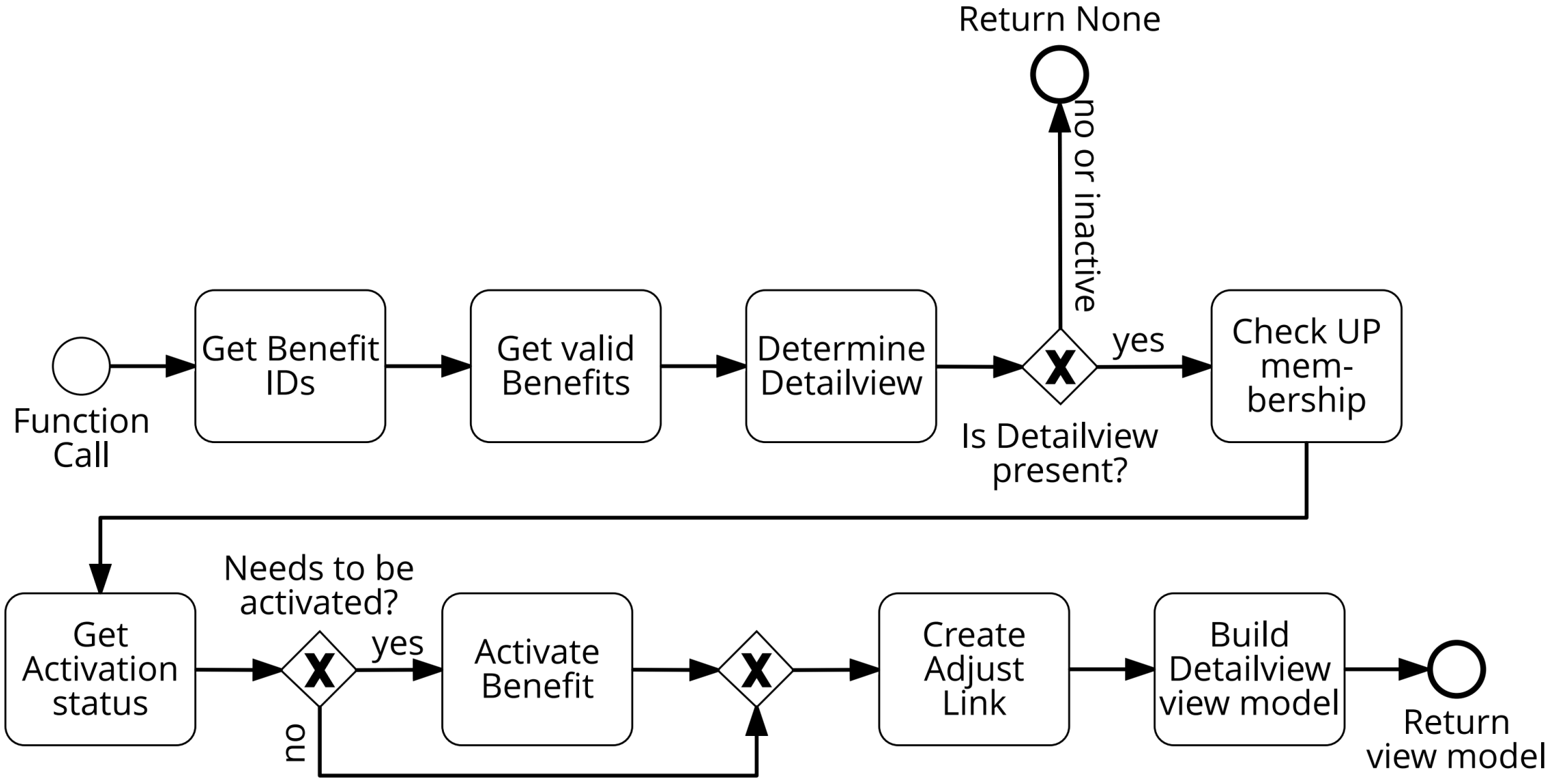

**Figure 2:** BPMN process model of the FT9 Detailview Service, selected as first use case, which is a single function with a long and linear sequence of operations that need to happen in order

When the service is called, the benefit IDs are first retrieved, and then used to collect the valid benefits. From those benefits, the one that will be displayed is selected. If there was no benefit found or the selected benefit is inactive, the service returns a `None` value early, depicted at the first `XOR`-gate in Figure 2. Otherwise, the membership status of the customer in the "OTTO UP" bonus program gets checked. Next, it is checked whether the customer activated the benefit or not. If this isn't the case and the selected benefit needs to be activated automatically, it will be activated by the service. Otherwise, this step is skipped, which can be seen at the `XOR`-gate labeled "Needs to be activated?" in Figure 2. Then the "Adjust" link, which is a link navigating mobile users to the OTTO app, gets created. Finally, the Detailview view model is built with the data retrieved so far and gets returned.

### 4.1.1 Relevance

The Detailview service can be seen as representative for many services used throughout the codebase at FT9. Services are commonly used to execute certain business logic, retrieve data and request other services. They are usually invoked by handlers, which return the result of the service to the requesting instance. Because this is a common architectural pattern at FT9, the Detailview service will be used to represent the use case **Service** at FT9.

### 4.1.2 Problems of the status quo

First of all, the main method in the service called `create_detailview_view_model` takes 7 parameters, which increases complexity [12]. The high amount of parameters causes difficulties in testing, as it requires many mocks and different combinations of values for covering all test cases [12]. This way, the number of test cases required quickly becomes large and hard to manage. Also, dependencies like `detailview_cache` make testing more difficult, as they need to be mocked or instantiated [16]. Even when dependencies like this are not needed for a specific step, every test of a condition or invariant requires mocking all dependencies [7]. Furthermore, because the function is defined on the struct `DetailviewService`, which requires multiple dependencies to construct, the function cannot be tested without mocking or setting up all those dependencies. Extensive test setup makes testing more difficult than it needs to be, so parameterized tests or property tests are currently not possible [6]. Overall, **C6 Testability** is seen as negatively in the current status quo.

Moreover, the function is notably long with 103 LoC, making it hard for the maintainer to understand the function and reason about isolated parts of the function. According to expert A, it is hard for the developer to understand the whole context of the function, which reduces **C4 Analysability** [12]. Experts B and D agree that the length of the function makes it difficult to grasp what operations are made and what the function actually does [16], [7]. Furthermore, the imperative programming style and high amount of conditional blocks and booleans makes it harder to read [6].

Additionally, the function handles too much logic at once [7] and has too many responsibilities. This causes a mix of concerns like Input / Output (I/O), business logic and logging [12], [16], [6], violating the Single Responsibility Principle (SRP) and therefore degrading **C2 Modularity**. According to expert A, different steps like database access and filtering should take place in separate modules [12]. Another concern is that the function is tied too tightly to this use case, which makes it hard to reuse in other modules, according to experts A and B [12], [16]. As a result, the **C3 Reusability** of the current code is degraded.

Also, navigating and changing the code becomes a major challenge, since the code lacks a clear isolated flow and logic is described as intertwined by expert C [6], affecting **C5 Modifiability** negatively. This is confirmed by expert D, who states that making changes carries some risks because each change in the beginning of the function might affect everything downstream, increasing the chance of unintended consequences [7].

Lastly, the function contains a range of if-conditions and invariants that are only checked during runtime, but not during compile time. This makes the function prone to errors like changing the order of some of the logic, which isn't detected until running the code [6], [7]. Without explicit enforcements of function order, **C7 Faultlessness** is compromised.

### 4.1.3 Proposed solution

This use case will be refactored using the *Typestate* pattern, adhering to the suggestion of expert C to model the domain logic as a state machine [6]. The concept of typestate was first introduced by Strom and Yemeni, who define it as a refined version of the concept of types [SY86]. Under the premise that syntactically incorrect programs and semantically incorrect programs are different, the goal of typestate is to detect operations that are semantically undefined, but syntactically correct and therefore remain unnoticed by the compiler. The authors state that typestate can be used to enforce invariants at compile time, so that the behaviour during runtime is constrained to allowed operations for that type. For example, take a type `File` that can be read and closed. Without typestates, the file might be closed, then read again, because the type system only defines which operations are possible, not which are allowed. When defining typestates, a program where a file is closed and subsequently read, can no longer compile. Closing the file will transition the program to a state where the read operation can no longer be performed.

The concept of typestates has been applied to practice through the design of a programming language with native typestate support called "Plaid" [Ald+09]. The authors argue that while typestates are possible in Java, the complexity of generic types makes typestates difficult in practice. Rust's language features however make typestates idomatic and practical, mainly because of ADTs and the ownership concept. The first way of encoding a typeside is by using a generic field on a struct, which is a marker for the current typestate. The struct can contain data, but if the used struct is empty, it is a zero-cost abstraction that is optimized away by the compiler [17]. The second way uses structs directly to represent each state: the structs define data of the state and allowed transitions can be implemented on them, whereas enums can be returned to indicate multiple possible resulting states. Enums are typically handled through a `match` statement, which is exhaustive by default, ensuring that all states are handled when the state is consumed. Furthermore, Rust's borrow checker enforces that when a state transition consumes `self`, it cannot be invoked multiple times, thereby preventing invalid operations. Aldrich et al. primarily focus on library and API design, whereas the application of typestate to isolated business logic remains largely unexplored [Ald+09], making this use case a novel problem domain.

Applying the typestate pattern will first of all address issues in **C7 Faultlessness** by defining the correct sequence of operations and encoding the assumed invariants in the type system. This way, errors like a wrong sequence of operations or wrong assumptions about the state of data will be prevented. The logic will be split down into state transitions, which explicitly define the input and output data. Therefore, **C5 Modifiability** will be addressed because changes can be narrowed down to the corresponding state and transition, which limits effects on the rest of the code. Since every state can theoretically be constructed from inside a unit test, **C6 Testability** will be enhanced significantly, because only the dependencies required for each specific state transition are needed. It will no longer be necessary to pass all dependencies to the single, large main function. The state's data will not contain external dependencies, making the construction of test values trivial.

A possible shortcoming is that the code necessary for defining states and possible transitions might cause conceptual overhead, as already described by Aldrich et al. [Ald+09]. Thus, **C4 Analysability** might not be unanimously seen as improved after the refactoring. **C3 Reusability** is also not a main aspect of the refactoring, because more code outside of this use case needs to be investigated for finding reusable code parts.

## 4.2 Use case 2

The second use case is chosen from FT9 as well. This use case is the "Tag Component" handler. Tags are small indicators seen on products and usually display discounts. The Tag Component handler directly handles requests from other teams who integrate the Tag on the shop's website. The handler returns a data object that is used to enrich the HTML template.

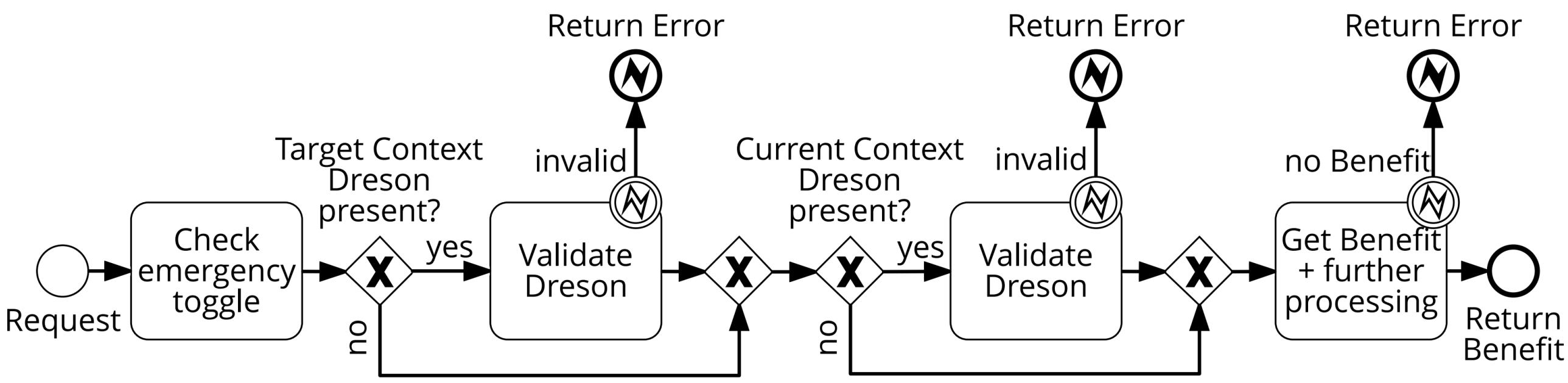

**Figure 3:** BPMN process model of the FT9 Tag Handler, selected as second use case, which is a sequence of operations that depends on validated "Dreson" data and should otherwise fail

When a request is received, the handler first checks the emergency toggle, which can be used to quickly disable the handler in case there is an issue. After that, the input parameters "Target Context Dreson" and "Current Context Dreson" are read. A Dreson is a domain-specific expression used in the OTTO shop to describe product selections. It serves as a declarative selection rule that can reference individual products, sets of products, or entire product categories. Each of the Dresons, if it is present, is validated, and the handler returns an error if validation fails. This can be seen in and around the two "Validate Dreson" steps in Figure 3, where invalid Dresons are depicted as error events. The rest of the handler is responsible for retrieving the benefit and returning it. If no benefit was found, an error is returned by the handler.

### 4.2.1 Relevance

The Tag Component handler represents **Handlers** in the code base of FT9. Handlers are invoked when their corresponding endpoint is called, and are responsible for handling incoming and outgoing data. Every endpoint requires a handler, and business logic is usually delegated to services, which are already represented through the first use case.

### 4.2.2 Problems of the status quo

After the Dreson parameters are validated, the information that they are valid is not encoded in the type system. As a result, the strings might be redundantly re-validated later, or validation might be forgotten entirely without a notice from the compiler [6]. Due to the usage of raw strings, developers that extend the code can't make assumptions about the data's state [7]. These issues have a negative impact on **C7 Faultlessness**. Also, the parameters should semantically already be valid because they were passed through the struct `ValidatedParams`, so the Dreson values should be validated at construction as well instead of the beginning of the handler [16], [7]. Besides the Dreson, expert B mentions that the "Sheet URL", which gets constructed in the end of the function, is based on raw strings instead of a builder pattern or typed logic, causing poor **C7 Faultlessness** and **C5 Modifiability** [16].

Overall, the use of raw strings and a manual validation violates the principle "Parse, don't validate", which states that input values should parsed into a domain type as soon as possible. If parsing fails or the input is invalid, construction of the type fails entirely, which eliminates the need for a manual validation later [18].

Furthermore, the two Dreson input values have very similar structures for their validation, indicating code repetition, which also means low **C3 Reusability** [12]. The validation across different handlers can also be inconsistent because the logic is not encapsulated in a type, but duplicated for each validation [6]. Expert D agrees with the poor encapsulation and reusability, and adds that this also affects **C2 Modularity** negatively [7].

The function is seen as poorly testable by expert A due to the high number of possible execution paths that need to be covered [12]. Not only the logic itself, but also the validation of the Dreson itself needs to be tested, resulting in degraded **C6 Testability**.

Finally, the validation logic for the Dreson values suffers from poor **C4 Analysability**, because it is deeply nested and represented through repeated if-conditions and mapping calls [16], [6], [7].

### 4.2.3 Proposed solution

As shown in the previous Chapter 4.2.2, an issue repeatedly mentioned by experts is the current validation of the Dreson input values. For encoding the information that the Dreson is valid in the type system, a newtype can be used. The newtype pattern in Rust is a concept where a tuple struct with a single field is introduced, and a definition can be found in the community-maintained "Rust Unofficial Book" [9]. This way, the original type is wrapped, hiding its functions and enabling new functions to be defined on the wrapper. According to the book, the newtype cannot be confused with the original type, as they are not type compatible, while being a zero-cost abstraction, so it doesn't add runtime cost. Moreover, the newtype pattern is a way of implementing the "Parse, don't validate" principle, as it allows defining a more concrete type with a construction method that only succeeds when the value is valid. Thus, the subsequent lines of code can operate on the newtype and rely on the fact that it must be valid. Thereby, the main issues mentioned for **C7 Faultlessness** are addressed, as invalid states regarding the Dreson cannot exist, and redundant validation is no longer possible.

Introducing a centralized newtype for Dreson values also eliminates code duplication, and allows other handlers to import and use the newtype with the same validation logic, which increases **C3 Reusability**. Moreover, **C6 Testability** is increased, because the validation logic can be tested in isolation, thereby eliminating complicated setup and mocking that were previously required for testing the validation as part of the handler function. The type logic then only needs to be tested once instead of for every handler, an advantage mentioned by expert C while discussing current code smells [6].

## 4.3 Use case 3

The third use case is taken from team Boxfish at OTTO. Boxfish maintains backend services for returned orders, including REST endpoints for retrieving the return status. Some of Boxfish's endpoints are served by AWS Lambdas, which are serverless functions that run in the cloud and are executed on every incoming trigger. The lambda handler called `update_return_status_to_announced` will be analyzed. Because Boxfish hasn't migrated their full codebase to Rust yet, their developers are seeking more concrete coding guidelines and practices for writing idiomatic Rust code.

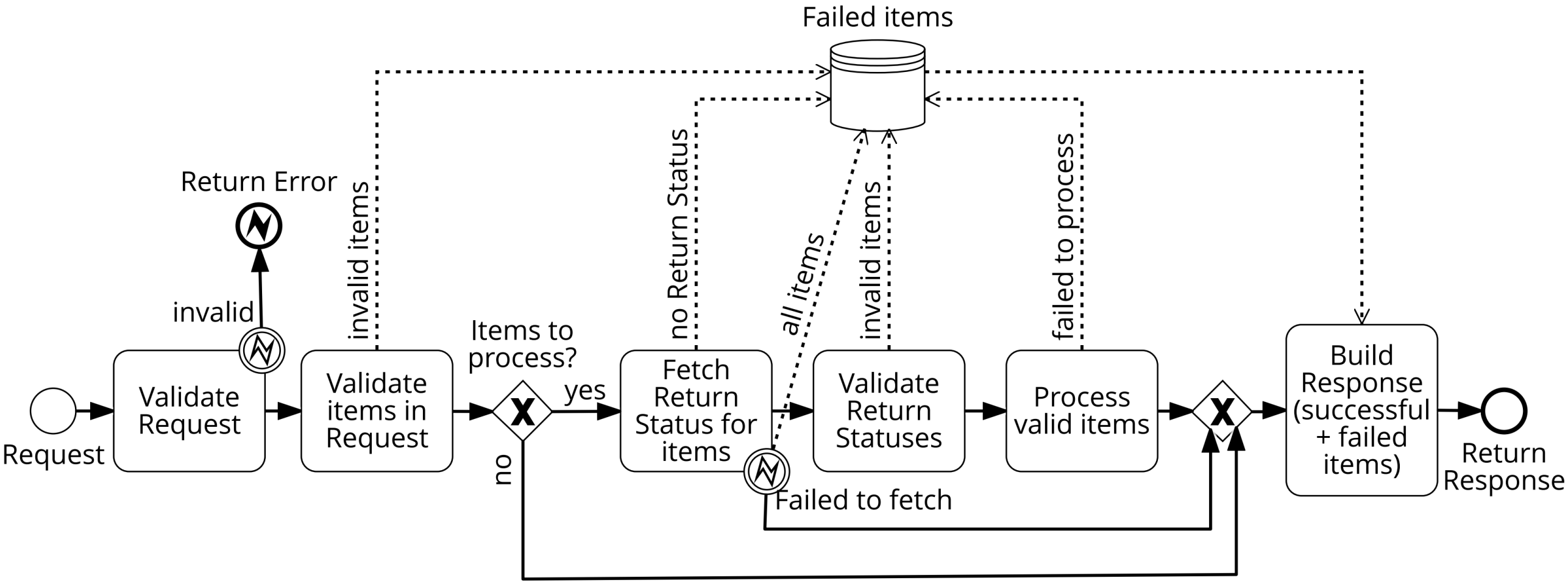

**Figure 4:** BPMN process model of the Boxfish Lambda Handler, selected as third use case, which is a sequence of operations that continuously append a shared data structure, read by the final operation

When a request triggers the lambda handler, the request first gets validated. If it is invalid, the handler directly returns an error. Otherwise, the items transmitted in the request are validated. Besides the valid items, the invalid items are also stored inside an array, depicted as "Failed items" data storage in Figure 4. If there are no valid items, the following processing steps are skipped and the final response is built, as seen at the long arrow pointing to the merging `XOR`-gate on the right of Figure 4. In case there are valid items, the return status for each item gets fetched. Again, if there are items that have no return status, they are added to the array of failed items. If the whole request fails, all items are considered as failed and the following steps until the final response building are skipped, jumping to the merging `XOR`-gate in Figure 4. This is followed by another validation, which validates the return statuses, with invalid items being added to the failed items array. The last processing step is to process the valid items. Similar to the other steps, items that couldn't be processed are added to the failed items array. Lastly, the successful items and all failed items from the array are used to build the final response, which is returned as response to the original request.

### 4.3.1 Relevance

In contrast to use cases 1 and 2, this use case covers all aspects from handling the request, validation, executing business logic, to responding. At team Boxfish, lambdas are a common way of serving endpoints with relatively simple logic, so that no advanced architecture like in the long-running services at FT9 needs to be used. Furthermore, a developer from team Boxfish stated in the expert interview that they are currently building a template for their lambdas, so that a reference example with clean code structure is needed [7]. This use case will be representative for the **Lambda Handlers** used at Boxfish.

### 4.3.2 Problems of the status quo

To begin with, the main function owns two mutable vectors `results` and `api_errors`, which are always passed to each function in the processing step as a mutable reference. In Rust, mutable references indicate that a function has side effects. Those side effects decrease **C6 Testability** [12], and especially the testing of the whole function gets difficult because the right interactions of these side effects need to be ensured [16]. Expert D adds that side effects make it harder to verify and test the logic in an isolated manner [7]. Another indicator for poor testability is the complete absence of any unit tests [12].

Furthermore, the shared mutable references are described as unidiomatic in Rust [16], [6]. They make it harder to follow the logic and reason about the state and correctness of the code at any given point [16]. This is because the passed mutable data needs to be tracked down to understand where it's modified. In addition, expert C expects them to cause errors, for example when concurrency is introduced, and the expert thinks they can be easily replaced by a different approach [6]. These arguments indicate that runtime behaviour will be unpredictable, which decreases both **C4 Analysability** and **C7 Faultlessness**.

Some of the functions have complex return types, with one of them returning a `Result` wrapped by another `Result`, which is noted negatively by expert a [12]. This impedes **C4 Analysability**. Another point are functions that accept too generic parameters, for example a whole Request or the whole `ProcessorDependencies` struct, although they only need parts of it. Expert B finds that this obscures what is actually required by the function [16], therefore contributing to poor analysability. It has also been identified that it is hard to follow which function can result in what kinds of error cases. Some error cases are only represented implicitly through states like empty arrays, for example.

Another code smell that affects **C4 Analysability** negatively are deeply nested control structures, like loops and if-conditions [6]. They are especially present in the function that builds the final response, which covers different error cases, making the logic harder to understand [6]. The code also consists of deep function call hierarchies that are hard to follow, resulting in poor traceability of the control flow [7].

It is also noted by an expert that too much logic is concentrated inside a single module [12], which signals bad **C2 Modularity**. Confirming this argument, expert C adds that business logic is mixed with technical logic such as I/O, which results in a decreased separation of concerns [6]. Tight coupling additionally decreases **C3 Reusability** [7].

Lastly, invalid state is representable in the code because the transitions between the phases are not encoded in the type system and therefore not enforced by the compiler. For example, validation is achieved through a separate function call, but there is no compile-time guarantee that the validation actually happened [7]. Similar to use case 1, a change by a developer, like changing the order the functions are called in, might introduce new bugs that stay unnoticed and are only noticed during runtime. This violates the criterion **C7 Faultlessness**. There might also be assumptions about the code not explicitly defined, which might break if changes are made in another code section, resulting in degraded **C5 Modifiability**.

### 4.3.3 Proposed solution

In summary, the main shortcomings of the current code that need to be solved are impure functions with side effects due to shared mutable state, missing separation of concerns like validation, processing and building the result, deep function call hierarchies, implicit states and errors, as well as invariants that are only represented through procedural calls rather than compile-time guarantees.

An architectural design principle that addresses multiple of the named issues is Functional Core - Imperative Shell (FC-IS), which was introduced first by Bernhardt in 2012 [19]. The principle aims at applying functional programming for domain logic, which can be encapsulated in pure functions without state or side effects. Their results are predictable and straightforward to unit-test. Everything necessary for interactions with external dependencies, such as handling requests, making database calls or I/O is dealt with by the imperative shell, which calls the functional core and is interchangeable. [20], [21]

Firstly, applying the FC-IS principle to this use case would eliminate side effects, addressing issues of **C6 Testability** and **C4 Analysability**. Secondly, it separates different concerns like validation and processing (functional core) from external calls like database calls (imperative shell), which improves **C4 Modularity** and potentially **C2 Reusability** of the functional core. Moreover, it replaces deep function hierarchies with a one-directional functional pipeline and clearly defines the inputs and outputs of each function, enhancing **C5 Analysability**. However, states would still be represented implicitly instead of explicitly through enum variants or result structs, which is compatible with the principle but not required by it. The invariants still wouldn't be enforced, allowing to skip or swap crucial operations, for example.

When using the typestate pattern, which was introduced in Chapter 4.1.3, states are represented through structs that contain the data necessary for each step. The resulting struct from each transition becomes the struct or enum that explicitly encodes the state. Many of the refactoring steps involved when applying FC-IS are also compatible with the typestate pattern, such as using pure functions (the state transitions) and replacing deep functional hierarchies with a functional pipeline. But additionally, the typestate pattern enforces invariants, thus making refactorings safer and shifting error detection to compile time, thus addressing **C5 Modifiability** and **C7 Faultlessness**. Thereby, the typestate pattern will be applied, while aiming for pure functions and encoding multiple possible resulting states in the type system in the form of enums.

# 5 Implementation

This section will explain how the proposed solutions from Chapter 4 have been realized in the code. Design decisions made in the refactoring process will be justified, and any tradeoffs or drawbacks will be documented. Selected code snippets will be used to point out some of these aspects. The refactored code can be found in Appendix A.1.2 in full length.

## 5.1 Use case 1

The logical steps from the status quo code, like getting benefits, filtering valid benefits, determining the Detailview benefit and checking the UP membership were extracted into separate state transitions. Each activity documented in the BPMN model in Figure 2 was encoded as state transition. Each transition results in a new struct, for example `DetailviewDetermined`, which then defines the next allowed operation, like `check_for_up_contracts`. This way, the logical order is enforced, so that it becomes impossible to swap operations like checking the UP membership and determining the activation status, or to leave out the activation of auto-activate benefits, for instance.

Rust

**start_state.rs**

```rust
1 async fn get_benefit_ids(variation_service: &VariationService, variation_id: VariationId) ->
  BenefitIdsFetched {
2     let benefit_ids = variation_service.get_benefit_ids_by_variation(&variation_id).await;
3     let span = Span::current();
4     span.record("benefit_ids_on_variation", debug(&benefit_ids));
5     BenefitIdsFetched { variation_id, benefit_ids, span }
6 }
7 struct BenefitIdsFetched {
8     variation_id: VariationId,
9     benefit_ids: Vec<BenefitId>,
10     span: Span,
11 }
12 impl BenefitIdsFetched {
13     async fn get_valid_benefits(self, detailview_cache: &'static DetailviewCache) ->
  ValidBenefitsFiltered { /* omitted */ }
14 }
```

**Listing 1:** First function that marks the entry point of the state machine. The function returns the first state of the state machine, which is a struct on which the next state transition is defined.

As the state machine needs an entry point, the `get_benefit_ids` function shown in Listing 1 is taken to return the first state `BenefitIdsFetched`. The resulting state contains immutable data, and the external dependency `variation_service` is passed into the function once, but not carried over to the next state. This clearly separates dependencies for each step and indicates which data is needed for the next step, as seen in ll.7-11 in Listing 1. The next state transition `get_valid_benefits` is then implemented on the struct `BenefitIdsFetched` and consumes `self` in l.13, which means that the old data cannot be used again, and the developer is forced to operate on the new resulting struct `ValidBenefitsFiltered`.

One drawback is that simple steps require a disproportionately high amount of boilerplate, as show in Listing 1. The original operation is only three lines long (ll.2-4), but now requires 11 lines due to the function signature and the new struct. However, for more complex processing steps, the amount of additional lines becomes marginal in relation to the logic itself, and the definition of the structs is purely declarative, which could possibly reduce the mental effort for understanding them.

Rust

**example_state.rs**

```rust
struct DetailviewDetermined {
    variation_id: VariationId,
    detailview: DetailviewBenefit,
    ...
}
impl DetailviewDetermined {
    fn check_for_up_contracts(self, ...) -> (UpContractTaskSpawned, JoinHandle<bool>) {
        let is_up_member_join_handle = /* omitted */;
        (UpContractTaskSpawned { /* omitted */ }, is_up_member_join_handle)
    }
}
```

**Listing 2:** An example state transition that not only returns the next state, but also a handle on the asynchronous background task as well. This shows how state transitions can produce outputs besides the next state itself.

Listing 2 shows a design choice that was made when applying the typestate pattern. The transition produces a `JoinHandle`, which encapsulates the eventual result of a background operation. In the status quo code, this allows the application to maintain non-blocking execution, performing other tasks while running the database access for the UP membership in the background. An `await` can be called on the `JoinHandle` later to retrieve the result, and wait longer in case the task is still running. This state transition doesn't produce new data directly, but it is a type-level guarantee that the background task has been started, and the following states can safely assume that the `JoinHandle` exists without knowing about it. A later state transition that requires the result of the `JoinHandle` can then require it as function argument.

All data required for constructing the `DetailviewViewModel` at the end is accumulated in the structs representing the states. On one side, each struct grows larger and more boilerplate code is needed for structs and function signatures. On the other side, no temporary variables for each transition's produced data are necessary at the invocation point, which makes invoking the state machine less complex.

The function `get_valid_benefits` couldn't be fully integrated into the state machine because it's used by another handler that has different states. While handlers generally shouldn't depend on other handlers, and the function should best be extracted into another common module, this also indicates a potential limit of the pattern. State transitions in the typestate pattern are tightly tied to their states, so reusing them in other state machines becomes difficult. As a result, some domain logic must remain outside of the state machine and be invoked from within state transitions, which preserves type safety, but limits full integration of shared functionality.

Changing the order of states or introducing a new state requires refactoring previous and subsequent states, as the data passed between consecutive states has to remain compatible. The consequences of this will be evaluated in Chapter 6.

## 5.2 Use case 2

Except for the Dreson validation logic, the code remains mostly unmodified. As the changes focus solely on the newtype pattern, the newtype `ValidatedDreson` has been created and used in the beginning of the handler method. The two displayed validation activities in the BPMN model in Figure 3, including their surrounding `XOR`-gates, have been replaced by the newtype.

Rust

**newtype.rs**

```rust
1 pub(super) struct ValidatedDreson(String);
2 impl ValidatedDreson {
3     pub fn from_string(dreson: String, exceptions: &[String]) -> anyhow::Result<Self> {
4         if is_valid_dreson(&dreson) || exceptions.contains(&dreson) {
5             Ok(ValidatedDreson(dreson))
6         } else {
7             Err(anyhow::anyhow!("Invalid dreson format"))
8         }
9     }
10 }
```

**Listing 3:** Newtype introduced for the data type Dreson. The newtype directly validates the input and constructing the newtype will fail if validation fails.

The first line of Listing 3 shows the newtype `ValidatedDreson`, which wraps a string and provides the `from_string` method in the implementation block in l.4, which allows constructing the newtype from a raw string, and accepts a list of exceptions as a second parameter. It was necessary to allow certain exceptions, as some consumers of the endpoint use special `Dreson` values for testing purposes. However, the need for these exceptions could also be eliminated in the future if consumers use formally valid values. Every string in the `exceptions` list will be treated as always-pass. Because the newtype exists within a separate `mod` (Rust module) and is not made public, the struct can only be created through the `from_string` method, making it impossible to otherwise construct a `ValidatedDreson`.

An alternative to allowing exceptions through an additional parameter in the current `from_string` method would be implementing the `TryFrom` trait. Through implementing the `TryFrom` trait, the code could potentially become slightly more idiomatic and practical. Since the trait doesn't allow a second input parameter, a static vector for the exceptions could be defined, for instance. On the other hand, this could be less reusable in case more exceptions are added later, which is why the `TryFrom` variant was not chosen.

For compatibility with subsequent processing steps, the wrapped string can be accessed through the `inner` method. This allows other dependencies or libraries that don't accept the `ValidatedDreson` newtype to be provided with the original string. However, the type safety gained through the newtype pattern will be lost once operating on the inner string. Another possible way to achieve this is by adding the derive macro `#[derive(Deref)]` to the newtype, which instructs the compiler to automatically deconstruct the type when a string is needed. However, compilation won't fail when the newtype is passed into a function that accepts the wrapped type, thereby places that require refactoring might be overlooked. For that reason, an explicit `inner` method was chosen.

To ensure compatibility with code that used a raw string previously, it is possible to implement corresponding traits, such as `PartialEq<str>`, on the newtype. For this use case, the method `contains` was additionally implemented. As a consequence, migrating to the newtype does not require access to the wrapped string.

A common practice in tests is the usage of dummy values to test edge cases of a system. To support dummy values, the method `new_unvalidated` has been introduced and scoped with the macro `#[cfg(test)]` to make it available to test code only. This method skips the validation and directly constructs a newtype.

Rust

**handler_call.rs**

```rust
1 let target_context_dreson = request_params
2     .target_context_dreson
3     .map(|dreson| ValidatedDreson::from_string(dreson, &[]))
4     .transpose()
5     .map_err(|_| StatusCode::BAD_REQUEST)?;
6 let current_context_dreson = request_params
7     .current_context_dreson
8     .map(|dreson| ValidatedDreson::from_string(dreson, &[String::from("(test.articlelist)")]))
9     .transpose()
10    .map_err(|_| StatusCode::BAD_REQUEST)?;
```

**Listing 4:** Usage of the newtype ValidatedDreson in the handler. The construction happens through a map call on the option value, and the error is extracted through transposing and mapping the error to a bad request response code.

Finally, the newtype's usage is displayed in Listing 4. The construction is done via a mapping in l.3 and l.8, respectively. In l.8, it can be seen how an exception is passed inside the construction method. For extracting the error from the `Result` type, which is wrapped in an `Option`, the method `transpose` is called to make `Result` wrap `Option` instead, allowing to handle the error through subsequent map calls and the `?` operator.

## 5.3 Use case 3

A main issue that has been identified in Chapter 4 to influence multiple criteria negatively, are the mutable vectors `results` and `api_errors`, which were eliminated in the refactoring. Instead of passing a shared mutable reference to all functions, which caused side effects in the status quo code, the results and errors are now accumulated in the states of the typestate pattern. All of the processing steps from Figure 4 that access the shared data storage are now encoded as states, which are each represented through structs with fields for result and error cases. Similar to the first use case, the typestate pattern was applied with structs containing data instead of a zero-sized type marker, so that the accumulation is possible without a shared mutable state.

Rust

**invocation.rs**

```rust
1 enum Either<L, R> { Left(L), Right(R) }
2 async fn handle_valid_request(valid_request: ValidAnnouncedPositionItemsRequest, ...) ->
  FinalResponse {
3     let fetch_return_statuses_result = valid_request
4         .validate_announced_items()
5         .fetch_return_statuses(...).await;
6     let items_with_return_statuses = match fetch_return_statuses_result {
7         /* omitted cases */ => items,
8         FetchReturnStatusResult::NothingToFetch(items) =>
9             return FinalResponse::from(Either::Left(items), ...),
10    };
11    let process_items_result = items_with_return_statuses.process(...).await;
12    FinalResponse::from(Either::Right(process_items_result), ...)
13 }
```

**Listing 5:** Generic type `Either` that can represent an outcome of two possible states, or a state that accepts two possible preceding states. The displayed function shows how the state machine is invoked, with functional pipelining in ll.4-5 and making use of the `Either` type.

As starting point, the code snippet in Listing 5 shows how the state machine is invoked by the handler. The function only accepts a `ValidAnnouncedPositionItemsRequest`, indicating that the state machine can only be entered once a valid request was received. On the valid request, subsequent operations are called in a functional pipeline in ll.4-5.

Furthermore, Listing 5 shows how the generic type `Either`, which can contain either one data type or another, is introduced. This design choice was inspired by the state machine pattern from Crichton's typed functional design patterns [Cri23]. In this example, either an early returned items array gets passed as left state (l.9), or the processed items as right state (l.12), both of which are accepted by the `FinalResponse` state. Nevertheless, it is also possible to represent two converging states as an enum sum type. An enum encodes the joining of two states more explicitly and is potentially more extensible for future changes, whereas the `Either` type is reusable for cases where other branches are joined, which an enum tied to the specific use case is not. Therefore, the `Either` type was chosen for minimal overhead and reusability.

Another issue of the status quo code were complex return types, like a `Result` wrapped by another `Result`, and implicit states like empty arrays instead of explicit encodings. Because enums can be used to return multiple resulting states of the state machine, it was decided to encode these implicit states and complicated return types in enums instead. While it is typical to represent errors through `Result` types in Rust, it was chosen to represent all variants in enums instead, so that different kinds of errors or success cases become more clear for the developer. Rust's exhaustive `match` requires all variants to be handled by default (except when using the _ operator), so states that were implicit before now need to be handled explicitly.

Rust

**enum_variants.rs**

```rust
struct ItemsWithReturnStatuses {
    valid_position_items: Vec<ValidAnnouncedPositionItem>,
    api_errors: Vec<ApiError>,
    return_statuses: Vec<ItemReturnStatus>
}
struct ItemsWithoutReturnStatuses {
    valid_position_items: Vec<ValidAnnouncedPositionItem>,
    api_errors: Vec<ApiError>,
}
enum FetchReturnStatusResult {
    NothingToFetch(ItemsWithoutReturnStatuses),
    Success(ItemsWithReturnStatuses),
    Failure(ItemsWithoutReturnStatuses),
}
```

**Listing 6:** An enum type that encodes various possible return states. Return states can contain different data, namely `ItemsWithReturnStatuses` and `ItemsWithoutReturnStatuses`. This serves as an example of how implicit application states can be made explicit and possibly handled separately later.

An example of this can be seen in Listing 6, where the `FetchReturnStatusResult` can have multiple resulting states, which can hold different kinds of data, namely items with a return status and items without a return status. These states can then be accepted by different subsequent states to ensure every operation leads from a valid state to the next valid state. For example, Figure 4 showed that a “Failed to fetch” error, represented as intermediate error event in the middle of the diagram, directly leads to building the response. In the refactored code, the type-level encoding guarantees that no further transition from that error state can be made, and that only the final error building function accepts this state. This has been achieved through accepting an `Either` type with the states `ItemsWithoutReturnStatuses` or `ProcessItemsResult`. They are passed into the construction method of `FinalResponse`, which encodes the identified error case based on the accumulated `results` and `api_errors` vectors.

In the status quo code, there was no distinction made between new items that were created during processing and old valid items. With the application of the typestate pattern, these two types have been separated to encode this semantic difference of the items in the type system. However, when integrating this with the final response logic from the status quo, the types must be converted back to match the existing logic. This issue could be addressed in the future, but would also involve changing the shape of the returned response.

During the application of the typestate pattern, the “Parse, don’t validate” principle has been applied as well to the `AnnouncedPositionItemRequest` struct. The logic of the separate validation method has been included in the construction method `try_from`, making the separate validation step obsolete. Additionally, the `validate_announced_items_request` function violated the SRP in the status quo, since it constructed http responses, which are the responsibility of the handler, and not of validation steps. The function now returns an enum that encodes the error state, which the handler can match on accordingly.

# 6 Results

In this chapter, the refactored code that was outlined in Chapter 5, will be assessed using the criteria listed in Table 1. Both a qualitative and quantitative analysis will be performed, using the methods defined in Table 2 in Chapter 3. The measurement process for the metrics will be described briefly. Their outcomes and a detailed evaluation of the conducted expert interviews will be presented. For each use case, an interpretation containing recommendations for the applied pattern will be made. Finally, the threats to validity of this thesis will be displayed.

## 6.1 Measurement

As explained in Chapter 3.2 previously, the criterion C1 measures time behaviour and will be evaluated through the microbenchmarking tool "criterion-rs". Criterion has been configured to run with:

- Sample size: 100
- Warmup time: 30 seconds
- Measurement time: 4 minutes

The full configuration used for running criterion can be found in Appendix A.1.5. For each use case, an existing happy-path integration test was copied and adapted. For external dependencies that are invoked in the code, such as databases and AWS services, local instances have been used based on the test setup present in each repository. The experiment setups that instantiate the testing data and environment for each use case can be found in Appendix A.1.5. The benchmarks have been run on a MacBook Pro (2021) with the following specifications:

- Apple M1 Pro chip (8-core with 6 performance cores and 2 efficiency cores; 14-core GPU)
- 32GB memory

All benchmark results, including the raw data and full reports generated by Criterion, can be found in Appendix A.1.6.

The other software metrics, which have been identified and assigned to appropriate criteria in Chapter 3.3, will be measured by using "rust-code-analysis", an open source static code analysis tool written in Rust [Ard+20]. It supports static analysis for various programming languages, including Rust. Currently, it supports 16 different source code metrics that can be computed across all supported languages. The exported results from the tool can be found in Appendix A.1.4.

## 6.2 Use case 1

This section presents and interprets the result of the refactored version of the first use case. Additional plots and data for the time behaviour measurement can be found in Appendix A.8.1. The static code analysis results can be found in in Appendix A.7.1 in Table 7 and Table 8 for the status quo and refactored code, respectively.

### 6.2.1 Time behaviour

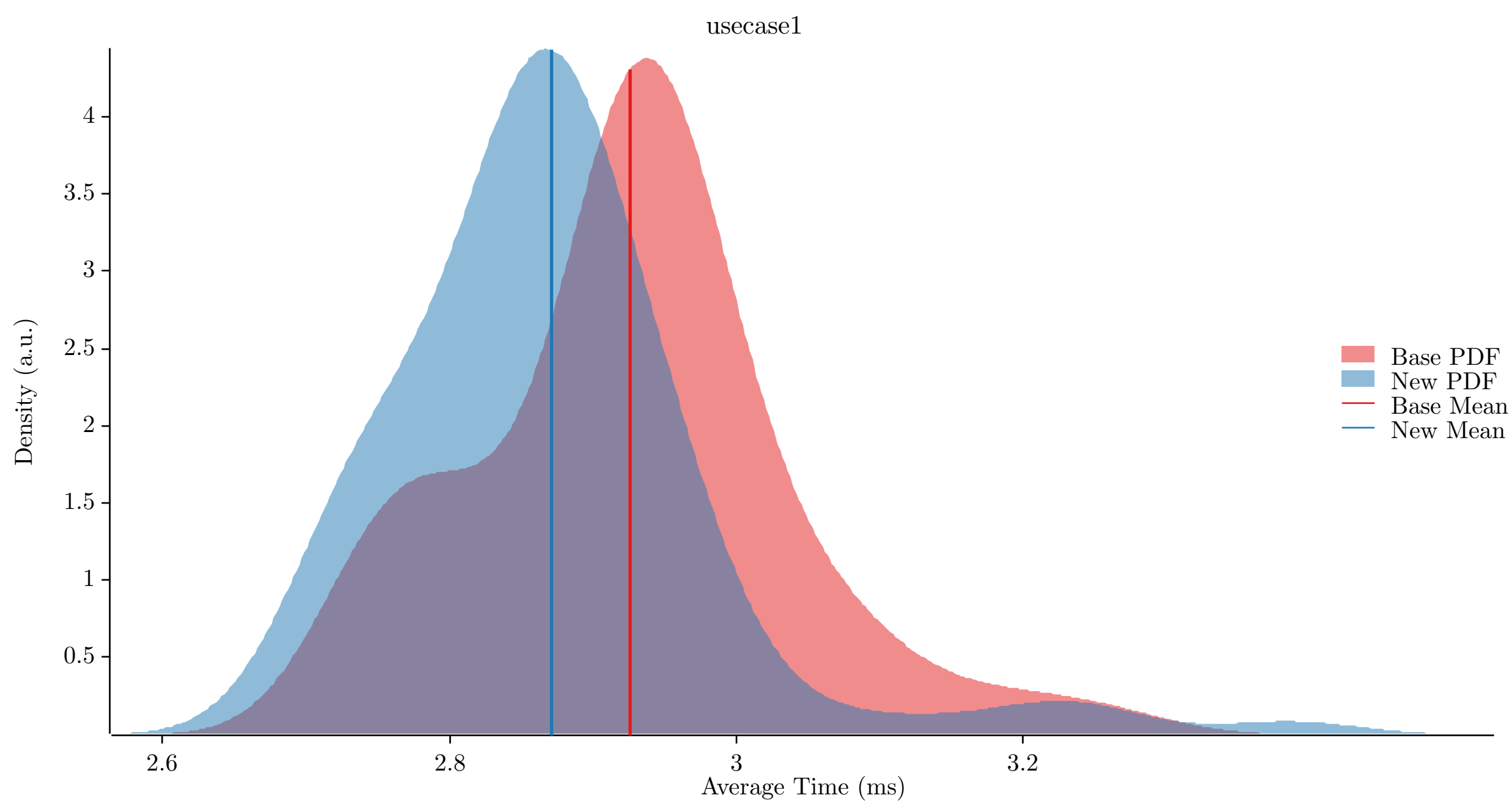

**Figure 5:** Compared Probability Density Functions (PDFs) of the execution time between status quo and refactored code for use case 1. The x-axis represents average time in ms, while the y-axis represents the density. The red and blue distributions belong to the status quo and refactored code, respectively. Each distribution's mean execution time is displayed in the plot. The distribution and mean time of the refactored code lie slightly more on the left.

| Metric | Estimate | 95% confidence interval |
|---|---|---|
| Mean | 2.871 ms | [2.85 ms, 2.894 ms] |
| Std. Dev. | 111.68 µs | [80.703 µs, 140.10 µs] |
| Change in time | −1.86% | [−2.856%, −0.76%] |
| $R^2$ | 0.874 | [0.869, 0.866] |

**Table 3:** Statistical evaluation of the benchmark for use case 1. The point estimates for each measure, as well as lower and upper bound of the 95% confidence interval are displayed.

As the plot in Figure 5 shows, there is only a small difference between the execution times. The refactoring is around 1.86% faster according to Table 3, which is statistically significant, but within the noise threshold of 2%. Because the experiment setup executed in the benchmark involves I/O operations, such as interactions with a locally running database or docker containers, a slight difference between each benchmark run is expected. Moreover, it is possible that the machine the benchmark was run on executed other background tasks, which can slightly interfere with the benchmark. The coefficient of determination $R^2$ is relatively high at 0.874, indicating that the linear regression applied fits the data points well. This can also be seen in detail in the regression plot in Figure 9, Appendix A.8.1. Because the dimension of the mean is milliseconds, the standard deviation of 111.68 $\mu s$ is low in comparison. It can be seen that some outliers are present on the right side of the distribution plot.

Overall, it can be said that there was no change in time behaviour beyond the noise threshold, and if at all a slight improvement. This indicates that the additional structural code for the states doesn't significantly impact runtime performance, which is due to the absence of logic or actual operations in the added code.

### 6.2.2 Modularity

The modularity of the refactored code is seen as neutral to slightly positive among the experts. Two experts observed increased modularity because each step is represented through an own state and state transition now, which separates the steps from one another and breaks the linear flow into smaller modules [12], [7]. On the other hand, expert B pointed out that the original code was not modular, and that the refactoring introduces significant boilerplate, while performing the same logic as before, which leaves the modularity same [16]. Because the states strictly depend on one another and changing one of them might require changing subsequent states, expert B has an impression of high coupling and low cohesion. The remaining expert has a neutral impression, stating that there are no major differences in modularity [6].

While the expert's judgements point to an overall slightly positive improvement, the counterpoint made by expert B about the tight coupling of the states [16] remains valid. According to expert C however, it is an inherent feature of the typestate pattern that the states rely on one another, and when a major change to one state is made, it is intended that other states require changes as well [6].

### 6.2.3 Reusability

Most experts agree that there is a slight improvement in reusability [12], [6], [7]. The steps from the former large function that have been extracted to state transitions are very specific for this use case, which lead to all experts' judgement that they cannot be directly reused for other use cases. However, splitting up the steps is seen as a possible starting point for later abstractions, and a definite improvement when compared to the old code [6], [7]. While expert B states that it is still hard to reuse due to a mix of technical and business related code in each function [16], two other experts now find it easier to extract the logic and identify parts that can potentially be reused [6], [7]. As stated by expert A, more generalization would possibly be needed when abstracting the states for reusage [12]

Furthermore, there is disagreement on whether the specificity of the states is positive or not. On one hand, reusing the code requires creating intermediate states if only a specific part of the logic is needed, which is seen as complex by expert B [16]. On the other hand, this enforces logical constraints and prevents misusing logic that doesn't suit other use cases [6]. According to expert C, if part of the code needs to be reused later in a different context, the business logic should be extracted further, which is seen as an intended feature of the typestate pattern [6].

Overall, the refactoring has been limited to one specific use case. For covering more use cases and abstracting state machines that are applicable to multiple use cases, more use cases need to be considered from the beginning, as suggested by expert B [16]. Another expert also describes it as difficult to separate what belongs to the state machine and what not, suggesting that external calls such as database and other services should remain outside of the state machine, while only processing and validation are extracted via the typestate pattern [7]. Moreover, the expert D describes it as a general difficulty to reuse code produced by the pattern due to its rigidness [7]. If two use cases require just a slightly different data type or detail, the whole code might need to be copied and all states adapted.

### 6.2.4 Analysability

The criterion readability has seen major disagreement among experts. On one hand, two experts dislike the naming of states and functions, stating that it is inconsistent and doesn't always make sense from a domain logic perspective, leading to possible confusion [12], [16]. However, this is not an issue caused by the pattern itself, but rather an implementation choice made in this study. Naming can be changed easily later, and expert D even describes the naming as positive, stating that it closely resembles the business functions [7]. This indicates that naming is rather subjective and unrelated to the pattern itself. Therefore, naming will not be considered for the results.

A feature that has been criticized sharply is the increased number of lines, described by expert B as boilerplate code that is only used to pass parameters along and hides business logic [16]. While the expert argues that the code now takes longer to scan and is less readable, other experts don't see this as a problem [6], [7]. Expert D sees boilerplate as small disturbance in relation to the long-run benefit it provides for understanding the code [7]. Moreover, both experts C and D agree that the additional code, consisting mainly of structs and function signatures, is purely descriptive and therefore neither complex, nor hard to understand [6], [7]. However, they both disagree regarding navigating the code: expert C finds it easy to navigate the structs in modern IDEs [6], while expert D finds navigating the split-up code more difficult, even though understanding the code on a higher level is easier now [7]. A possible solution for the boilerplate could be macro rules or existing libraries, as suggested later in the interview with expert C [6].

Furthermore, the pattern forces the developer to find suitable names for each state and transition, which is seen as a major advantage by two experts [6], [7]. They agree that this is a feature of the pattern, and expert C particularly sees the advantage of identifying issues in the domain model that were hidden earlier, which now need to be discussed with business stakeholders to model the steps explicitly and correctly [6]. The expert also stated that while the pattern might look unconventional for developers with a Kotlin or Java background, understanding the domain logic becomes easier because of the explicit flow. Expert C furthermore points out that learning the pattern will be a challenge due to the unfamiliarity [6]. Additionally, the location where errors occur and which data is needed for each step are now seen as more transparent by expert D [7].

Pipelining the functions as a result of the refactoring makes it easier to understand the code for two experts [12], [7]. To further enhance analysability, it is suggested by experts A and C that each state and their transitions could be extracted to separate files [12], [6].

The highest CoC in the status quo code was present in the method `create_detailview_view_model`. As seen in Table 7 and Table 8, its value of 8 has decreased to 0 in the refactoring, indicating that the refactored version imposes significantly less cognitive effort. In the refactored code, the highest CoC value of 4 was measured in the state transition `determine_activation_status`. Since no function has a value above the defined threshold of 15, and the maximum value has decreased to 4, the refactored code is regarded as an improvement.

As for HDiff, the method `create_detailview_view_model` had the highest value of 32.22 before. The same method's HDiff has decreased to less than half in the refactored code, now being at 15.11. Besides that, some state transitions like `determine_detailview_benefit` and `determine_activation_status` in the refactoring have been measured at 24.76 and 22.5, respectively. These are also the functions with the highest cognitive complexities, indicating that they simply contain more logic. It is not possible to draw a concrete conclusion for them, since no comparable functions existed before.

Both metrics don't reflect the experts' comments about boilerplate of the structs and only allow isolated assessments of each method. As shown previously, the function pipelining in `create_detailview_view_model` is seen as easier to understand, reflected in the lower CoC and HDiff for that function. Overall, the metrics reflect that the difficulty of understanding is now distributed over multiple functions, making each unit easier to understand.

### 6.2.5 Modifiability

Most experts state that certain aspects of modifiability have improved after the refactoring [16], [6], [7]. This can be attributed mainly to smaller and and more manageable functions [16] and faulty changes now being caught at compile time [6], [7], which makes it easier to perform changes without introducing new bugs. According to two of the experts [16], [7], the changes are now isolated within the modules, so it's less likely to break other states when changing one of the states, and the clearly defined input and return types give developers a better overview when implementing changes. Moreover, the increased readability of the code is assumed to make developers more confident about new changes, easing modifiability [6].

On the contrary, expert A finds it more complicated to make changes now, since one change to a function signature or struct might require changing all related states across the typestate pattern as well. Thus, changes like refactoring are now seen as more complex, while other changes like modifying or sorting are still easy to implement [12]. However, this barrier when making changes is seen as positive by the other experts [16], [6], [7], because it forces the developer to carefully plan their changes and think about their correctness, which is more time-intensive but still seen as positive.

When the logic doesn't fit the existing state machine, it signals that something in the domain logic is likely incorrect, and that a clarification together with business stakeholders is necessary [6]. Expert D added that the extra effort necessary for changes is outweighed by the other benefits, but if the original code would have already been split up, the benefit of the pattern would have been smaller [7].

In general, it must be ensured that the pattern is understood and applied correctly by all team members. Developers can still write poor quality code, and just like using `unsafe` in Rust, there are ways to bypass the existing rules [12], [6].

The MI measured for the status quo version of the `create_detailview_view_model` function indicates poor maintainability with only $50.20 < 65$. For the refactored code, it has increased to $65 < 83.67 < 85$, indicating moderate maintainability. Otherwise, only two functions have an MI of under 85 in the refactored code, with no function being difficult to maintain. This could possibly be explained because the functions with the lowest MIs contain the most logic. Generally, the metric reflects what experts noted about changes being easier due to smaller, more manageable functions. The noted complexity in modifying due to the need of adapting multiple states cannot be reflected clearly by the metric. Overall, the MI measurements indicate an improved modifiability of the refactored code, with room for improvement for some larger functions.

### 6.2.6 Testability

All experts agree that the testability of the refactored code is higher. Especially, the changes have enabled higher test coverage, with each test needing less setup like mocks and less verifications for different conditions [12], [16], [7]. Multiple experts report that the state transitions isolate the logic, so that unit tests only focused on that isolated part are possible, when compared to the massive function from before [12], [16], [7]. Moreover, the tests now only require mocks for each specific state, whereas before all dependencies needed to be mocked, even if the logic that was tested didn't require that dependency [16], [7]. Expert A supposes that fewer test cases are now necessary in total, which is supported by expert C's statement that the invariants are now part of the type system, so that developers can rely on it rather than having to test these invariants [6].

While testability is already seen as positive compared to the old code, expert B assumes that it could still be improved in general [16]. Supporting this, the testability could further be enhanced by only extracting the functional core, representing the business logic in the state machine, while leaving side effects and other dependencies like databases or services in the imperative shell [6]. The concept FC-IS has previously been explained in detail in Chapter 4.3.3. According to expert C, the current refactoring limits the potential benefits of the pattern [6].

In the status quo code, the CyC of the function `create_detailview_view_model` was at 17 and therefore exceeded the defined threshold of 10. In the refactoring, the same top-level function only has a CyC of 3, indicating that it is now easier to test. This is reflected by the experts' stating that the broken down, isolated functions are easier to test. Most functions in the refactoring have a CyC between 1 and 4. There is only one outlier function `determine_detailview_benefit`, which also has the highest other metrics, indicating that this function could be broken down further. All in all, since there are no threshold validations anymore, the refactored code is regarded as more testable.

### 6.2.7 Faultlessness

The last criterion faultlessness is seen neutral by some and significantly better by other experts. Two of the experts see no significant changes in faultlessness [12], [16], with one of them indicating that the pattern just split up the original logic in multiple functions without additional type-level constraints [16]. Although the types restrict the possible breaking changes, expert A is of the opinion that there is no guarantee that the developer won't bypass them [12].

In contrast, the remaining experts see the faultlessness as significantly improved, since the risk of accidentally introducing bugs is reduced and correctness guaranteed by the compiler [6], [7]. The execution order is now strict and only the correct set of data can be used at each step [7], which makes invalid state unrepresentable [6]. This is also ensured by the fact that each state transition consumes the state, which means that it takes ownership, so that the Rust compiler doesn't allow calling the same state transition again or out of sequence. Additionally, expert C thinks that the correctness of the program could now theoretically be proven in a formal way [6].

Despite that, expert B thinks that the typestate pattern would have been more suitable for cases with branching logic, and that this specific use case's linear flow doesn't leverage the typestate pattern's full potential [16]. Other use cases, such as a builder with different possible subtypes, could benefit more from the pattern.

### 6.2.8 Interpretation

Overall, the typestate pattern leaves runtime performance unaffected. Complexity metrics indicate that complexity was redistributed from a single, large function to smaller units, which aligns with the experts' judgments on improved analysability. Small state transitions make isolated unit testing possible, and less mocks are required since each transition only accepts selected dependencies. In comparison, the logic was concentrated to a large function before, where more mocks were required for all dependencies. The execution order and correctness of data are enforced by the type system, which reduces the risk of invalid states during runtime and shifts error detection to compile time to a certain degree. Smaller functions make changes easier to localize, but the strictly encoded invariants and states require more effort when implementing changes and make reusability hard. Modularity is only improved marginally, since the main function wasn't modular before, but the extracted states are coupled tightly and depend on each other. While the typestate pattern mainly offers benefits in correctness and testability, a significant amount of boilerplate code is introduced, causing differing subjective views on analysability.

The benefits of the typestate pattern could be possibly higher when

- refactoring multiple code sections that have similar logic and enable reusing,
- complex branching logic is present instead of a linear flow, affecting a higher number of invariants,

- state transitions are kept functional only without accessing external dependencies like databases (FC-IS principle),
- and state transitions are extracted into separate modules.

## 6.3 Use case 2

This section presents and interprets the result of the refactored version of the second use case. Additional plots and data for the time behaviour measurement can be found in Appendix A.8.2. The static code analysis results can be found in in Appendix A.7.2 in Table 9 and Table 10 for the status quo and refactored code, respectively.

### 6.3.1 Time behaviour

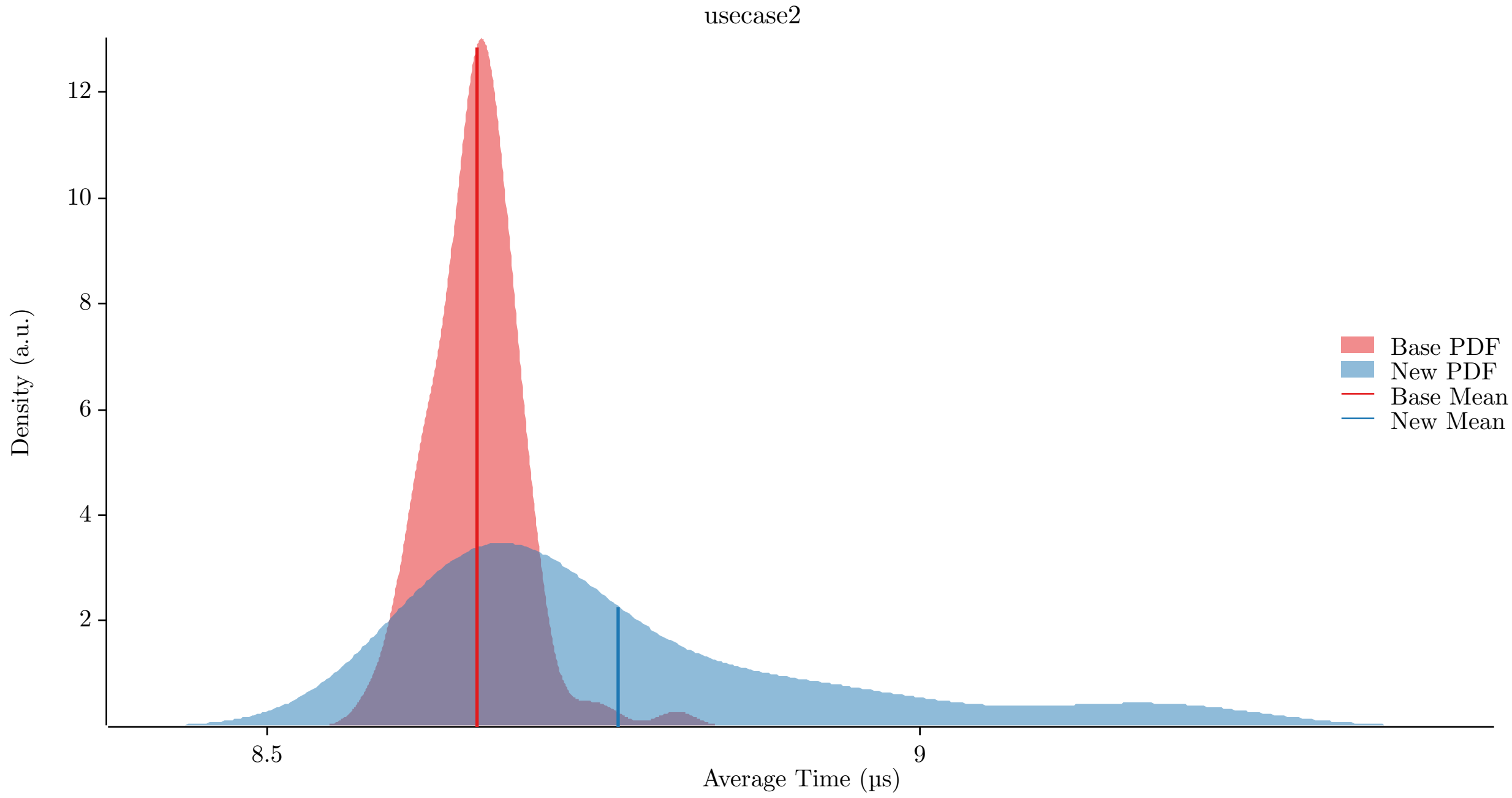

**Figure 6:** Compared PDFs of the execution time between status quo and refactored code for use case 2. The x-axis represents average time in ms, while the y-axis represents the density. The red and blue distributions belong to the status quo and refactored code, respectively. Each distribution's mean execution time is displayed in the plot. The distribution of the status quo code is more concentrated, with a significantly higher peak that lies left of the refactored code's lower peak. The refactoring's distribution is much flatter and stretched out.

| Metric | Estimate | 95% confidence interval |
|---|---|---|
| Mean | 8.769 µs | [8.739 µs, 8.802 µs] |
| Std. Dev. | 161.20 ns | [130.17 ns, 186.87 ns] |
| Change in time | +1.252% | [+0.905%, +1.653%] |
| $R^2$ | 0.964 | [0.962, 0.962] |

**Table 4:** Statistical evaluation of the benchmark for use case 2. The point estimates for each measure, as well as lower and upper bound of the 95% confidence interval are displayed.

For the second use case, the distributions look significantly different in shape, as seen in Figure 6. While the status quo code had highly concentrated execution times and has one large peak, the execution times of the refactored code are less concentrated and the distribution is long and flat towards the right side. The average execution time of the refactored code, which is displayed in Table 4, increased by 1.252%. While this indicates a slightly worse performance, the numbers are within the noise threshold of 2%. Thereby, the difference can be attributed to background tasks or overhead through test setup, similar to use case 1. The data points are more linear than in use case 1, as the $R^2$ value of 0.964 indicates. It should be noted that according to Table 4, the lower and upper bound indicate an $R^2$ value of around 0.962, indicating that the point estimate is not fully accurate. The mean execution time is only 8.769 $\mu s$, indicating that this use case involves fewer overhead and I/O operations than use case 1, which had an average execution time of 2.871 ms. Moreover, the small standard deviation of only 161.2 ns indicates relatively high concentration of data around the estimated average.

In summary, it can be said that while a statistically significant difference has been measured, it is not significant enough to exclude system noise influencing the benchmark. Therefore, the refactored code is seen as having no negative impact on time behaviour.

### 6.3.2 Modularity

Modularity is seen between neutral and improved by the experts. While expert A sees no significant change in modularity [12], expert C sees the refactored code perfect in terms of modularity, since the introduced newtype is the strongest form of isolation [6]. An improvement is also seen by expert B, who argues that the validation logic is now decoupled from the handler and encapsulated within the newtype [16]. The newtype serves as an interface, so that the handler is independent from the underlying implementation, which can be modified independently now. An improvement is seen by expert D as well, but when compared to the status quo, they classify the change as less significant than for the previous use case [7].

### 6.3.3 Reusability

All experts agree that the refactored code is more reusable [12], [16], [6], [7]. Expert A argues that the code is already being reused within the same place, which indicates good reusability [12]. Another expert states that the validation logic, which is now centralized in the newtype, can theoretically be used in other handlers as well [16]. The sharpest contrast is seen by expert D, who argues that the validation logic went from not reusable at all to fully reusable, and agrees that other handlers can easily reuse the newtype without having to copy-paste the validation logic [7].

### 6.3.4 Analysability

The analysability is seen positive in general. According to expert A, it's easier to read the code because the developer can ignore the validation logic and rely on the correctness of the newtype while examining other parts of the code [12]. Another expert added that downstream functions are also easier to understand now, since they rely on the newtype and allow the developer to assume it is valid [6]. Furthermore, expert C states that the newtype acts as a hint, making it clear that the developer is not dealing with any kind of string, but with a Dreson [6]. The code is also seen as more concise and clearer now than before, according to expert B [16].

Regarding the code section where the newtype is constructed, experts disagree on whether it is more readable. One expert is worried that developers who are unfamiliar with Rust might have difficulties understanding the chaining of `map`, `transpose` and `map_err` with the operator `|_|` inside the lambda [12]. On the other hand, this is something expert D points out as particularly helpful, as it makes the code easier to understand for seasoned Rust developers, when compared to the nested `let` statements from before [7]. However, this aspect will not be considered in the evaluation of this refactoring, as it is described more of a coding style than related to the pattern itself by expert B [16].

There have also been suggestions for further improvement of the refactoring. Two experts suggest that the struct should be initialized even earlier inside the `ValidatedParams` struct instead, which guarantees that the function is only called when the newtype has already been constructed successfully [16], [6]. It was also suggested to implement the trait `Deref` instead of providing an `inner()` function by expert B [16], and the `TryFrom` trait to make construction easier and more idiomatic by expert C [6]. Expert D finds the list of exceptions passed inside the `from_string` method confusing, because the exceptions are now less clear, and switching to a `TryFrom` implementation is hindered [7]. A minor critique is to rename the newtype from `ValidatedDreson` to `Dreson`, as it should be inherent to the type that it is valid once it was constructed [6].

Especially the CoC has sharply reduced in the `handle_request` method: from 17 before down to 8 in the refactored version, which is less than half before, as seen when comparing Table 9 and Table 10. As a result, the threshold of 15 is no longer violated. Likely, this is due to the higher penalties for nesting that CoC imposes, so the two validation steps had a high impact on CoC. Another factor is the functional chaining which is now used and doesn't add additional nesting. However, this has already been pointed out as a coding style rather than part of the pattern itself, according to the previously summarized expert opinions. The `from_string` method has a CoC of 3, but is now reused two times, so the complexity for understanding that part in isolation is relatively low. The other trivial functions defined on the `ValidatedDreson` have a CoC of 0 and are negligible. Overall, the fact that no functions violate the threshold of 15 anymore can be seen as an improvement in analysability.

HDiff of the `handle_request` function has slightly decreased from 31.07 to 27.93, indicating that the function is easier to understand now. The `from_string` method on its own has a Difficulty of 10, which is only a third compared to the previous HDiff of the handle method. The other submethods have negligible HDiff values between 2.67 and 6.75. This also indicates an overall improvement in analysability.

### 6.3.5 Modifiability

All experts agree that modifiability of the code has improved [12], [16], [6], [7]. It is noted by two experts that changes, particularly to the validation logic, are now easier, because the logic is contained inside a smaller unit [16], [7]. In comparison, there were multiple locations that needed to be adapted before, potentially introducing errors if changes are not made synchronously in all places [7]. Expert A sees the modifiability more nuanced, depending on what changes are made [12]. If the validation logic is changed, the developer needs to understand the implications on all usages, making changes more complex. Other modifications to the newtype like error handling are now seen as easier by the expert. Moreover, it is now no longer possible to accidentally switch up string arguments, for example, and no wrong value can be accidentally passed to functions that accept the newtype [6]. Expert C also mentions it's harder to introduce mistakes, because they are caught by the type system instead of unit tests [6]. According to expert B, the modifiability for the handler itself hasn't changed much [16].

When migrating the codebase towards the newtype, the compiler will indicate where more attention is needed, imposing no major difficulty when making modifications according to two experts [12], [6]. One of the experts doesn't see it as a disadvantage that adapting code that uses the newtype requires more initial effort, because it makes the code more explicit and readable later [6].

The MI of the status quo `handle_request` function was $52.61 < 65$, indicating that it was difficult to maintain. However, the metric value has only gone up to 54.89, a neglegible improvement that still indicates poor maintainability. All newly introduced methods have a MI of over 110, which is well above the threshold of 85, so they are considered highly maintainable. Overall, this metric doesn't quite reflect the statements by the experts that the handle function becomes easier to modify. However, it confirms that the validation logic itself is easy to modify. Moreover, it reflects expert B's view that the handler's modifiability itself didn't change significantly.

### 6.3.6 Testability

Every expert agrees that the testability of the validation logic itself, now part of the newtype, has improved, because it can be tested in isolation [12], [16], [6], [7]. The validation logic can now be tested without needing a setup for the whole handler function, requiring less mocks for the dependencies [7].

On one side, some experts agree that the business logic is now easier to test because the validation logic doesn't need to be tested again [12], [6]. Validation can be tested separately and only once, while its usages don't need to be covered again. On the other side, this is seen critical by expert B, who argues that the handler function still needs to test the validation separately. Otherwise, wrong modifications like removing the newtype usage won't be noticed through tests [16]. As a result, the expert sees no change in the testability of the handler itself, only of the newtype.

The function `new_unvalidated` was introduced to ease test setup, which is pointed out positively by expert D [7]. On the contrary, another expert sees this as a code smell, which should be avoided, and suggests the use of the `Default` trait instead [6].

It is noted by expert B that modifiability and testability might correlate in this use case [16]. A function that is easy to modify is also easy to test.

The CyC metric before and after the refactoring is exactly the same at 16, both violating the threshold of 10. The newly introduced method `from_string` has a CyC of 3, indicating that the newtype logic is easily testable. The handler itself is still difficult to test, reflecting what expert B already pointed out. Even though the validation logic is now in another module, the handler still needs to ensure that path is covered. However, it doesn't reflect that less tests are required for the handle method itself, as some experts stated. Overall, the metric therefore indicates no significant improvement or degradation in testability.

### 6.3.7 Faultlessness

In terms of faultlessness, multiple improvements are noted by the experts. First of all, other functions being called by the handler now accept a newtype, which means the compiler guarantees that any data reaching them is valid [16], [7]. Verifying once at the beginning prevents errors for the entire system [7], and any failure is detected at the entry point [6], [7]. This way, runtime errors are prevented [16], and invalid state cannot occur anymore [6]. Validation deep inside the function call hierarchy becomes obsolete, because the type system already enforces correctness [6]. Arguments also cannot be accidentally passed to another function in the wrong order, because the ValidatedDreson cannot be confused with a regular string, according to expert C [6].

The remaining expert A sees similar benefits in theory, but practically sees less changes to faultlessness, as the match statement from before already covered all cases. It was already assumed before that the Dreson is verified at the beginning, so safety was already assumed for downstream functions before, and the type is converted back to a regular string at the end of all processing, making the change less relevant to this business case in the eyes of expert A. [12]

### 6.3.8 General

During the interview with expert C, a few improvement suggestions were made to enhance the newtype itself and make it ready for being integrated across other code sections. Firstly, the traits `Deref`, `Into`, `TryFrom` and `TryInto` are recommended to implement or derive for the newtype, which makes construction and type conversion easier, more idiomatic and eases integration in other places. This could potentially make calls to functions like `transpose` obsolete during construction. Generally, newtypes have already been used in other places of the codebase, and the newtype `PSR`, which is semantically very similar to a `Dreson`, could be combined in the future. [6]

### 6.3.9 Interpretation

The newtype pattern causes no practical degradation in runtime performance. While the effects on the whole handler function itself were smaller, the validation logic itself clearly improved in most regards. The centralized validation logic decouples it from the handler, and therefore makes it more reusable and modular. Code duplication is reduced, and other code sections can reuse the standardized domain type. Furthermore, modifying the validation logic can be localized easier, and modifying the handler is safer due to compile-time guarantees regarding the state of the newtype. The validation logic itself is easier to test in isolation without mocking, while testability of the original function is unchanged. Lastly, the faultlessness is increased through guaranteeing the validation of the type after construction, which prevents invalid states from reaching downstream functions at compile time. However, newtypes only improve faultlessness if adoption is enforced consistently at each function, otherwise developers can bypass it.

Overall, the newtype pattern combined with the principle “Parse, don’t validate” offers many advantages at relatively small changes. It is therefore recommended to use for primitive data that underlies invariants, such as validation. When implementing a newtype, it is recommended to:

- prefer implementing Rust’s traits like `TryFrom` for construction rather than a manual function,
- use traits such as `Deref`, `Into` or `TryInto` to ease adoption of the newtype,
- don’t include prefixes like “validated” in the newtype’s name when they are inherent to the newtype,
- and use the newtype as early as possible in the function hierarchy to maximize the amount of code benefiting from the compile-time guarantees.

## 6.4 Use case 3

This section presents and interprets the result of the refactored version of the third use case. Additional plots and data for the time behaviour measurement can be found in Appendix A.8.3. The static code analysis results can be found in in Appendix A.7.3 in Table 11 and Table 12 for the status quo and refactored code, respectively.

### 6.4.1 Time behaviour

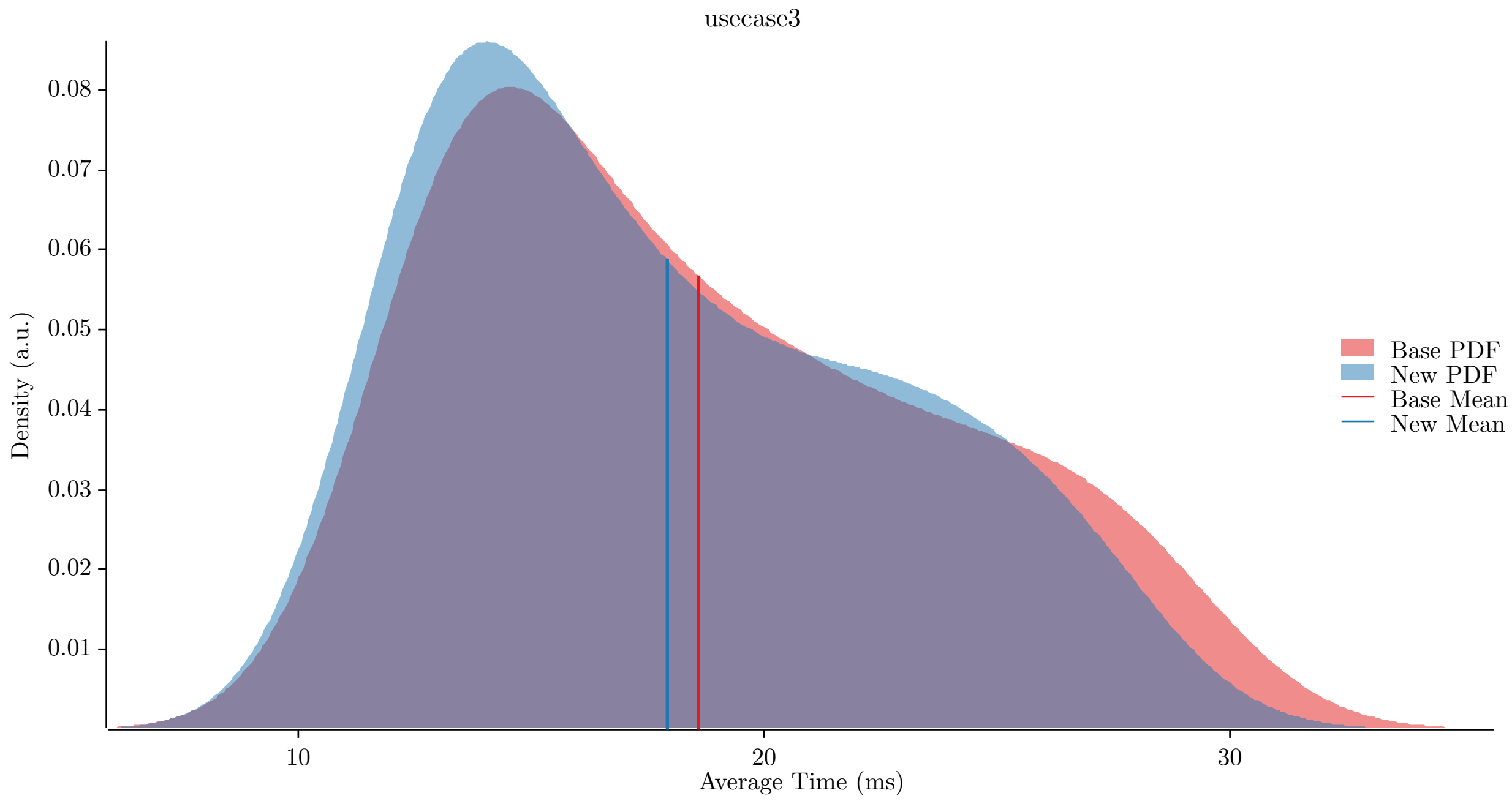

**Figure 7:** Compared PDFs of the execution time between status quo and refactored code for use case 1. The x-axis represents average time in ms, while the y-axis represents the density. The red and blue distributions belong to the status quo and refactored code, respectively. Each distribution's mean execution time is displayed in the plot. Both distributions have a very similar shape, being centered around their mean and having their peak left of the mean execution time.

| Metric | Estimate | 95% confidence interval |
|---|---|---|
| Mean | 17.926 ms | [17.015 ms, 18.858 ms] |
| Std. Dev. | 4.723 ms | [4.203 ms, 5.147 ms] |
| Change in time | −3.692% | [−10.668%, +3.876%] |
| $R^2$ | 0.368 | [0.353, 0.357] |

**Table 5:** Statistical evaluation of the benchmark for use case 3. The point estimates for each measure, as well as lower and upper bound of the 95% confidence interval are displayed. The $R^2$ value of 0.368 indicates a low correlation of the regression model with the actual data. The standard deviation is relatively high compared to the mean.

As the plot in Figure 7 indicates, the distribution of the status quo and refactored code largely overlaps. Although the execution time according to Table 5 has improved by 3.629%, the tool Criterion reports no statistically significant change. Although the change in time is higher than for the other use cases and higher than the noise threshold, the confidence interval is noticeably bigger, and the $R^2$ coefficient is only 0.368, indicating that the linear regression model doesn't fit the benchmarked data points. This can also be seen in the regression plot in Figure 13, Appendix A.8.3, which shows a non-linear trend, possibly pointing towards a superlinear increase. The increasing execution time can probably be attributed to an accumulation of data in the docker containers running in the background. For a future test setup, a cleanup after each sample could be considered, which would in turn increase benchmark execution time throughout all runs evenly.

As a result of the low fit of the linear regression model, the estimated change in time is assumed to be inaccurate. The mean execution time of 17.926 ms is the highest among the use cases, which was expected because the scope of the refactoring is significantly larger, covering more code in the benchmark. The standard deviation of 4.723 ms is high, but also possibly inaccurate due to the low $R^2$ coefficient. Overall, due to the statistically insignificant outcome and the almost full overlap of the PDFs in Figure 7, it is assumed that the refactoring did not impact time behaviour significantly.

### 6.4.2 Modularity

Regarding modularity of the third use case, expert A reports that the code has become more modular because the handle function is now only responsible for setup and delegates core logic to the states and transitions [12]. Expert D agrees that the modularity has increased because the long control flow has been broken down into individual isolated states, which the expert finds similar to the improvements of the first use case [7]. These arguments are supported by expert C, who states that the elimination of shared mutable state makes each state affect other states less, resulting in slightly improved modularity [6]. However, the expert doesn't see modularity as the main improvement in this use case, and the code was not modular to begin with.

On the contrary, expert B sees the refactored code as neither better, nor worse [16]. The expert argues that changes to one state still affect multiple other states, so they are not completely isolated from one another, which is not seen as modular.

### 6.4.3 Reusability

In terms of reusability, most experts didn't see major improvements [12], [16], [6]. Firstly, the status quo was not very reusable, and the refactoring didn't change that fact significantly [12]. Secondly, the reusability is limited to this use case, since the states cannot be reused outside of the state machine due to their tight coupling to their preceding and succeeding states [16].

Another expert sees potential for reusing the code because it is more contained, but the “boilerplate” that comes from the states themselves makes it harder to reuse the logic directly [7]. Future potential is seen by expert C, who argues that the logic is now more explicit, making it easier to identify potential for reusage or adapt similar logic elsewhere [6]. Moreover, expert B states that the resulting states from each function could be destructured into its values, but classifies it as an unusual and unlikely way of using the pattern [16].

### 6.4.4 Analysability

Analysability is seen as improved by all experts. The fact that the code is now broken down into smaller parts that can be understood in isolation makes reading easier, as developers can focus on one part, without needing to understand all other aspects of the code [12], [7]. Expert B adds that the logic is now split into distinct phases, such as validation and processing, making it easier to comprehend [16]. Moreover, each section is now self-contained according to expert B, which confirms that they can be understood in isolation. The removal of shared mutable state, which has already been mentioned for modularity, is also seen as beneficial for analysability [16], [6]. This is probably because developers don't need to reason about side effects when focusing on a specific code section.

Another advantage of the refactored code is the explicit encoding of the state of the code, which helps the developer reason about where in the code they are [6], and provides a clear overview of edge cases [7]. According to expert B, encoding different results in enums gives developers the chance to trace whether the logic is correct [16]. Particularly the final result enum is pointed out by expert D, who finds it helpful for getting an overview of all possible outcomes, which existed only implicitly before [7]. Expert D adds that `match` statements in Rust are exhaustive by default, which means every enum variant needs to be handled, unless the `_` (underscore) operator is used explicitly.

However, the enums used for modeling success and error cases are seen critical by two experts. One of them finds it unusual to have success and error cases in one enum, as it would be more idiomatic to use `Result` types in Rust [12]. Expert D agrees with this and adds that error handling with the `Result` type would be more convenient, as other features such as the `?` (question mark) operator would normally be used, and without the `Result` type, scanning the code becomes more difficult for developers used to Rust [7]. Instead of the current refactoring, the expert suggests only modeling recoverable, expected exceptions that are part of the business logic, such as an empty array or a wrong property, inside the enums. Non-recoverable exceptions, which are usually technical exceptions like a crashed database or unavailable service, should be returned as an error inside a `Result` return type.

Furthermore, an effect of the pattern is that it forces the developer to find a name for each state and transition. If the naming is done correctly, this is seen as an advantage for readability [16], [7]. Due to the naming, expert A finds the code more traceable [12]. One negative aspect of the naming is that the states often end in "…Result", which is ambiguous with the Rust built-in type `Result`, as pointed out by expert C [6].

The `Either<Left, Right>` type, which was used to model two possible paths in the state machine that come together again later, is criticized by two experts. Expert A worries that it might be unfamiliar for new developers, who would need a certain level of experience to understand the construct [12]. Furthermore, expert D finds it more difficult to understand, since it's not obvious which type would be `Left` or `Right` [7]. In the expert's opinion, this is contradictory to the rest of the code, where everything is made more explicit, so they suggest to define an enum with the two possible variants.

A general aspect, that is not specific to the typestate pattern, is the unfamiliarity with the new code structure, according to experts A and C [12], [6]. However, both experts don't see this as a disadvantage, because patterns can be learned quickly by developers.

Lastly, expert A sees the combination of the principle "Parse, don't validate", as introduced in Chapter 4.2.2, with the typestate pattern as positive. An improvement suggestion by the expert is to extract each state with its transitions into an own file, because the increased lines of code make it more difficult to understand them inside one single file, where the developer needs to scroll more. [12]

In the status quo, the function with the highest CoC was `build_final_response` at 8, which is still within the threshold of 15. In the refactoring, most of the nesting logic from the former `build_final_response` function was moved to the `FinalResponse::from` function, which decreased CoC by 1. The former `process_position_items` logic is now in `ItemsWithReturnStatuses::process`, which has a decreased CoC from 6 to 5. This indicates that the large functions have been optimized, while the small functions remain at similar levels at around 0-2 CoC, overall indicating a slight improvement in analysability.

None of the functions have an overly high difficulty in the status quo, when compared to the previous use cases. The highest HDiff was measured for the functions `validate_announced_items_request` at 20.06, `build_final_response` at 19.83, `fetch_item_return_status_for_valid_items` at 19.45 and `handle` at 19.38. While the `validate_announced_items_request` and `handle` functions saw a decrease in difficulty to 14.73 and 14.17, respectively, the difficulties of the other two functions increased. This is probably due to a higher volume of variables introduced. However, there hasn't been criticism from the experts for these specific methods, so it is interpreted that the values indicate a similar analysability after the refactoring.

Both metrics fail to capture aspects such as the `Either` or `Result` arguments made by experts. Advantages of the enum variants in explicitness and increased readability are also not captured. Especially the function that builds the final response `FinalResponse::from` was pointed out as positive for making the possible error states more visible, which is penalized by the HDiff metric.

### 6.4.5 Modifiability

The criterion modifiability is mostly seen as increased, with a few restrictions. Overall, the now more clearly separated phases like validation and processing make it significantly easier and safer to modify the code, according to expert B [16]. This is confirmed by expert C, who points out that the type system prevents many mistakes during modification, which makes it harder to introduce new errors [6]. Expert C furthermore sees the state machine as an enforcement that changes are made in the correct place. Additionally, because the code is now separated into smaller steps, modifications are seen as easier by expert D, when compared to the deep function call hierarchy from before [7].

Moreover, two experts see the eliminated mutable state as one of the main reasons for increased modifiability [16], [6]. The reasons are similar to the ones for readability. Because the functions are now pure, changes can be limited to a single unit, according to expert B [16].

Some changes, such as adding parameters or states, are now seen as more complicated by expert A [12]. Still, the expert assumes this can be an advantage because it forces the developer to put more thought into each change. This is under the assumption that the developer is motivated to invest more time, as code quality will otherwise suffer.

One counterpoint for modifiability is the `Either` type, which can be bothersome to modify if a new business requirement neccesitates a third possible outcome besides `Left` and `Right`. It's unclear what the developer should do in this situation and which name would be appropriate for the third generic. Expert D thinks this would be easier if the developer was dealing with an enum instead, so that a third variant can just be added [7]. As business requirements are prone to unexpected or frequent changes, `Either` is not seen as the ideal solution for this use case.

In the status quo, six functions are moderately maintainability, while in the refactored version, five functions are moderately maintainable. The functions `handle`, `validate_announced_items_request` and `validate_announced_position_items` have shown an increased MI of over 85 for their corresponding functions in the refactored code, indicating that they are now highly maintainable. However, the function `add_missing_position_item_errors` (renamed to `build_missing_position_item_errors`) has a decreased MI, now being moderately maintainable at 81.86. The function `build_final_response` was already moderately maintainable before at 78.13 and is now split into two functions on the struct `FinalResponse`, which have an MI of 72.97 and 77.1. This introduces one more moderately maintainable method, but also made the old method `build_unprocessable_entity_error_response` in exchange. Overall, no function is difficult to maintain in either the status quo and refactoring, but the number of moderately maintainable functions decreased from 6 to 5, indicating a slight improvement. This is reflected in the expert's opinions, as they don't mention any decreased modifiability in any single functions. Yet, the improvements they noted are not reflected in this metric. The facts that the type system prevents introducing errors when making changes, or that the eliminated shared mutable state makes introducing new bugs less likely, are not captured by the metric.

### 6.4.6 Testability

The refactored code is seen as more testable by all experts. Contributing to this improvement is the reduced number of required mocks. Functions no longer depend on complex or global state [16], can be called with certain dependencies instead of one big depenency object that was passed before [6], [7], and some units don't require mocks at all anymore [7]. Each function also has less parameters and conditions, as well as one single responsibility, which makes testing them simpler according to expert A [12]. Additionally, the elimination of side effects and functional purity contributes to an improved testability [16].

While expert C finds it hard to assess testability because there were no unit tests present in the status quo, they find that the improved analysability helps understand what needs to be tested, which ultimately has a positive effect on testability [6]. The explicit states encoded in enums are not seen as better for testability than the implicit states before, because both are easy to construct and compare [16].

CyC in the status quo code saw the highest values in the functions `build_final_response` at 8 and `AnnouncedPositionItemsRequest::try_from` at 6, but these values are still within the threshold of 10. Besides that, the problems pointed out in Chapter 4.3.2 about mocks being required for all processor dependencies and that many functions have side effects are not captured by the metric by design. In the refactoring, the CyC mainly remains the same for most functions, in a range between 1 and 4. The CyC of the function `ItemsWithReturnStatuses::process` has increased to 6, which is an increase by 1 compared to before. While the number for the function `AnnouncedPositionItemsRequest::try_from` hasn't changed, `build_final_response` has been split up and seen an increase. Their CyC measurements result in the numbers of 8 and 9, indicating that splitting the final response building has made testability more difficult, with one more function that has a comparatively high value. Nevertheless, all functions are still within the threshold of 10. The benefits in testability noted by the experts were mostly seen in eliminating side effects and mocks, which is not captured by this metric. In terms of control flow that needs to be tested, the metric indicates that there has been a slight degradation.

### 6.4.7 Faultlessness

Finally, improvements in faultlessness are pointed out by all of the experts. One reason for that are the enums used as return types, which make success and error states more explicit [12], [16]. Another expert sees the same benefits as for the other use cases, where error detection is shifted from runtime to compile time, so that errors can be detected earlier [6]. Expert D sees faultlessness as the biggest advantage of the typestate pattern, because it makes invalid state impossible to exist from the beginning [7]. Errors can be identified directly where they originated, and it's not possible to call functions in the wrong order or mess them up otherwise.

### 6.4.8 Interpretation

Runtime performance is unaffected by the typestate pattern, since the difference in execution time is statistically insignificant due to high variance and a poor fit of the linear regression model. Analysability has improved because each logical step is now separated more clearly, and the pattern makes it easier to reason about the state locally. The states are dependent on each other and tightly coupled, meaning that the code is not much more modular than before. For the same reasons and the fact that the logic is still bound to the specific use case, reusability also remains limited. Modifiability improves for local changes inside states, but structural changes require more effort. Functions are now largely pure, which eliminates side effects, and fewer required mocks further ease testability. Lastly, faultlessness improves significantly, as the correct execution order is enforced, and the handling of different error states is ensured through enums and exhaustive match statements. The data's state is explicitly encoded in the type system, further contributing to compile-time safety. Overall, similar benefits compared to the first use case are noted, with the largest improvements being in faultlessness and testability. More than for the first use case, the analysability improved, while the criteria Modularity, Reusability and Modifiability remain neutral to slightly improved.

In addition to what was stated in Chapter 6.2.8, it can further be recommended to:

- use Rust's `Result` type for unrecoverable errors and enum variants for expected errors that are part of the domain logic,
- and to refrain from using an `Either` type and instead use explicit enum variants, which ensures readability and extensibility in the future.

## 6.5 General remarks by experts

In addition to the criteria evaluation, there has been general feedback from some of the experts. First of all, expert B thinks that the typestate pattern was not the ideal choice for both use case 1 and 3. This is due to the lack of branching logic that the typestate pattern excels in. In the presented use cases, the flow was mostly linear, which doesn't justify the overhead and boilerplate caused by the pattern, according to expert B. [16]

On the contrary, expert C sees strong value in the patterns, and suggests to work together more closely with business stakeholders to model the logic conceptually and then transfer it correctly to the code. The code after applying the typestate pattern is described as more formal and mathematical, but at the same time closer to the business domain. [6]

Similar advantages are seen by expert D, who suggests to incorporate a state machine or model during the planning phase of new features or changes, so that business stakeholders can clarify their needs and developers can identify easily which states need to be modified, added or removed. It would also be possible the other way round, when a diagram is generated out of the state machine represented in the code, which would make it more accessible to non-technical stakeholders. The expert also reported positive feedback when presenting the changes to use case 3 in their team, except from the enum variants mentioned before, which the expert suggested to partially handle through the `Result` type. According to the expert, the team will adapt some of the changes, while not strictly adhering to the typestate pattern as rigidly as presented in this study. [7]

## 6.6 Threats to validity

This section aims to identify possible weaknesses of the validity of this study. The findings will be classified in the four sections *Construct validity*, *Internal validity*, *External validity* and *Reliability*, following the structure suggested by Wohlin et al. [Woh+24, pp. 101-102] for case studies in software engineering.

### 6.6.1 Construct validity

Firstly, the selected criteria from the ISO 25010 SQuaRE model might not fully capture all facets of code quality influenced by design patterns. It was tried to focus only on criteria that are typically addressed through design patterns. However, it is possible that design patterns have other implications, such as unknown effects on runtime performance that cannot be captured without observing the application for a longer period of time.

Secondly, the selected criteria might be interpreted differently than what was intended by each experts. To minimize this risk, the literal definitions from the ISO catalogue were provided in the expert interview guide, which the experts received prior to the interview to familiarize themselves with the criteria and prepare accordingly. Moreover, if a diverging answer was noticed during the interview, the interviewer clarified the ISO definitions to ensure consistency. Nonetheless, the definitions still leave room for improvement, which might cause different judgements from the experts.

### 6.6.2 Internal validity

While the use cases were mainly refactored using the proposed design patterns, some aspects of the refactoring are also general practices such as naming, simplification, or the principle *Parse, don't validate*. While benefits or drawbacks are attributed to the pattern itself, they might stem from other changes. Thus, the additional refactoring techniques were explicitly mentioned and included in the recommendations on how to apply the pattern. Factors like naming were ignored in the analysis of the expert interviews.

Additionally, experts who are already familiar with a certain use case, for example because they implemented it or frequently modify it, might have a positive or negative bias in certain criteria. For example, the status quo might be seen as more readable than the refactoring simply because the former is more familiar. On the other hand, experts who are more familiar with a use case might know more of its shortcomings and therefore prefer the refactoring to a higher degree.

Furthermore, the benchmarks were executed on a laptop device with a regular M1 pro chip. As already pointed out before, background tasks, CPU scheduling and other external factors might influence the benchmarks and interfere with the results and conclusions. Specifically, it could be seen in use case 3 that the execution time became progressively longer, possibly due to an accumulation of containers running on the system. The interference of these factors has been tried to minimize by stopping all other running processes that are not vital to the system and expanding the measurement time to 4 minutes, and cleaning up all containers and data after each run.

### 6.6.3 External validity

The results were intentionally focused on business oriented code in backend Rust applications, a specific domain that needed further investigation at OTTO. While this is a common category of code, the results might be less applicable to libraries, SDKs and such, because they involve less business specific code and more general code that is integrated in code bases the authors themselves do not know.

Moreover, the selected experts have high experience levels, so their judgements might not translate well to less experienced developers. Since the number of participants in the case study is small, experienced developers were selected who are able to reason about how a junior developer thinks.

### 6.6.4 Reliability

The conducted interview was semi-structured, so there might be questions and answers that are not reproducible in other interviews. Depending on the interviewees, different discussions might occur. This was tried to mitigate through moderating the interviews and keeping the questions close to the original guide.

# 7 Conclusion

This chapter displays the most relevant findings of this study, and discusses their implications and limitations. Possible areas of future research in the field of design patterns in Rust are suggested.

## 7.1 Summary

This study identified that for the programming language Rust, there is little empirical evidence on which design patterns are beneficial. Most design patterns known and used in modern software engineering originate in OOP, and their effects have mostly been studied for the language Java. Novel design patterns exist [Cri23], but haven't been evaluated in terms of their benefits on code quality. Consequently, this study formulated the goal of identifying design patterns that improve code quality, while making use of Rust's specific language features like its powerful type system.

To address the goal, a case study on three use cases selected from the company OTTO's backend applications has been conducted. Each use case was refactored by applying a selected design pattern. Criteria were selected from the SQuaRE quality model, and the refactored code has been evaluated for each criterion using benchmarking, static code analysis and expert interviews. A mixed-method approach was necessary to mitigate the known shortcomings of static code analysis on its own and to capture human-centric quality attributes such as readability and perceived complexity, which can be captured more realistically by qualitative expert interviews.

All patterns were found to have no additional runtime cost. The typestate pattern was found to significantly improve faultlessness and testability, while different perceptions of readability exist, especially regarding the additional boilerplate code, which can cause cognitive overhead. Although the typestate pattern effectively shifts error detection to compile time, the additional boilerplate prevents it from being universally beneficial, especially when the application flow is linear, or when no complex invariants or domain rules exist. The newtype pattern, combined with the principle "Parse, don't validate", was found to introduce minimal downsides and code changes, while offering improvements in almost all quality dimensions. It is broadly applicable and enforces compile-time invariants wherever it is used. Specific implementation recommendations for the patterns have been formulated.

This study yields a better understanding of the benefits of the two design patterns typestate and newtype in the context of real-world backend Rust applications. It became evident how the application of patterns alters different aspects of software quality, aiding decisions on future refactorings where similar challenges as in the presented use cases exist. Furthermore, this study developed a reusable methodology for evaluating design patterns' benefits on software quality, which can be used for other languages, design patterns and use cases. Yet, the findings suggest that much knowledge on how to use Rust's unique language design to enhance quality and compile-time correctness remains to be discovered and encoded in novel design patterns.

## 7.2 Future work

In this study, only three use cases and two design patterns have been studied, and future research into other use cases and a broader range of patterns is necessary. Use cases from a variety of domains should be investigated, such as embedded systems or CLI tools, to improve the generalizability of the identified benefits. To decrease the results' subjectivity and bias, future studies should also recruit a high number of participants instead of a handful of experts. Participants with varying backgrounds and experience levels could lead to better insights on how different seniorities and developers unfamiliar with the code view the changes.

Furthermore, the metrics used for evaluation have known limitations. As soon as more human-centric metrics based on correlation with real cognitive effects are available, more rigorous assessments will become possible. DevOps metrics like the Four Key Metrics, often also referred to as DORA metrics, which originated in large-scale empirical studies on software delivery performance [FHK18, p.51], could be measured over an extended period of time. This would not only allow researchers to capture more aspects of software quality, but also how it affects developers and a whole team, without having to interview them.

In addition, performance measurements, such as benchmarking, should be executed on hardware with fewer external factors and higher stability. A CPU with a stable clock rate and a system free of other background processes and operating system overhead would decrease the interference with benchmarks. These measures would further enhance the benchmark's validity and reproducibility.

Another possible direction of future research could be the interaction of business requirements with source code, as with the typestate pattern, it is possible to encode some of the requirements in the type system. Not only the transfer of business requirements to source code, but also the extraction of the currently encoded business requirements in the source code should be investigated, for example through diagrams generated from the encoded state machine, that can be discussed with stakeholders.

# Online References

# A Appendix

## A.1 Digital appendix

Parts of the appendix are not included in the print-out and are available as digital attachments only.

### A.1.1 Status quo code

Contains the status quo source code of each use case before refactoring.

### A.1.2 Refactored code

Contains the refactored source code of each use case, after the design patterns have been applied.

### A.1.3 Expert interview transcripts

Contains the original audio files from the interviews, which were conducted in German, and an automatically generated transcript. The transcripts for Experts A, B and D were generated by Microsoft Teams, while the transcript for Expert C was generated by the model OpenAI Whisper.

### A.1.4 Static code analysis raw results

Contains the raw data calculated by the tool rust-code-analysis.

### A.1.5 Benchmarking setup

Contains the source code for configuring the tool Criterion and setting up the test environment and data for each use case's benchmark.

### A.1.6 Benchmarking raw results

Contains the raw result data and reports generated by the tool Criterion.

## A.2 Selected ISO criteria

| Functional Suitability | Performance Efficiency | Compatibility | Interaction Capability | Reliability | Security | **Maintainability** | Flexibility | Safety |
|---|---|---|---|---|---|---|---|---|
| Functional Completeness | **Time Behaviour** | Co-existence | Appropriateness Recognizability | **Faultlessness** | Confidentiality | **Modularity** | Adaptability | Operational Constraint |
| Functional Correctness | Resource Utilization | Interoperability | Learnability | Availability | Integrity | **Reusability** | Scalability | Risk Identification |
| Functional Appropriateness | Capacity | | Operability | Fault Tolerance | Non-repudiation | **Analysability** | Installability | Fail Safe |
| | | | User Error Protection | Recoverability | Accountability | **Modifiability** | Replaceability | Hazard Warning |
| | | | User Engagement | | Authenticity | **Testability** | | Safe Integration |
| | | | Inclusivity | | Resistance | | | |
| | | | User Assistance | | | | | |
| | | | Self-descriptiveness | | | | | |

**Table 6:** Criteria and subcriteria of the SQuaRE quality model according to the ISO/IEC 25010:2023 norm. The main criteria are in the header of the table, and each column contains its subcriteria. Selected criteria for assessing design patterns are marked in bold.

## A.3 Correspondence with Crichton via email

**Date**: 2025/11/21, 18:05 CET
**From**: Leon Heuer
**To**: Will Crichton
**Subject**: Questions Regarding Your "Typed Design Patterns for the Functional Era" Paper

Dear Will Crichton,

I am a computer science student from the German university Nordakademie, currently working on my bachelor's thesis on the benefits of applying software design patterns to backend Rust applications. My thesis is being supervised by professor Jan Haase and advised by my company supervisor Falk Woldmann.

Your workshop paper Typed Design Patterns for the Functional Era inspired us to explore how design patterns could improve our Rust codebase. We work at OTTO, a major German online retailer with 12.2 million active customers. In our department, we develop part of the online shop's backend in Rust and optimize it for low cost and latency. For our business domain, Rust is still a niche choice compared to Java, or Node or Python. But we believe that due to Rust's rapid adoption, good software engineering practices must be established early on.

As you already pointed out in your paper, there is a lack of design patterns for functional programming. As far as I know, there is also a lack of research specifically on design patterns for Rust. In my thesis, I apply design patterns to selected examples from our Rust code and assess their impact on code quality. There are two questions that I would like to ask you as an expert in functional design patterns:

1. We started using the FC-IS ("Functional Core, Imperative Shell") approach and found improvements in readability and testability of the code, among others. I plan to verify the observed benefits in my bachelor's thesis. Is FC-IS a design pattern in the traditional sense, or would you consider it an architecture approach? Furthermore, I didn't find any literature on FC-IS yet. Do you know of any prior work that discusses it, maybe under a different name?
2. For assessing the benefits of design patterns in my thesis, I would prefer to use quantitative code quality metrics. But when asking experienced developers in our department about this, almost none knew any beyond Cyclomatic Complexity. While measures like the Maintainability Index or Halstead measures exist, developers don't seem convinced by them. Are there quantitative metrics you consider meaningful and commonly used for evaluating the impact of design patterns?

We would greatly appreciate your opinion on these questions.

Kind regards, Leon Heuer

- - - - - - - - - - - - - - - - - - - - - - - - - - - - - - - - - - - - - - - -

**Date**: 2025/11/21, 23:52 CET
**From**: Will Crichton
**To**: Leon Heuer
**Subject**: Re: Questions Regarding Your "Typed Design Patterns for the Functional Era" Paper

Hi Leon et al., thanks for reaching out, I'm glad you found inspiration in the paper. To your questions:

> Is FC-IS a design pattern in the traditional sense, or would you consider it an architecture approach?

I don't think there's an authoritative distinction between patterns and architectures, but I'll give you my personal views. The difference between patterns and architectures is principally a matter of scale. Both are about describing patterns of computation that can't be neatly encapsulated into a particular reusable code component or library. Patterns are about a smaller scale, like on the type of code in the GoF book or my FUNARCH paper. Architectures are about a larger scale, such the "virtual DOM architecture for reactive web apps" or "single-page-application architecture for full-stack web apps". From this perspective, FC-IS is more of an architecture since it tends to be an orienting principle for an entire codebase. Another great example of FC-IS is the Codemirror 6 API design [1].

> Do you know of any prior work that discusses it, maybe under a different name?

I don't know of any academic work on FC-IS. I think it's primarily circulated in industry circles since Gary Bernhardt first put a name to it.

> Are there quantitative metrics you consider meaningful and commonly used for evaluating the impact of design patterns?

Sadly, I would also agree that we lack good metrics for evaluating design patterns. Halsteads / cyclomatic / etc. are not much better than just measuring lines of code [2]. I even wrote about this on the SIGPLAN blog [3]. For example, if you use a pattern to make your code more extensible (e.g. a visitor), how do you measure extensibility? If you use a pattern to make your code more correct (e.g. the typed design patterns in my paper), how do you measure bugginess?

The lack of good metrics is a serious problem in our community, one I hope to address in due time through my research. But for now, I don't have much useful advice beyond "don't leave too heavily on metrics". I would focus on qualitative analysis through case studies.

[1] https://codemirror.net/docs/guide/
[2] https://ieeexplore.ieee.org/document/9402005
[3] https://blog.sigplan.org/2024/11/21/evaluating-human-factors-beyond-lines-of-code/

## A.4 Correspondence with Campbell via email

**Date**: 2025/11/25, 13:45 CET
**From**: Leon Heuer
**To**: G. Ann Campbell
**Subject**: Questions regarding your white paper on Cognitive Complexity

Dear Ann Campbell,

I am a computer science student from the German university Nordakademie, currently working on my bachelor's thesis on the benefits of applying software design patterns to backend Rust applications. My thesis is being supervised by professor Jan Haase and advised by my company supervisor Falk Woldmann.

During my research on measurements for code quality, I came across the Cognitive Complexity measure from Sonar defined in your white paper [1]. I believe it is an interesting measure to complement other mathematical quality models, since it was derived from practical experience and is based on assumptions on which code structures increase perceived complexity for developers.

In the white paper, you listed a few exceptions for language specific differences. Do you consider CC applicable for Rust code without further exceptions? There are static code analysis tools that can calculate Cognitive Complexity for Rust [2]. However, Rust comes with a few language structures that can make code easier or harder to understand, such as Algebraic Data Types, newtypes, and the Ownership concept. Have you already taken these language specifics into account in the current version of your white paper?

We would appreciate hearing back from you on this question.

Kind regards,
Leon Heuer

[1] https://www.sonarsource.com/resources/cognitive-complexity/
[2] https://mozilla.github.io/rust-code-analysis/metrics.html

- - - - - - - - - - - - - - - - - - - - - - - - - - - - - - - - - - - - - - - -

**Date**: 2025/11/25, 18:05 CET
**From**: G. Ann Campbell
**To**: Leon Heuer
**Subject**: Re: Questions regarding your white paper on Cognitive Complexity

Hi,

I'm not familiar enough with Rust to give you a definitive answer, only the principles of one. Are these language-specific structures related to control flow? Based on the names, I would guess not, and thus say the answer is likely 'no.'

In formulating Cognitive Complexity we made a conscious decision to look only at control flow. Why? Every language has "hard" features. Some people thought Java generics were hard. People new to C think pointer indirection is hard, but seasoned C programmers are well used to one level of pointer indirection, and two is still routine. So then when does it get "hard"? At 3 levels of indirection? 4? When you're pointing to a function? At that point, you're reduced to arguing personal opinion, rather than measuring. So we simply don't in Cognitive Complexity.

Does that help? Do you have any other questions?

Ann

---

**Date**: 2025/11/25, 23:34 CET
**From**: Leon Heuer
**To**: G. Ann Campbell
**Subject**: Re: Questions regarding your white paper on Cognitive Complexity

Dear Ann,

thanks for your response. The language-specific structures I listed are indeed not directly related to control flow. Right now, I can think of two other additional language features that affect control flow:

1. The "?" operator, which propagates an error to the caller of a function [1]. The closest equivalent in Java, for instance, would be adding a "throws" to the method signature for a checked exception that is not handled by a try/catch statement. Rust has neither try/catch, nor exceptions and relies on a wrapper type for returning and handling errors [2]. But in analogue to how your white paper argues that a return doesn't increase complexity, one could argue that a "?" operator makes the program easier as well, even though being a possible exit point of a function. How would you argue in this situation?
2. Rust has "match" statements that are essentially a switch with more abilities like pattern matching and match guards [3]. Pattern matching supports tuples as well. For example, one could match on a tuple consisting of three booleans, where the cases cover different combinations of true and false values. Clearly, the statement "a switch - which compares a single variable to an explicitly named set of literal values" from the CC white paper is not true for Rust. How would you calculate complexity for Rust's match statements?

I am curious to hear your stance on these language specifics.

Kind regards,
Leon

[1] https://doc.rust-lang.org/book/ch09-02-recoverable-errors-with-result.html#a-shortcut-for-propagating-errors-the–operator
[2] https://doc.rust-lang.org/book/ch09-00-error-handling.html
[3] https://doc.rust-lang.org/reference/expressions/match-expr.html

---

**Date**: 2025/12/01, 14:14 CET
**From**: G. Ann Campbell
**To**: Leon Heuer
**Subject**: Re: Questions regarding your white paper on Cognitive Complexity

Hi,

Regarding `?`, if you look at it as an early return-analog, then I would certainly argue to ignore it.

Regarding `match` and other enhanced `switch` statements… I've kinda been waiting for someone to ask me that ever since `switch`es started getting more sophisticated. You're right that the expanded capabilities of modern `switch`es and `match` don't correlate to the "single variable to an explicitly named set of literal values" premise, and thus I think they should probably be treated as `if`-trees.

Ann

## A.5 Expert interview guideline

The expert interview guide was originally sent to the interviewees in German. This is a translated version.

---

The following questions will be asked to the expert during the interview. The expert is asked to think about the questions in advance and, if necessary, prepare notes to ensure a smooth interview process.

**Please review the code only after answering these questions.**

**1 - Introduction**

1. How long have you been working with Rust, and in which domain or on which types of products?
2. In your opinion, what distinguishes Rust from other established languages at OTTO, such as Java or Kotlin, with respect to maintainability and fault prevention?
3. From your perspective, what are the main causes of technical debt in large Rust codebases?

**2 - Current Problems in the Code**

1. What are the biggest pain points in the current codebase? What bothers you when maintaining the code or implementing new changes? *(e.g., redundancies, complexity, insufficient use of the type system or language features, poor readability, missing documentation)*
2. How easy or difficult do you think it is for new developers to understand how the domain is modeled in the code, and why?
3. When you think about poorly maintainable code, do concrete code examples come to mind?
4. The following criteria will later be used in the expert interview to evaluate the refactored code:
    1. Modularity
    2. Reusability
    3. Analysability
    4. Modifiability
    5. Testability
    6. Faultlessness

    Which methods do you know to quantify these properties, and how do you assess their practical relevance in everyday development as well as their explanatory power?

Now the focus shifts to the concrete use cases that were selected and refactored:

1. **FT9 Detailview Service**
    - GitHub repository: `otto-ec/ft9_benefit`
    - File: `services/tag/src/domain/detailview/detailview_service.rs`, entire file except tests
    - Applied pattern: **Typestate**
    - New version on branch `bachelor-project-leon-heuer`
2. **FT9 Tag Component Handler**
    - GitHub repository: `otto-ec/ft9_benefit`
    - File: `services/tag/src/api/tag_component_handler.rs`, lines 112-207

- Applied pattern: **Newtype**
- New version on branch `bachelor-project-leon-heuer`

3. **Boxfish Return Status REST API Handler**
   - GitHub repository: `otto-ec/boxfish_return_status`
   - File: `lambdas/rest_apis/src/update_return_status_to_announced/function_handler.rs`, entire file, as well as the related modules `incoming_request.rs`, `validation.rs`, `api_error_response_builder.rs`, lines 66-81
   - Applied pattern: **Typestate**
   - New version in the fork `leonheuer/boxfish_return_status`

**The following sections are evaluated separately for each use case.**

**3 - Status Quo**

Please review the current code of the use case. Which code smells do you notice? Are there parts that make the code hard to maintain or error-prone? Which of the criteria mentioned in Section 2 are negatively affected as a result?

**4 - Refactoring**

Now review the new, refactored code. Evaluate the following criteria, which were selected from the ISO 25010 SQuaRE model for software quality:

1. **Modularity**
   - **Definition:** "Capability of a product to limit changes to one component from affecting other components."
   - To what extent has the modularity of the code worsened or improved due to the refactoring?
2. **Reusability**
   - **Definition:** "Capability of a product to be used as assets in more than one system, or in building other assets."
   - To what extent has the reusability of the code worsened or improved due to the refactoring?
3. **Analysability**
   - **Definition:** "Capability of a product to be effectively and efficiently assessed regarding the impact of an intended change to one or more of its parts, to diagnose it for deficiencies or causes of failures, or to identify parts to be modified."
   - This criterion can also be understood as readability. Has the code become easier or harder to read, analyze, and understand as a result of the refactoring?
   - How has the cognitive effort required to understand the examined code changed?
4. **Modifiability**
   - **Definition:** "Capability of a product to be effectively and efficiently modified without introducing defects or degrading existing product quality."
   - Has it become easier or harder to modify the code without introducing new defects as a result of the refactoring?
5. **Testability**
   - **Definition:** "Capability of a product to enable an objective and feasible test to be designed and performed to determine whether a requirement is met."
   - To what extent has the testability of the code worsened or improved due to the refactoring?
   - Are individual components now easier or harder to test, for example due to changes in the number of required mocks or reduced side effects?
6. **Faultlessness**
   - **Definition:** "Capability of a product to perform specified functions without fault under normal operation."
   - This criterion was selected with regard to Rust's compile-time guarantees. For example, faultlessness is already improved by enforced memory safety through the borrow checker at compile time, thereby reducing runtime errors. Another way to make code more "correct" at compile time is to use Rust's type system to encode rules at compile time, for example through the applied Typestate and Newtype patterns.

- Has the faultlessness of the code actually improved? Does the refactoring prevent invalid application states, and has the risk of bugs or incorrect function calls been reduced?

## A.6 Expert interview summaries

### A.6.1 Expert A

**Expert**: B.Sc. Nico Schmidt
**Date**: January 5, 2026, 1:00 pm
**Location**: Online meeting
**Background**: Software Developer, Otto GmbH & Co. KGaA, 2 years Rust experience

The following is a summary of the expert interview, which was created with the support of large language models. The summary has been confirmed to be correct in its content by the expert.

**1 - Introduction**

**Q1**: How long have you been working with Rust, and in what area or on what kind of products?

**A1**:

- Working with Rust at FT9 for approximately 2 years
- Replaced Spring Boot microservices (previously Kotlin) and Lambdas (previously Python, TypeScript, etc.) with Rust
- The team aims to use a single language for all production code; they are currently at 98-99%
- Only one batch task remains in Python, which is also planned to be migrated to Rust

**Q2**: And what, in your opinion, distinguishes Rust from other languages at Otto, such as Java or Kotlin, in terms of maintainability and faultlessness?

**A2**:

- Very little in principle, but the forced redevelopment allowed for reducing dependencies on other teams
- This drastically improved maintainability, development speed, and technical debt
- Owning the crates locally enables changes without waiting for or informing other teams
- For example, the self-written toggles crate is much simpler and easier to maintain than the previous feature-heavy Spring Boot implementation
- Developers are forced to adhere to certain patterns, such as the absence of null values, which leads to higher faultlessness when applied correctly
- Features like out-of-the-box linting and the integrated dependency management via Cargo are significant advantages
- The necessity of rewriting everything built up internal team expertise and reduced overall system complexity

**Q3**: Would you say that these specific language features of Rust, which ensure faultless development, also bring difficulties, especially when coming from a Java or Kotlin context?

**A3**:

- At the start of the migration from Spring Boot to Rust, many developers had little Rust experience
- While team members had read the Rust Book or done personal projects, none had written productive Rust applications before
- This led to the development of varying code styles among developers
- As is common in software development, there are many ways to achieve a goal, which allows individual experience to play a significant role
- Deadlines and the pressure to finish tasks often resulted in a lack of refactoring to make the code more homogeneous
- There was a lack of clarity regarding what constitutes “correct” code
- Despite having some guidelines, they were insufficient, leading to different development approaches

- Individual opinions further contributed to the resulting mess

**Q4**: What do you think are the main causes of technical debt in large Rust codebases?

**A4**:
- The business context is very complex, requiring a deep understanding of extensive functional logic before it can be translated into high-quality software
- Achieving this translation is not a simple task
- The team consists of people from different generations who have learned different approaches during their studies and professional practice
- It is difficult to align developers on a consistent code style, especially since many come from a Kotlin and Java background

---

**2 - Problems in the current code**

**Q5**: What are the biggest pain points in the current codebase you work with daily at FT9? What bothers you regarding code maintenance and implementing new changes?

**A5**:
- Different states of customer benefits, such as valid, cancelled, or aborted, are not well-represented, especially the transitions between them
- Functions are heavily nested, calling one another in a chain that follows an object-oriented approach
- This structure is reminiscent of writing code in Spring Boot, where services call other services, which does not result in clean code in a Rust environment
- Rust development should ideally lean more toward a functional style
- A major pain point is the adoption of code styles from Spring Boot, Kotlin, and similar backgrounds

**Q6**: In your estimation, how easy or difficult is it for new developers to understand how the domain is modeled in the code, and why?

**A6**:
- Understanding the domain is difficult, especially for those who lack full business context
- Developers must first grasp the complex business cases before they can comprehend the corresponding code
- The team includes a mix of experience levels, including junior developers who are just starting their careers after university or vocational training
- Documentation is often unreliable as it is frequently outdated, which is a general industry problem rather than a Rust-specific one
- Challenges are mitigated through pair programming for new team members and structured quality assurance processes with regular feedback
- Despite these measures, the learning process remains demanding

**Q7**: You mentioned several general factors, such as pair programming and the difficulty of understanding business logic due to its complexity. Regarding the code itself, are there other factors besides the adopted Java and Kotlin styles that make it difficult for new developers to understand?

**A7**:
- The code contains extensive branching with numerous if-statements and special cases derived from the business logic
- These edge cases significantly increase complexity as each must be explicitly represented in the code
- It is common to write 200 additional lines of code to handle a single edge case that rarely occurs and has minimal impact on the user

**Q8**: Can you think of specific examples when you consider code that is difficult to maintain?

**A8**:

- Input validation is a frequent issue, as security guidelines require validating all data entering services and lambdas, such as URL parameters and request bodies
- A problematic pattern has emerged where controllers or handlers start with numerous lines of code dedicated solely to validation
- The handler for the "Sheet" service accepts an excessive number of parameters, approximately 15 to 20
- This high parameter count makes maintenance difficult, as changing one parameter requires updating it across many functions
- The "Cinema" service uses various types with inconsistent filtering, sorting, and flattening logic
- These components utilize different data sources and fields for their operations, which increases complexity and reduces the maintainability of the code

**Q9**: Which methods do you know for quantifying the seven criteria, and how do you assess their relevance and validity in daily development?

**A9**:

- Regarding correctness, testing across all levels of the test pyramid up to functional acceptance is essential, with regular reviews of unit, integration, and end-to-end tests
- For modularization, there are no specific key performance indicators, while for reusability, IDEs can help identify duplicated code or test setups like mocks and fixtures
- Analysability depends on how quickly a developer understands the code, which is influenced by their familiarity with language-specific features like the question mark operator or complex if-let conditions in Rust
- The McCabe metric provides an initial indication of code complexity at the function or class level, though it is not without its critics
- Modifiability could be assessed by measuring the time required to implement a change while ensuring that test results and values remains consistent
- Testability can be evaluated through several metrics, including test coverage, the distribution of tests according to the test pyramid, and the duration of test execution
- Further indicators for testability include the number of mocks, the number of parameters in method signatures, and the degree of coupling to other classes
- Functional purity is important, as side effects or writing to global error lists are signs of poor design
- For faultlessness, the primary requirement is some form of formal acceptance or sign-off process

**Q10**: You mentioned McCabe Complexity, or cyclomatic complexity. Do you also know it in the context of testability, to determine how many different test paths need to be covered?

**A10**:

- I have never applied it in that specific context before

**Q11**: How relevant are such metrics - for example, cyclomatic complexity, lines of code, or the maintainability index - in the daily life of a developer? How often are they actually measured, and how meaningful are they?

**A11**:

- I have never measured lines of code and do not consider it a particularly good metric
- However, an excessively long class is generally seen as a negative indicator
- I have McCabe complexity metrics running permanently; when the value exceeds a certain threshold and turns red, I usually take action
- IDE warnings regarding an excessive number of function parameters are also very helpful
- Rust offers several high-quality linting tools that are effectively used in this regard

---

**Use Case 1 - Status Quo**

**Q12**: Which code smells did you notice in the Detailview Service, and which criteria are negatively affected by them?

**A12**:

- The first issue is the high number of parameters, with eight different values being passed in, which increases complexity
- Passing these parameters through various functions and sub-functions is problematic, especially regarding the ownership model in Rust
- This negatively impacts criteria such as maintainability and modularization
- Although the function name `createDetailviewViewModel` is appropriate, the internal logic lacks proper separation of concerns
- The process should be split into distinct steps: fetching a DTO from the database, filtering it in a separate function, and finally converting it into a view model
- The current implementation performs too many tasks simultaneously: database queries, filtering, triggering activations, writing metrics, and building the final view model
- Additionally, some functions are used outside of the Detailview Service similar to static functions in Java, which suggests poor architectural design

**Q13**: Can you name specific criteria that are negatively affected by the code smells you mentioned?

**A13**:

- Most of the criteria are affected; while the code is functionally correct, it suffers from poor testability
- The high number of parameters and conditions requires an excessive amount of tests to cover the entire function
- Readability is significantly reduced because developers must keep too much context in mind while trying to focus on sub-functions
- There is very little reusability, as the function is tailored to exactly one specific use case, making it a poor module
- Regarding modifiability, personal bias plays a role; while the code feels easy to change for someone who wrote it, it is likely difficult for others to read and maintain

---

**Use Case 1 - Refactoring**

**Q14**: Regarding functional correctness, does the refactored code still produce the same results without affecting the original logic?

**A14**:

- I assume so, although I have not deployed or tested it myself
- Based on my understanding of the code, it should produce the same results

**Q15**: To what extent has the modularity of the code deteriorated or improved as a result of the refactoring?

**A15**:

- In my opinion, it has improved
- Many functions were added that return named structs, and I am a proponent of function pipelining
- Individual functions and their return values could be moved to separate classes to reduce file length and further enhance readability

**Q16**: To what extent has the reusability of the code improved or deteriorated through the refactoring?

**A16**:

- Reusability is only slightly better, as the functions serve a very specific use case that is unlikely to be used in other projects
- Certain components, such as the `get_benefits` function, are well-written and could potentially be reused
- Other parts remain too specialized for broader application

**Q17**: Do you agree that the Typestate pattern makes reusability difficult because the specific structs required to call a function might contain fields that are not needed elsewhere?

**A17**:

- Agreement that this approach likely necessitates the introduction of generic types
- One would have to consider using traits or inheritance, which can increase complexity and reduce readability
- Naming the various structs is a significant challenge in this context
- While a name like `UpContractTaskSpawned` fits the specific use case, it may not be easily understood by others
- Clear naming requires careful consideration to ensure the code remains comprehensible for other developers

**Q18**: Regarding analysability and readability, is the code easier or more difficult to read, analyse, and understand after the refactoring?

**A18**:

- Has a personal bias because the original code is familiar, making it naturally easier to read
- A developer unfamiliar with the codebase might find the new structure more advantageous
- Understanding often depends on the individual's prior knowledge of the specific implementation

**Q19**: Do you notice differences in readability because the Typestate pattern forces you to name individual states and transitions?

**A19**:

- The response is mixed; while the code sequence is logical, certain implementations like the activation status and "Up Contract" do not necessarily belong together from a business logic perspective
- These logical discrepancies can negatively impact readability for developers who understand the domain
- Minor adjustments to these connections could lead to overall positive effects on the code's readability

**Q20**: How do you view fields in structs that aren't necessary for a specific step but are passed through anyway?

**A20**:

- Metadata, such as the responsible team or the request origin, is often included in structs because it is needed by a specific function further down the line
- I do not necessarily believe that including this information in the main structs is the correct approach

**Q21**: Is this inherent to the Typestate pattern, or its specific application here? Could it be applied more effectively?

**A21**:

- I assume it could be improved
- Theoretically, the pattern does not need to be applied quite so strictly

**Q22**: Has the refactoring made it easier or more difficult to modify the code without introducing new bugs?

**A22**:

- It has become more complicated in some respects; adding or changing a parameter now requires updating all return values across the Typestate pattern
- Refactoring tasks, such as renaming or introducing newtypes, are more complex than before
- However, specific logic changes, such as modifying the sorting of benefits, may have become easier to implement

**Q23**: Do you see it as an advantage that the pattern forces you to think more deeply about changes? For instance, when adding a state, you are required to explicitly define the structs and transitions. Is this a benefit or a drawback?

**A23**:

- I would characterize that as an advantage
- While adding new logic becomes more time-intensive, it can have positive effects if the developer is committed to doing it correctly
- However, you must ensure the implementation is handled properly, as a developer could still simply add code to an existing state without following the intended design, which would be counterproductive

**Q24**: To what extent has the testability of the code improved or deteriorated through the refactoring?

**A24**:

- Testability has improved dramatically because the functions are simpler and adhere to the Single Responsibility Principle (SRP)
- I expect that writing tests for individual states requires fewer test cases per function
- This leads to significantly better test coverage and smaller, more manageable tests with less setup and fewer conditions to verify

**Q25**: Has there been a change in the number of required mocks or the functional purity (freedom from side effects)?

**A25**:

- Yes, both have improved positively

**Q26**: Finally, let's look at faultlessness. This differs from functional correctness by focusing on runtime behavior and how many errors the compiler can catch before execution. Has the faultlessness of the code improved?

**A26**:

- I believe it has changed very little because the Rust compiler is already excellent at catching errors
- Tools like Clippy and Cargo were already detecting issues in the original code
- Without diving deep into the specific algorithms, I don't think the code changes have a significant impact here

**Q27**: If a developer tries to change the order of functions, is the code now safer from incorrect modifications? Does the compiler protect the developer by ensuring that certain filtering can only be called at a specific point?

**A27**:

- Theoretically, yes, because the developer must make active decisions
- While the previous version might have allowed reordering list calls, your changes likely prevent that
- However, there is no guarantee that a developer won't simply modify the structs to make it fit
- If the compiler flags a change as "red," the developer will have an active thought behind the adjustment and may just adapt the structs accordingly, which offers little additional safety

---

**Use Case 2 - Status Quo**

**Q28**: Moving on to the next use case: the FT9 Tag Component Handler, where the Newtype pattern was applied. Which code smells do you notice, and which criteria are negatively affected by them?

**A28**:

- The primary issue is input validation; numerous parameters must be validated both technically and functionally to ensure data integrity
- The first lines of the function immediately show two very similar structures: `Target Context Dreson` and `Current Context Dreson`
- Both must be either valid or unset, which is currently poorly implemented and lacks reusability
- Overall, the code suffers from poor readability

**Q29**: Which specific criteria are negatively impacted?

**A29**:

- Reusability, modularity, and readability are all negatively affected
- Testability is somewhat compromised; while the code is testable, it requires a high volume of tests to cover all possible execution paths

**Q30**: You already mentioned validation. Would you say this affects faultlessness? For instance, because a validated string currently remains a plain string, providing no verification that it actually represents a valid "Dreson"?

**A30**:

- Yes, that is also a direct consequence.

---

**Use Case 2 - Refactoring**

**Q31**: Regarding functional correctness: has the original functionality changed, or does the code still produce the same results?

**A31**:

- It works exactly as before; the results remain identical

**Q32**: To what extent has the modularity of the code improved or deteriorated through the refactoring?

**A32**:

- I don't think it has changed significantly, as it only involved a few lines of code
- I wouldn't say there is a major improvement

**Q33**: But it hasn't significantly deteriorated either?

**A33**:

- Correct, it hasn't deteriorated either

**Q34**: To what extent has reusability improved or deteriorated through the refactoring?

**A34**:

- Reusability has improved, as evidenced by the new `ValidatedDreson` Newtype
- You used the `ValidatedDreson` for both input values, replacing what was previously duplicated code—a clear sign of increased reusability

**Q35**: Do you agree that the Newtype pattern requires more changes during refactoring? If we wanted to use this validated type throughout the entire project, it would involve a high volume of modifications.

**A35**:

- Yes, absolutely; that is the consequence, but it offers the advantage of immediately highlighting where changes are necessary
- Consider a Customer ID: if you need to adjust it, you want that change reflected everywhere
- You are forced to update all instances and verify the implementation accordingly, which actively prevents errors

**Q36**: How has the analysability changed? Is the code easier or more difficult to read and understand?

**A36**:

- Readability improves if you trust the individual Newtypes; you can mentally "filter out" the validation logic and focus on the fact that a value is valid, rather than how it became valid
- This allows you to ignore complex helper cases and branching paths during analysis, which is a clear positive
- However, the refactoring introduced more advanced functional programming patterns that could be a barrier
- Specifically, functions like `transpose`, `map_err`, and the use of the underscore operator (`_`) in closures might be confusing for developers less familiar with these idioms
- While the overall structure is better, the reliance on these specific Rust patterns means a developer must be familiar with them to truly understand the function's execution

**Q37**: Has the refactoring made it easier or more difficult to make changes without introducing new errors?

**A37**:

- It depends on the type of change being made
- Adjusting the `Dreson` (or similar core types) is relatively complex because it is used in multiple places throughout the code, requiring an understanding of all affected areas
- Conversely, other modifications like updating error handling, adding logging, or implementing metrics are now relatively simple to execute

**Q38**: Regarding testability, even if it's not the primary focus of the Newtype pattern, do you think it has improved or deteriorated?

**A38**:

- It improves because testing is split into two distinct areas: specific tests for the Newtype validation (e.g., `ValidatedDreson`) and separate tests for the actual business functions
- The business function tests become significantly smaller and reduce "mental overload" since they no longer need to handle complex validation logic
- With fewer required mocks and smaller test fixtures, the overall testability is enhanced

**Q39**: In other words, we no longer need to test the "Dreson" logic everywhere it is used, but only where it is actually validated.

**A39**:

- Exactly.

**Q40**: Apart from that, is there a change in the number of required mocks or the freedom from side effects?

**A40**:

- There is more freedom from side effects in a sense, because the main handler tests no longer need to guarantee that all input values are correct
- Theoretically, you could even mock the Newtype if necessary, though the impact of that is likely relatively minor

**Q41**: We've reached the final criterion: faultlessness. Has the code's reliability actually improved? Does the refactoring prevent invalid application states or minimize the risk of bugs and incorrect function calls?

**A41**:

- I don't believe so, for the same reasons mentioned previously
- Rust is already very effective at error detection; the original code used `match` statements to cover all cases that are now handled by the Newtype
- You are still required to perform error handling, such as using `map_err`, so the overall safety hasn't improved significantly

**Q42**: Would you say that other functions using the Newtype benefit from increased safety? Specifically, the fact that a function can now be certain the "Dreson" is validated rather than just dealing with a primitive string?

**A42**:

- Theoretically, yes; practically, we already assume that inputs are validated and have the correct type
- In this specific use case, we do very little with the "Dreson" itself other than passing it to a client for another system
- At that exit point, it is converted back into a string anyway, and responsibility shifts to the next team
- Therefore, I don't see a major impact in this specific instance, though I can imagine it being beneficial in other use cases

---

**Use Case 3 - Status Quo**

**Q43**: Now for the final use case: the Boxfish Return Status REST API Handler. Since this isn't part of your daily development work, what code smells did you notice in the current status quo? Which parts make the code hard to maintain or error-prone, and which criteria are negatively affected?

**A43**:

- I skimmed it briefly, and the first thing that stands out are the highly complex return types
- For example, you have a `Result` containing an `announced position item request`, which then contains another `Result` with a `Response` in the body

- These nested `Result<Result<...>>` structures make readability difficult; you have to think three times just to understand what is actually happening
- Like in our previous examples, there is too much logic in a single class—the original had around 290 lines of code
- Interestingly, your refactored version grew to over 400 lines, which isn't necessarily an improvement in terms of volume
- I also immediately noticed that the test coverage is not very good

**Q44**: Could you comment on the mutable state regarding API errors and results?

**A44**:

- You're referring to the API errors and value position items at the top of the handler. Yes, that is quite "unclean"
- We actually have similar patterns in our own code, like in the Kafka Importer, where we handle batch processing
- When you receive 500 events and a few fail, you need a way to track them for retries; using mutable lists passed into functions is one way to handle that
- Personally, I don't find it unreadable because I'm familiar with the pattern and use it myself, but it is a point of contention
- It could be solved differently, perhaps by using global lists or, as you likely did, by using structs to manage the state more effectively

**Q45**: How do you think this affects testability?

**A45**:

- I don't think the impact is very negative, though it does change the testing process
- You have to pass empty lists into your functions and, at the end, verify more than just the return type
- Because mutable references are passed in, you are forced to check the state of the input objects after the function executes to ensure they were updated correctly
- It is manageable, but whether it is truly intuitive is another question
- If the parameters are well-named, for example `api_errors` or `valid_items`, then I believe it remains understandable for a developer

**Q46**: What do you think regarding the freedom from side effects?

**A46**:

- The code is not truly free from side effects
- Error handling is often a special case that runs alongside the main logic
- The function does not just perform one task but must also manage failures
- One could solve this by using return values for successful and failed items
- Retries could be implemented within the function but the errors still need handling
- This approach is a simple solution even if it is not the most elegant one

**Q47**: Are there any other specific criteria you notice that are particularly affected, such as analyzability?

**A47**:

- Understanding the code took some time initially because it is heavily split up
- The handle function serves as the entry point and makes many calls to other functions
- The functions are well-named, allowing a developer to understand the process by looking only at the handle function
- Detailed analysis is easy by using IDE features to jump into specific functions for deeper tracking
- The code is quite understandable after some initial thought
- While improvements are always possible, the current state has reached an acceptable level

---

**Use Case 3 - Refactoring**

**Q48**: Has the functional correctness been affected by the refactoring, or does the code still produce the same results?

**A48**:

- I cannot give a definitive answer because there are no tests for me to run and I am not familiar with the code
- Naively, I would assume it is correct, but I cannot confirm it 100 percent

**Q49**: To what extent has the modularity of the code deteriorated or improved through the refactoring?

**A49**:

- Modularity has improved, similar to the first case discussed today
- The original code had a single handle function that performed every task
- Now the handle function handles initialization while a separate handle request function processes the core logic
- This structure ensures that input parameters are validated and errors are caught before the main processing starts
- The code is much more modular because different handlers could theoretically reuse the handle valid request function

**Q50**: I would argue that using these specific structs as inputs and outputs makes them very tailored to this use case, which might negatively affect reusability.

**A50**:

- I was going to mention that, but reusability was not any better before the refactoring
- The original functions were already very specifically tailored to this single use case
- The refactoring has not made the situation worse than it already was

**Q51**: So regarding reusability, would you say it has neither significantly deteriorated nor improved?

**A51**:

- Yes

**Q52**: Regarding analyzability, since you were not familiar with this code, can you judge if the refactoring made it easier or harder to read and understand?

**A52**:

- This is the main point: you do not have to look at everything anymore
- You can follow only the paths that interest you because you can trust that certain results are returned correctly
- Good naming of functions and types significantly helps in making the logic traceable

**Q53**: How has the cognitive effort required to understand the code changed?

**A53**:

- The effort has increased because the lines of code have grown, which requires more scrolling
- This could be improved by moving logic into different classes or files with descriptive names
- Including tests in those files would also help readability
- Despite the increased volume, it is a step in the right direction

**Q54**: Alright, then let's move on to modifiability. Has the refactoring made it easier or harder to make changes to the code without introducing new bugs?

**A54**:

- I cannot say much because I have never made changes to either the status quo or the new version
- I can imagine that certain changes might actually become more complicated depending on what you want to do
- For instance, adding additional parameters or states is not very simple in this structure
- However, this complexity forces the developer to invest the necessary time rather than just making a quick, potentially messy fix as might happen in the original version

**Q55**: This is probably the same point as with the first use case, the detailview service, where you have to think more carefully about the structures and state transitions, right?

**A55**:

- You really have to give it thought because if you do not, the code quality will suffer
- That risk exists here, but it was exactly the same in the status quo version

**Q56**: So would you say overall that modifiability has slightly deteriorated?

**A56**:

- I would say it has slightly improved
- This assumes the developer is motivated to invest the time and is given the necessary resources for the project to write high-quality code
- Under those conditions, the developer is forced to create a better solution

**Q57**: To what extent has the testability of the code deteriorated or improved through the refactoring?

**A57**:

- Testability has improved because tests can be divided more easily
- Individual tests become smaller with fewer parameters to manage and fewer assertions to verify
- Since the functions now have a single responsibility, you usually only have one value to compare at the end to see if it works
- If these functions and their corresponding tests are well-encapsulated in separate files, it has a very positive effect on testability

**Q58**: How has the number of required mocks and the freedom from side effects changed?

**A58**:

- Smaller tests are created which are hopefully more free of mocks

**Q59**: I forgot one point regarding readability: the refactoring used enums for return types to represent various error states that were previously handled implicitly, such as leaving API errors empty. Do you think this improved readability?

**A59**:

- This is a core feature of Rust enums often used to cover success and error cases
- You also used the Either pattern in some places
- These patterns bring both advantages and disadvantages
- They can make the code harder to read for a new developer who might not understand why an Either type is used
- These specific Rust language features are powerful but require a certain level of experience to be fully understood by team members

**Q60**: To summarize, would you say that the Type State Pattern and the concepts associated with it require more prior experience or familiarity with the pattern?

**A60**:

- Yes, as is often the case with design patterns
- Just like the Builder pattern in Kotlin or Java, you have to understand the general concept first
- I would not say it is overly complicated, but you must know the pattern to apply it
- Learning it does not take long, but it naturally requires more effort than simply writing code sequentially
- However, it brings corresponding advantages to the codebase

**Q61**: Finally, let's look at faultlessness. Has the reliability of the code improved? Does the refactoring prevent invalid application states and has the risk of bugs or incorrect function calls been minimized?

**A61**:

- Reliability has improved because using enums for return types covers specific successful cases

- However, the Fetch Return Status Result enum includes both success and error cases like NotSent and Success
- I believe this could be modeled better, for example by using a Result type where only the success cases are in an enum
- Such a structure would allow the IDE, Clippy, and the compiler to better verify that every case is handled correctly
- If implemented consistently, this approach has a positive impact on faultlessness

**Q62**: Is there anything else you would like to add regarding the refactoring?

**A62**:

- I noticed that in Incoming Requests, a TryFrom implementation was added instead of a manual validate function
- The implementation checks specific conditions, such as whether a list is empty
- This is a good design choice because conversion and validation are often linked
- Implementing standard Rust traits like TryFrom is beneficial because every Rust developer understands their purpose
- This approach provides better IDE support and is more idiomatic than custom validation functions

**Q63**: This could perhaps be described as the "parse don't validate" principle, which we often use. Would you say this principle works well in combination with the Type State Pattern?

**A63**:

- It works well with Type State, but also with other patterns like New Type
- For example, a string that must follow a specific regex format can be validated and converted into a new type simultaneously
- While it is not a requirement for Type State or New Type, it complements both patterns very effectively
- This principle is generally versatile and fits many different architectural approaches

**Q64**: I just remembered one more thing: initially, there was a function that validated the request and had a complex return type consisting of a Result with the validated request and a body response. In the refactored version, the function now returns an enum instead of the response body. I did this to follow the Single Responsibility Principle, so the validation function only returns semantic errors instead of handling HTTP errors. Would you say this improved modularity?

**A64**:

- Not necessarily, as there is usually a reason for using a Result type with an explicit error value
- As mentioned before, I personally do not find it ideal to include error cases within the same enum as success cases
- Error handling is essentially an exception to the Single Responsibility Principle because a function must always be able to handle failures
- In Rust, using a Result is the standard syntax for this instead of throwing exceptions

**Q65**: Alright, thank you for your assessment. I have no further questions. Do you have any general remarks?

**A65**:

- No, but I find some of the changes really good and we should see what we can adopt in the future

### A.6.2 Expert B

**Expert**: M.Sc. Dirk Rusche
**Date**: January 6, 2026, 1:00 pm
**Location**: Online meeting
**Background**: Senior Software Developer, Freelancer under contract at Otto GmbH & Co. KGaA, 2.5 years Rust experience

The following is a summary of the expert interview, which was created with the support of large language models. The summary has been confirmed to be correct in its content by the expert.

**1 - Introduction**

**Q1**: How long have you been working with Rust and in which fields or products?

**A1**:

- Started using Rust in mid-2023 which totals about two and a half years of experience
- Developed functions for Shopify shops that compile to WebAssembly and run on servers
- Worked on backend development at FT9 and various private projects
- Developed full-stack applications involving HTML rendering with Rust

**Q2**: What distinguishes Rust from established languages like Java or Kotlin regarding maintainability and correctness?

**A2**:

- From most to least important, and under the assumption that no unsafe code is used within the Rust implementation:
- Higher number of compile-time checks including the ability to validate SQL queries against live database instances
- Absence of runtime reflection ensures static code execution and prevents runtime failures caused by dynamic method calls
- Integrated Option type eliminates the possibility of null pointer exceptions
- Powerful pattern matching with enums that allow variants to hold different data types and structures
- Exhaustive match blocks require explicit handling of all possible cases which prevents errors during refactoring
- Integrated Result type replaces exceptions and enforces explicit error handling for every case
- Default immutability and explicit mutable references provide clarity on side effects without needing to inspect function bodies
- Generic implementations for types satisfying specific traits allow for reduced code duplication
- Type system utilizes Send and Sync traits to prevent various concurrency bugs at compile time

**Q3**: What are the primary causes of technical debt in large Rust codebases?

**A3**:

- The primary causes of technical debt are often independent of the specific programming language
- Implementation of flawed or leaky abstractions that increase unnecessary complexity
- Failure to write idiomatic Rust code which affects long-term maintainability
- Neglecting the use of specific types or the Newtype pattern in favor of primitive types like strings or UUIDs
- Lack of defensive programming such as using catch-all patterns in match blocks instead of listing cases explicitly
- Missing compile-time enforcements that would otherwise guide developers when extending the codebase

---

**2 - Problems in the current code**

**Q4**: What are the main pain points in the current FT9 codebase regarding maintenance and new changes?

**A4**:

- Poor testability of individual functions necessitates extensive mocking of dependencies and complex initialization of various objects
- Test methods become excessively long and disorganized, making it difficult to identify the actual logic being verified at a glance
- Understanding a single test case requires as much time and effort as understanding the actual function being tested
- High barriers to testing lead to weak validation logic or the complete absence of test cases for critical code sections

**Q5**: Do you think these testing difficulties are due to the test setup or because unit testing in Rust is less effective than in other languages?

**A5**:

- Issues are not inherently caused by the Rust language but by the specific way functions are currently structured in the codebase

- Heavy reliance on asynchronous functions implies the presence of external dependencies that require complex mocking
- Adoption of a functional core architectural pattern would significantly improve testability by separating logic from side effects
- Proper application of functional programming principles would lead to more isolated and easily verifiable code units

**Q6**: Are there any other pain points in the code?

**A6**:

- Long functions and methods that perform too many distinct tasks and are often defined as asynchronous by default
- Difficulty in testing arises from these large asynchronous blocks, which should instead be decomposed into smaller, testable units
- Tight coupling between core business logic and non-functional requirements such as database access, metrics, logging, and tracing
- Cluttered function bodies due to the mixing of infrastructure concerns with the actual application logic
- Infrequent use of defensive programming patterns, leading to situations where new enum variants are overlooked during codebase extensions

**Q7**: Does defensive programming refer only to the example with enums or does it apply generally to other areas?

**A7**:

- Application of defensive programming principles is relevant across various parts of the codebase beyond enums
- Mapping between different data structures presents a risk when new attributes are added to a source struct
- Use of explicit destructuring during the mapping process ensures the compiler flags missing attributes immediately
- Failure to use destructuring can lead to silent errors where new data fields are ignored during transformation
- Manual mapping without compiler-enforced checks makes it difficult to locate all code sections that require updates

**Q8**: How easy or difficult is it for new developers to understand how the domain is modeled in the code and why?

**A8**:

- Basic domain logic is relatively simple as it involves managing benefits, user selections, and activations without complex transaction requirements
- Significant complexity arises from the fragmented system landscape and the specific way components like the Importer and Model Creator interact
- Unclear boundaries exist regarding where data mapping occurs, leading to confusion about whether logic belongs in the Model Creator or the Service
- Fragmented responsibilities make it difficult for newcomers to predict the location of specific business rules or data transformations
- Ambiguity surrounds the classification of different benefit types and why they are handled by different services or touchpoints
- Frequent local code changes without global refactoring lead to multiple redundant implementations for similar use cases
- Lack of a holistic architectural view prevents the unification of similar processes and complicates the onboarding process

**Q9**: When you think about poorly maintainable code, do any specific code examples come to mind?

**A9**:

- The activation service serves as a concrete instance of code that is difficult to maintain
- The use of generics to handle both Customer ID and Visitor ID is a specific point of concern
- Spontaneous recall of other examples is currently difficult without reviewing the codebase again
- A detailed list of further examples can be provided later after a more thorough review if required

**Q10**: Which methods do you know for quantifying these properties and how do you evaluate their relevance and validity in daily development?

**A10**:

- Quantification of code properties is difficult, whereas qualitative assessment is much easier
- Tools like Sonar and other code quality metrics have been used, but their value is questionable
- Metric comparisons are often problematic because it is unclear if they are absolute or in relation to lines of code
- Counting logic structures lacks consistency, such as whether an "if" condition or a "when" case carries the same weight
- Previous experience with metrics shows that it is rarely possible to derive actual improvements from the data
- Relevance in daily development is considered very low because the metrics lack sufficient validity and meaning

---

**Use Case 1 - Status Quo**

**Q11**: Which code smells do you notice in the status quo of the first use case (DetailView Service), and which criteria are negatively affected?

**A11**:

- Mixing of business logic with infrastructure concerns like tracing, logging, and metrics occurs within the DetailView Service
- The main method is overly long, making it difficult to grasp the sequence of operations at first glance
- Modularity is limited because the entry function depends heavily on specific helper functions like Get Valid Benefits
- Reusability of the outer method is low because it acts as an imperative shell tied to specific side effects
- Analyzability is negatively impacted by the length and complexity of the method, requiring significant effort to understand the flow
- Modifiability and Testability are hindered by the large function size, as smaller units would be easier to verify and change in isolation
- Dependencies like the DetailView Cache within helper functions make unit testing more difficult than necessary
- Faultlessness is already relatively high as the code already makes appropriate use of NewType patterns for domain safety

---

**Use Case 1 - Refactoring**

**Q12**: Regarding the refactoring of the DetailView Service, would you say the code still produces the same results in terms of functional correctness?

**A12**:

- Verifying the functional correctness is more difficult now because the code volume has nearly doubled from 170 to over 330 lines
- Increased line count makes it harder to trace whether the logic remains exactly the same as before
- A definitive statement on correctness would require a full review of all components and running the test suite
- Assumption is made that the code remains correct as long as the existing tests continue to pass
- Uncertainty remains regarding whether edge cases that were poorly tested before are still handled correctly in the new version

**Q13**: In what way has the modularity of the code worsened or improved through the refactoring?

**A13**:

- Initial impression of the refactoring points toward high coupling and low cohesion, suggesting a decrease in modularity
- The new structs created by the type-state pattern are not truly modular because they are strictly dependent on one another

- Changes made to one method require checking all subsequent methods in the chain to ensure the data flow still fits
- Modularity has not improved because the original code consisted of a single large function rather than distinct modules
- The overall assessment of modularity remains neutral or "the same" because the logic was simply split into smaller pieces
- The presence of significantly more boilerplate code is noted, though it does not directly change the modularity of the underlying logic

**Q14**: In what way has the reusability of the code worsened or improved through the refactoring?

**A14**:

- Reusability is viewed similarly to modularity, as the logic cannot necessarily be used directly elsewhere without overhead
- Reusing specific parts requires the creation of intermediate types, which adds complexity to any potential integration
- Presence of infrastructure-specific code, such as spans that are created and immediately dropped, complicates the reuse of business logic
- Splitting a method into individual steps technically makes parts accessible, but their actual utility remains questionable without a concrete need
- Evaluation of reusability should consider whether a better abstraction could cover multiple similar use cases rather than just splitting existing code
- Doubt exists regarding whether the current decomposition actually serves other existing or future use cases effectively

**Q15**: So you would say that if you tackle such a refactoring, you would need to include many more code locations and refactor them as well to make the code reusable?

**A15**:

- Approach to refactoring depends heavily on the specific goal of the changes
- Improving general service architecture or reducing duplication requires looking beyond a single service or handler
- Local refactoring of a single method is appropriate if the primary goal is to improve readability and testability within that specific area
- Enhancing maintainability can be achieved through isolated changes if the focus is strictly on making that one method easier to understand

**Q16**: Would you say the effects on modularity and reusability result from the nature of the pattern itself or from how the pattern was applied?

**A16**:

- Linked to the nature of the pattern and how transitions between different types are defined

**Q17**: Regarding Analyzability, is the code easier or harder to analyze and understand following the refactoring?

**A17**:

- Readability has decreased significantly because the total number of lines has doubled, requiring much more effort to scan
- The high-level entry method initially appears interesting because it outlines a clear sequence of operations like fetching IDs and determining benefits
- Complexity arises unexpectedly when a functional call spawns an "up contract task" and returns a join handle
- Duplicate naming conventions cause confusion, such as a method called "Get Valid Benefits" calling a separate function with the exact same name
- Detailed analysis is hampered by wrappers that exist solely to record a single field in a span before returning a new type

**Q18**: The context for that specific function is that it's used elsewhere; making it a state transition would break its reusability for other areas unless those were also refactored. Do you see this as weakness of the pattern or of the existing code structure?

**A18**:

- The observed issues are likely inherent to the pattern itself rather than just the implementation
- Initial feeling suggests that this specific pattern might not be the right fit for the problem being solved
- Readability and testability are perceived as worse than they were in the original code
- Analyzability is negatively impacted by methods like "Create Adjust Link," which performs significant overhead and validation despite its name
- High coupling and low cohesion are evident where methods handle multiple responsibilities like checking headermaps before calling the actual service
- The pattern leads to situations where business logic is buried under boilerplate that primarily serves to pass parameters along

**Q19**: Regarding Modifiability, has the refactoring made it easier or harder to make changes without introducing new errors?

**A19**:

- Modifiability has improved because the functions and methods are now smaller and more manageable
- Clearly defined inputs, type-specific data, and explicit return values provide a better overview for each step
- Isolated units reduce the cognitive load compared to the original long method, where the entire context had to be kept in mind to make a change
- Testability is enhanced alongside modifiability, as the smaller units can be placed under test more effectively
- Shifting certain functions away from being asynchronous significantly simplifies the testing and modification process

**Q20**: To touch on Testability again, would you say it has improved because we have these state transitions? What would you say regarding the number of required mocks or freedom from side effects?

**A20**:

- Freedom from side effects is a direct consequence of shifting away from asynchronous functions
- Absence of mutable references or global data in the new structure effectively rules out side effects
- Smaller methods and the lack of async calls naturally reduce the need for complex mocking in tests
- State transitions create naturally isolated units that are significantly easier to bring under a test suite

**Q21**: Could the testability be negatively affected because you have to construct the state (the struct) for every test instead of just passing input parameters? Or does that not matter?

**A21**:

- Assessment of testability depends on whether it is being viewed in absolute terms or relative to the status quo
- Relative to the original code, construction effort is reduced because you no longer need to satisfy all dependencies for one giant function
- Smaller, refactored units require only the specific data relevant to that transition rather than the entire context
- While constructing state structs adds some overhead, it is still an improvement over the previous monolithic structure
- Absolute testability is likely not at its peak, but the refactoring represents a step forward compared to the starting point

**Q22**: Regarding Faultlessness, meaning domain safety and the ability of the Rust compiler to catch errors at compile time, has the refactoring improved the representation of business logic in the type system?

**A22**:

- Domain safety has neither significantly improved nor worsened in this specific code example

- The refactoring mainly involved splitting functions rather than introducing new type-level constraints that restrict invalid states
- The Type State pattern is most effective when used for branching logic, such as a Builder pattern that yields different subtypes (e.g., HTTP vs. gRPC) based on previous choices
- In the current use case, there aren't distinct execution paths that provide different methods based on the current state
- Greater benefits for faultlessness would be realized in scenarios with complex state flows where only specific transitions are legally allowed by the compiler
- While the structure is different, the lack of conditional state transitions means the type system isn't doing significantly more work to prevent logic errors than before

---

**Use Case 2 - Status Quo**

**Q23**: Looking at the status quo of the second use case, which code smells do you notice, and which criteria are negatively affected?

**A23**:

- Validation logic is fragmented between the type system and manual checks for the target and current context traits
- Redundant validation occurs in multiple places, making the flow less efficient and harder to follow
- String concatenation is used to construct the sheet URL, which is highly error-prone and lacks type safety
- The manual assembly of URLs without a proper builder or type-driven approach increases the risk of runtime failures
- Logic within the handler function remains relatively simple, but the lack of structured URL handling is a notable weakness
- Analyzability and faultlessness are negatively impacted by the manual string manipulation and split validation logic
- Modifiability is hindered by the manual URL construction, as adding or changing parameters requires manual string manipulation rather than simple builder methods
- Analyzability is hindered by returning a generic `IntoResponse` trait, forcing the developer to scan the entire function to determine that it consistently returns a `Tag` in the success case

---

**Use Case 2 - Refactoring**

**Q24**: Regarding the refactoring of the second use case: does the function still do what it's supposed to, or were new bugs introduced?

**A24**:

- Functional correctness is maintained; the logic appears to perform the same actions as before
- The changes are manageable and straightforward, specifically regarding taking a string and creating a `ValidatedDreson`

**Q25**: How has the code's modularity changed following the refactoring, particularly considering the new `ValidatedDreson` struct?

**A25**:

- Modularity has improved because the validation logic is now decoupled from the specific handler and encapsulated within the `ValidatedDreson` struct
- The `FromStr` implementation serves as a clean interface, allowing the handler to remain indifferent to the underlying validation details
- The internal logic of the validation can now be modified or extended independently without affecting the handler's structure

- Implementing `Deref` was suggested to further enhance the design by providing "syntactic sugar" that allows the compiler to treat the struct like a string slice where needed, though its categorization (Modularity vs. Reusability vs. Analyzability) remains debatable

**Q26**: How has the code's reusability changed following the refactoring?

**A26**:

- Reusability has improved because validation logic is now centralized rather than scattered
- The new `ValidatedDreson` type can theoretically be utilized across various other handler functions, facilitating code sharing

**Q27**: How has the cognitive effort required to understand the code changed following the refactoring?

**A27**:

- Analyzability has generally increased; the respondent describes the new code as more concise and clearer
- While the use of `transpose` and specific mapping functions slightly "veils" the logic compared to imperative code, this is viewed as a characteristic of the coding style rather than a flaw in the pattern itself
- Readability is improved by the more compact structure, though questions remain regarding why validation occurs separately from the initial parameter deserialization
- A suggestion was made that implementing the `Deref` trait would further reduce cognitive load by removing the need for explicit `.inner()` calls, allowing the compiler to handle the conversion to a string slice automatically

**Q28**: Regarding Modifiability, is it easier to make changes without introducing errors?

**A28**:

- Modifiability for the handler remains largely unchanged; while the code is shorter, the use of `map`, `transpose`, and `map_err` makes the flow slightly less obvious at first glance than the previous `match` and `if/else` structure
- For the `ValidatedDreson` portion, modifiability has improved significantly because the methods are smaller, more focused, and easier to oversee
- Testability for the handler function itself has not changed, but it has improved "considerably" for the validation logic
- The refactoring allows for isolated testing of the validation logic, which was previously impossible when it was embedded within the larger handler
- The respondent notes a strong correlation between Modifiability and Testability, as both benefit from the same underlying structural improvements

**Q29**: One could argue that testability for the handler has also improved because you no longer need to test the validation logic within that specific context, as it has been moved to external unit tests.

**A29**:

- The argument is considered highly debatable; while offloading validation to unit tests simplifies the handler, it raises questions about how much integration testing is still necessary
- There is a risk that someone (e.g., a junior developer) could bypass the `ValidatedDreson` struct and pass a raw string slice instead, potentially re-introducing bugs if the handler doesn't enforce the type strictly
- If the type-safety isn't strictly enforced across all internal function calls, the enclosing function might still require its own tests to ensure nothing was broken during refactoring
- The trade-off between trusting isolated unit tests versus maintaining redundant coverage in the handler is a complex issue without a one-size-fits-all answer

**Q30**: Finally, regarding Faultlessness: Has the refactoring changed anything in terms of compile-time checks, correctness, or the prevention of runtime errors?

**A30**:

- Faultlessness has improved because functions like `get_benefit` now require the `ValidatedDreson` type as an input parameter

- Since instances of this type can only be created through the validation logic, the compiler guarantees that any data reaching the function is already valid
- This prevents runtime errors by moving the validation boundary to the type-system level, ensuring the function logic always operates on a valid state

---

**Use Case 3 - Status Quo**

**Q31**: Regarding the status quo of the third use case: Which code smells and issues affect maintainability or error-proneness?

**A31**:

- Heavy reliance on mutable references passed between functions creates side effects that make the logic hard to track
- Functions take the entire `Request` or `ProcessorDependencies` as arguments when they only need a small part (e.g., just the body or the Mongo client), obscuring actual dependencies
- The design uses an output parameter pattern (passing a mutable result vector) instead of returning a value, which is less idiomatic and harder to follow
- Analyzability is terrible because the interconnected side effects make it difficult to understand the state of the data at any given point
- Modularity is poor due to the tangled nature of the functions; changes in one place risk breaking hidden assumptions elsewhere
- Individual unit testing of small methods is possible, but testing the big picture is difficult because unit tests cannot easily guarantee the correct interaction of all mutable side effects
- The respondent questions the overall functional correctness because the high number of side effects makes it difficult to verify the exact behavior of the methods

---

**Use Case 3 - Refactoring**

**Q32**: Regarding the refactoring of the third use case: does the code still produce the same results, and is it still correct?

**A32**:

- Functional correctness appears maintained on first glance, though the respondent had not looked at this specific part in detail prior to the interview

**Q33**: How has the modularity of the code changed through the refactoring?

**A33**:

- Modularity has neither improved nor worsened; the respondent notes it remains similar to the previous state
- The implementation uses the Type State Pattern, which helps structure the flow, but adding a new step between existing ones still requires modifying multiple modules
- Because the individual steps are interdependent, they cannot be considered in complete isolation, a limitation that was also present in the original version

**Q34**: You mentioned that the Type State Pattern might actually make the code less modular and less reusable. Could you elaborate on that?

**A34**:

- Reusability is limited in the current implementation because the code defines a fixed sequence of steps
- Since each step is tightly coupled to a specific preceding and succeeding type, the individual parts cannot easily be reused in different contexts
- While the pattern theoretically allows for branching (e.g., transitioning from type A to type B instead of type C), the current linear fixed sequence doesn't take advantage of this
- One could potentially unpack the return value of a step to do something else with it, but the respondent finds it unlikely that a developer would actually use the code that way

**Q35**: How do you assess the Analyzability? Is the code readable, and how high is the cognitive effort required to understand it?

**A35**:

- Readability and comprehensibility have improved because the workflow is now clearly split into distinct phases: validation, fetching from MongoDB, processing, and the final response
- The removal of mutable side effects makes the code significantly more understandable, even if the total line count has increased
- While it might take more time to read through the entire file due to the increased volume of code, the individual sections are self-contained and free from global or shared mutable state
- The overall flow is much clearer and more manageable than the original tangled version

**Q36**: Would you say, and this is something I noticed, that the Type State Pattern forces you to introduce a name for every state and every transition, which might automatically help with comprehensibility if the naming is done correctly?

**A36**:

- If the naming is done correctly, then yes
- In the first use case, some names felt somewhat unfitting; for instance, a state might be named “A” but it was actually performing actions A, B, and C
- Correct naming is essential for the pattern to actually improve clarity

**Q37**: Naming strategy can be a problem with new patterns. I wondered whether to name a state after the result (e.g., “ItemsFetchedResult”) or the next logical step. Depending on the reader’s expectations, this affects readability. Would you say it improved for the third use case?

**A37**:

- Mhm (agreement)

**Q38**: Has the refactoring made it easier or harder to make changes without introducing new bugs?

**A38**:

- Modifiability has improved because final responses are now represented as an Enum, which encapsulates how they are composed and created
- Making changes is significantly easier because mutable references have been eliminated in favor of small, self-contained methods
- Functional purity contributes to better modifiability
- Separating the logic into distinct phases, like validation versus the actual processing of the validated request, allows for simpler and safer modifications

**Q39**: Has testability improved or worsened through the refactoring?

**A39**:

- Testability improved because input and output parameters are clearly defined and the elimination of mutable references prevents unforeseen side effects
- The shift to clear return values makes it significantly easier to write tests and ensures the functions are more predictable

**Q40**: I tried to use Enums for results. In the status quo, error states were often implicit, like an empty array or a None value. Does this use of Enums help with testability or other criteria, and how do you feel about the number of mocks or functional purity?

**A40**:

- The use of Enums has a positive effect on functional correctness and faultlessness because it makes success and error states explicit rather than implicit
- Analyzability is improved because it is easier to trace whether the logic is correct, though the expert is skeptical if it significantly changes testability compared to testing for empty arrays

- Testability and functional purity are linked; the reduced need for mocks in specific methods makes testing easier because functions no longer depend on complex global state or mutable references

**Q41**: Is there anything else you would like to add regarding this use case, or did you notice anything we haven't discussed yet?

**A41**:
- While the code is more readable and looks cleaner, the respondent believes the Type State Pattern is not the ideal choice for this specific scenario
- Because the implementation is just a linear flow from A to B to C without complex state transitions, the pattern doesn't offer unique benefits that couldn't be achieved with simpler architecture
- The respondent questions whether the benefits outweigh the overhead of the extra lines of code and the mapping complexity introduced by the pattern

### A.6.3 Expert C

**Expert**: B.Sc. Falk Woldmann Lu
**Date**: January 8, 2026, 1:30 pm
**Location**: Online meeting
**Background**: Senior Software Developer, Otto GmbH & Co. KGaA, 3 years Rust experience

The following is a summary of the expert interview, which was created with the support of large language models. The summary has been confirmed to be correct in its content by the expert.

**1 - Introduction**

**Q1**: How long have you worked with Rust and in what area or on what products?

**A1**:
- I have been dealing with Rust since winter 2022, totaling about three years, and have worked with it professionally since mid-2023
- I build web applications for Otto that handle significant load and therefore have strict requirements for performance and efficiency
- Parallel to this I maintain a relatively well-known HTTP mocking library for Rust

**Q2**: What distinguishes Rust from languages like Java or Kotlin regarding maintainability and faultlessness?

**A2**:
- Rust has a much more powerful type system compared to Java which allows embedding more logic that the compiler can formally validate
- Expressing logic formally improves maintainability because the code requires less manual maintenance over time
- Faultlessness is a negative correlation to maintenance work where using the compiler is much more powerful than using runtime if-conditions or logical operations
- This is very effective for our use cases because processes like filtering, sorting, and aggregating can be expressed well as state machines

**Q3**: What are the primary reasons for technical debt in large Rust codebases?

**A3**:
- A major issue is the lack of uniform standards and the fact that different approaches were tried during an evolution phase without consistently cleaning them up
- Many people I work with are not yet completely in a Rust-based way of thinking and still have a mindset influenced by Java or Kotlin
- This leads to attempts to encapsulate the application flow in class-like structures even where it makes no sense or using traits for abstraction when enums would be a better idea

**2 - Problems in the current code**

**Q4**: What are the biggest pain points in the current codebase and what bothers you when maintaining the code or making new changes?

**A4**:

- The primary issue is the attempt to map object-oriented thinking onto Rust at any cost, which leads to an immense amount of boilerplate code and unnecessary layers of indirection where services call other services and repositories
- This approach creates significant redundancies and forces the use of excessive mocking, resulting in a situation where a developer must read a lot of code to find a very small amount of actual logic
- There is a tendency to use traits for shared logic when enums would be a much better fit, and the use of static dispatch can result in bloated trait bounds that negatively impact the readability of the code

**Q5**: How difficult is it for new developers to understand how the domain is modeled in the code and why?

**A5**:

- It is very difficult because we lack a formal logical definition of the domain from the business side and have not yet developed one ourselves
- The codebase consists of approximately 120,000 lines of Rust code where logic is distributed across many locations and obscured by boilerplate and indirection
- While the current state is not a regression compared to previous Java or Kotlin implementations it is difficult for developers to maintain a large enough mental context to fully grasp the domain
- Since Rust provides more powerful tools for modeling we know that a better implementation is possible but limited time and resources often prevent us from achieving it

**Q6**: Can you specify what you mean by saying that Rust is more powerful, and how this helps developers understand code better?

**A6**:

- Rust allows developers to express large parts of the domain logic directly in the type system, using fewer constructs while making these aspects part of the overall design
- When the domain logic is modeled correctly, understanding the code becomes easier because state transitions and valid states are visible in the types
- This approach requires effort to design initially, but once established it simplifies reasoning about the system
- In Kotlin or Java, similar modeling is much harder and usually requires many additional classes and boilerplate code
- Developers often continue to use Java-style modeling patterns, which leads to flat type hierarchies and does not fully use Rust's type system

**Q7**: When you think about poorly designed code, can you think of concrete examples?

**A7**:

- Poorly designed code is often found in service layers where domain logic is mixed with imperative flow, async handling, database access, and technical concerns
- Additional structs are introduced mainly to enable mocking, even when they are not required by the domain itself
- Validation is done repeatedly on primitive types like strings instead of encoding invariants in the type system
- The actual business logic is scattered and hidden between boilerplate and infrastructure code
- This makes the code hard to maintain because developers must work through a large amount of irrelevant detail to find the important logic

**Q8**: You have the guideline with six criteria that will later be used to evaluate the refactored code. You do not need to address every single criterion, but can you describe methods you know to quantify such properties, for example maintainability or code smells, and how relevant and meaningful you consider them in everyday development?

**A8**:

- For modularity and reusability, there are metrics based on dependency graphs, cohesion, and coupling, which are known from Java and academic contexts, but their concrete definitions and usefulness are often unclear in practice
- For analyzability, tools like SonarCloud provide metrics such as cognitive complexity, but the expert is unsure how reliable or helpful these scores actually are
- For modifiability, the expert cannot name concrete metrics
- For testability, code coverage exists but is seen as limited and not a strong indicator of real testability or code quality
- A more meaningful view of testability is how much logic is implemented as pure functions following a functional core, imperative shell approach, enabling side-effect-free unit and property testing
- Beyond tests, encoding logic and invariants in the type system is seen as the strongest approach, as it shifts correctness towards formal verification at compile time

**Q9**: What is your view on metrics such as the Maintainability Index, cyclomatic complexity, or similar code metrics, and how meaningful do you consider them in practice?

**A9**:

- Cyclomatic complexity seems reasonable as a metric because it reflects control flow and helps avoid fragmented logic that is hard to follow
- Lines of code are not meaningful as a quality metric, especially in Rust, because more code can simply mean explicit modeling of states or data structures
- A large number of structs is not problematic, while many conditional branches in a single function are
- Lines of code can at best give a rough sense of project scale, such as distinguishing between small and very large codebases

---

**Use Case 1 - Status Quo**

**Q10**: Let us move to the use cases. We start with the first one, the Detail View Service, and first look at the status quo. What code smells do you notice, where is the code hard to maintain or error-prone, and which of the criteria from section 2 do you think are negatively affected?

**A10**:

- It is unclear why Detail View Service is a struct at all, and why related parts like Variation Service are also separate structs, since this seems mainly done for mocking rather than for domain reasons
- Mocking is seen as an anti-pattern here; if anything, only external edges like databases or HTTP services should be mocked, and even databases can often be tested with local instances instead of mocks
- The file contains a lot of boilerplate early on, where large parts could be removed without losing real logic, which makes the code unnecessarily long and harder to work with
- I/O and business logic are mixed, and the implementation is strongly imperative with large conditional blocks and many boolean operations, which is unpleasant to read and likely increases error-proneness
- The domain logic looks like it could be modeled more cleanly, possibly like a simple state machine, but instead it is tangled with indirection, side effects, and Spring-like structuring, rather than embedding transitions and constraints more clearly

**Q11**: Are there specific criteria that stand out to you, for example analyzability or faultlessness, that are negatively affected by this code?

**A11**:

- Analyzability can be treated as readability here, and the readability is reduced because I/O, non-I/O logic, domain concerns, and additional technical elements are mixed, forcing constant context switching and leaving a lot of technical slop between the actual domain reasoning
- This mixing also harms modifiability, because changes require navigating intertwined concerns and understanding many scattered steps instead of a clear, isolated domain flow

- Testability is weakened because testing relies on mocking and heavy setup, even though much of the behavior could be tested as pure logic with simple parameterized tests and ideally property tests
- Reusability is limited because the code mainly describes state transitions like filtering and iterating, but these transitions are tied to I/O; if the domain logic were separated from I/O, it could be reused without copying or extracting a large function
- Faultlessness is negatively affected because many runtime checks and many if-conditions make it easy to break invariants by changing lines or order, and the invariants are hard to keep in mind since the logic is spread across the file

---

**Use Case 1 - Refactoring**

**Q12**: Let us move to the refactoring. We will look at the new code and go through the selected ISO criteria, starting with modularity. To what extent did the refactoring worsen or improve the code's modularity?

**A12**:

- The refactoring encodes the state transitions of the flow in the type system in a more direct way, with cleaner and relatively pure, type-state-like code, which also makes it easier to reuse the flow elsewhere and move it into a library
- After clarifying the ISO definition of modularity as limiting the impact of changes between components, the expert does not see major changes at the file/component level, since the number of files stayed the same and the code remains encapsulated in the same place
- Still, the internal structure is seen as more coherent and more formalized, so invalid internal changes are more likely to be caught as compile errors or become unrepresentable, which the expert interprets as better protected inherent modularity, even if it also relates to faultlessness

**Q13**: On reusability, you already touched on it. Can you explain more how the refactoring affects reusability?

**A13**:

- By defining the individual state transitions more explicitly and using more pure functions, the logic becomes easier to extract, move elsewhere, and reuse in other places
- Parts of the logic might be reused even if the concrete cases differ, because the general flow could still be comparable and suitable for abstraction
- The refactoring also makes it easier to understand what the business logic actually does, and this clarity helps identify which parts can realistically be reused and which cannot

**Q14**: Could the applied type-state pattern also have negative effects on reusability, for example because you can only reuse an extracted state transition elsewhere if you already have the required state, which you might not be able to construct without going through other valid transitions?

**A14**:

- The expert does not see this as a drawback but as an intended feature, because using a transition without having the correct prior state should not be possible
- If some logic should be reused in a different context that does not fit the same type-state flow, it should be extracted further into a smaller function and then called from both places
- Being able to jump into the middle of a flow is considered undesirable, and if that is needed it suggests the business logic is modeled incorrectly

**Q15**: Regarding analyzability or readability, is the refactored code easier or harder to read, analyze, and understand?

**A15**:

- The refactored code is unconventional for the team because they are used to imperative code, but it is still considered clearly easier to understand
- Although the refactoring introduces more code, such as additional structs and explicit transitions, this code is simple and descriptive rather than complex
- The added code represents high-level structure and can be quickly grasped or mentally collapsed, unlike deeply nested logic and many if-conditions

- Understanding the overall business logic becomes easier because the flow is explicit and less cluttered with low-level control flow
- There is an initial learning curve because the style differs from Java or Kotlin, but once familiar, the code is much more understandable
- Improving analyzability in practice is especially important because most work is done in existing codebases where reading and modifying code dominates

**Q16**: Do you think that naming the states correctly becomes a bigger challenge because the pattern forces you to name every single state and transition, even when it doesn't always make sense?

**A16**:

- The expert sees this not as a drawback but as a feature, because if a state or transition cannot be named meaningfully, this indicates a problem in the domain model
- In a well-understood domain, every step should be describable, similar to modeling a process in BPMN
- Difficulties in naming suggest that the domain is not fully understood or is being modeled incorrectly
- This situation encourages closer interaction with business stakeholders to clarify the domain
- The pattern helps make implicit uncertainty explicit, whereas previously such issues were hidden in tangled code
- Making such problems explicit is seen as a general strength of Rust and of this modeling approach

**Q17**: Just to summarize, you said that the added boilerplate, which almost doubles the code size, does not have a negative effect on analyzability for you. Is that correct?

**A17**:

- The additional boilerplate is irrelevant for analyzability because it can be collapsed, navigated easily, or moved into separate files
- Metrics like lines of code are therefore meaningless, because the refactored code is better in most non-functional aspects despite being larger

**Q18**: Regarding modifiability, has it become easier or harder after the refactoring to change the code without introducing new errors?

**A18**:

- Modifying the code is much easier because invalid flows and wrong ordering of steps are prevented by the type system at compile time rather than by runtime checks
- Since the refactored code is also easier to read and analyze, changes can be made with more confidence and a lower risk of introducing new errors

**Q19**: Imagine you want to add a new feature, such as an additional filter, an aggregation, or another service call. Previously you could insert a line somewhere in the flow, but now you need to decide which state it belongs to, possibly introduce a new state, and adapt surrounding transitions. Do you see disadvantages in this, such as increased effort, frustration, or developers putting logic into the wrong state?

**A19**:

- The additional effort is intentional and beneficial because it forces developers to think about the correct placement of logic
- The approach follows the idea of catching errors as early as possible in the development process, ideally at compile time and not in production
- If logic does not fit into the existing state flow, this suggests a problem in how the domain is modeled; in such cases, the model should be revised, ideally together with business stakeholders
- The Rust compiler helps in this process by providing feedback when the code does not compile
- Like any guideline, the pattern can be bypassed, but if the team agrees on its benefits, it is likely to be applied consistently

**Q20**: Regarding testability, how did it change through the refactoring, and can you also comment on the number of required mocks and the freedom from side effects?

**A20**:

- Testability improves overall, but some parts still mix business logic with side effects, so the refactoring does not fully eliminate the original issues
- Using type-state without fully separating side effects adds some overhead and limits the potential benefits
- Type-state fits well with a functional core, imperative shell approach, because state transitions can be implemented as pure functions
- When logic and invariants are moved into the type system, fewer tests are needed and test setup becomes simpler

**Q21**: Finally, regarding faultlessness: has the error-freedom of the code improved through the refactoring, are invalid application states prevented, and has the risk of bugs or incorrect behavior been reduced?

**A21**:

- The expert explicitly calls this a suggestive question, because the prevention of invalid states is an inherent goal of the refactoring
- By formally modeling states, invalid states become unrepresentable and are checked by the compiler
- This shifts correctness guarantees to compile time and clearly reduces the risk of bugs and incorrect behavior
- The correctness of these guarantees could, in principle, also be argued formally with pen and paper

**Q22**: Do you think the question is too suggestive?

**A22**:

- Preventing invalid states is a core promise of the type-state pattern, so the question is easy to answer rather than misleading
- Faultlessness is a valid criterion from the standard, so the question is appropriate and justified

**Q23**: Do you want to add anything else about the refactoring, and do you also see concrete disadvantages of the type-state pattern in this use case?

**A23**:

- The type-state pattern works best when consistently combined with pure functions, and applying this more strictly could further increase its benefits
- The added boilerplate could be reduced with macros or existing libraries, and the main challenge is teaching the pattern and making the state flow explicit, for example with diagrams
- Initial unfamiliarity can make the pattern harder to grasp at first, but this improves with practice, while unrelated issues like indirection-heavy mocking should be questioned separately

---

**Use Case 2 - Status Quo**

**Q24**: Let us move to the second use case, the Tag Component handler. We again look at the status quo: what code smells do you notice, where is the code hard to maintain or error-prone, and which criteria do you think are negatively affected?

**A24**:

- The code uses plain strings (and option strings) for Dreson even though Dreson is a specific, important domain data type, which is especially inconsistent because the struct is called Validated Params but the type system does not reflect that the value is actually a valid Dreson
- Because Dreson stays a string, validation knowledge is not preserved, so the value may be revalidated later unnecessarily, or validation may be forgotten entirely, which makes the code error-prone and encourages operating on raw strings instead of constrained domain types
- Introducing a dedicated Dreson type with central parsing would improve faultlessness and modifiability, and could reduce test effort by testing parsing once and reusing it, for example by extracting it into a library
- In the status quo, reusability is low because validation and parsing are not encapsulated in a reusable domain type, and validated logic can be inconsistently applied across the codebase

- There is recurring indirection through services and caches that the expert finds unnecessary, suggesting the logic could be kept more directly in the handler or organized without excessive layering
- The code has awkward option mapping and multiple nested options and sums with additional if-conditions, leading to deeper nesting and increasing complexity and readability problems

---

**Use Case 2 - Refactoring**

**Q25**: If you have nothing else to add on the status quo, we move to the refactoring and the ISO criteria. Starting with modularity: to what extent did the refactoring worsen or improve the code's modularity?

**A25**:

- Modeling Dreson as a dedicated domain type is considered a clear improvement, because invariants are maintained through validated construction rather than leaving the value as a raw string
- The expert notes minor improvement potential by converting from string to the validated type as early as possible, but treats this mainly as a small refinement
- The naming Validated Dreson is criticized, because a proper domain type should only represent valid values and therefore should simply be called Dreson, ideally extracted into a library
- In terms of modularity, expressing such a domain concept as its own type is seen as the strongest form of isolation, so the change is viewed as essentially perfect for this aspect

**Q26**: Can you also comment on analyzability, not only for the newtype, but also for the code where it's integrated?

**A26**:

- Analyzability improves because the types act as code hints: when reading the code you immediately see that a value is a valid Dreson, rather than an unconstrained string
- When this typed value is passed down into deeper layers of the business logic, it becomes faster to understand what you are working with, because you no longer need to check again whether the string is valid before continuing

**Q27**: Do you think the call with transposed and map-error is good in this handler, or does it reduce readability?

**A27**:

- The expert finds it somewhat fiddly and would prefer implementing `TryFrom` and mapping into that conversion
- With `TryFrom`, error handling becomes straightforward: if conversion fails, you get the error back, and optional map-error logic can stay inside the conversion if needed

**Q28**: Regarding modifiability, did the refactoring make it easier or harder to make changes without introducing new errors?

**A28**:

- Modifiability improves because the typed parameters make it clear what each value represents, instead of passing multiple plain strings that can be accidentally swapped
- With distinct domain types, it becomes much harder to introduce such mistakes, because errors are prevented by the type system rather than only being caught in tests or in production

**Q29**: If you relate this to other functions that use this Dreson type: do you see it as negative or positive that you are now forced to operate with the new type, for example when adding a new function or integrating an existing function that expects a string?

**A29**:

- Being forced to use the domain type is considered a feature because it makes the code more explicit and more readable, even if adapting interfaces is initially annoying
- The compiler helps with the one-time migration work, and afterward it is easier to understand functions because the parameter type directly shows it is a valid Dreson and does not need revalidation
- If a string is required, the type can be made ergonomic via `deref` to string/slice and simple conversions like `Into`, so using the new type does not block practical integration

- Needing a raw string should mostly occur at boundaries such as database access, and even there using a string slice via `deref` is typically sufficient

**Q30**: Regarding testability, how did the refactoring affect the testability of the code?

**A30**:

- Testability improves because validation and parsing only need to be tested once at the boundary
- After that, the validated type can be reused throughout the codebase without duplicating parsing or validation tests

**Q31**: Did anything change here regarding the number of required mocks or side effects?

**A31**:

- Nothing changed regarding mocks or side effects, because no mocking is needed and a valid domain value can be constructed directly
- Constructors like `new_unchecked` are considered a code smell and should only exist, if at all, for a single test that verifies invalid values are rejected, but tests could instead rely on the `Default` trait

**Q32**: Finally, regarding faultlessness: how did the refactoring improve error-freedom, especially with respect to invalid states?

**A32**:

- Invalid states cannot occur anymore because invalid Dreson values can never enter the system if validation happens at the earliest possible point
- Removing validation later in the code cannot introduce invalid values, because the type system already enforces correctness
- Rust doesn't allow passing parameters in the wrong order, which would otherwise only be caught at runtime

**Q33**: Do you have any further general comments on the refactoring, including possible disadvantages or additional remarks?

**A33**:

- Newtypes are already used in our codebase, and using newtypes is appropriate for this use case
- Constructors like `new_unvalidated` should be avoided; conversions should be handled via `TryFrom` / `TryInto`, with ergonomic access through `Deref` to string slices, which removes the need for patterns like `transpose`
- The already existing "PSR" newtype is very similar to the Dreson and could be reused or unified to reduce duplication and improve extensibility in the future

---

**Use Case 3 - Status Quo**

**Q34**: For the last use case, the Boxfish API handler: what code smells do you notice, where is the code hard to maintain or error-prone, and which criteria from section 2 are negatively affected?

**A34**:

- The code has many nested constructs, including loops inside multiple if-conditions, which makes it hard to read and understand
- I/O is mixed with non-I/O logic, so technical concerns and business logic are intertwined
- There is a lot of slop code that is not immediately understandable or clearly necessary
- The `build_final_response` part is described as especially messy due to several nested if-conditions

**Q35**: Do you have an opinion on the mutable vectors for results and API errors that live in the handler function and are passed into other functions, and on the design where those functions do not return values but instead mutate these shared structures?

**A35**:

- This is criticized strongly as unidiomatic in Rust, since Rust is immutable by default and extensive mutability is seen as bad practice

- Passing mutable vectors around and mutating them across functions is expected to create problems, especially if concurrency is introduced, for example when moving work to other threads
- The expert does not see why this is necessary, because the logic looks like it mainly performs filtering and looping, which could be expressed without shared mutable accumulation
- The overall style is described as Java-like, and the pattern of looping and pushing into mutable lists is seen as harming correctness and faultlessness

---

**Use Case 3 - Refactoring**

**Q36**: Regarding modularity, did the refactoring worsen or improve the code's modularity?

**A36**:

- The original code was largely in one large handler function, so modularity was not strongly present to begin with and is not the main focus of this change
- The refactoring defines the business flow more formally, which the expert associates with slightly improved modularity
- The improvement is described as small compared to the earlier Dreson case, but it is clearly not worse and overall slightly better

**Q37**: Given the definition of modularity as limiting how changes in one component affect others, how do you assess that the refactoring removes mutable state that was previously modified across different functions? Does this affect modularity?

**A37**:

- Removing shared mutable state improves modularity because functions or state transitions are less able to affect each other indirectly
- Without `mut` and mutable references being passed around, changes in one part are less likely to unintentionally influence other parts of the flow

**Q38**: Regarding reusability, how did the refactoring worsen or improve the reusability of the code?

**A38**:

- The use case itself is fairly self-contained, so there is limited direct potential for reuse
- However, by expressing the logic and flow more explicitly, it would now be easier to reuse or adapt similar logic elsewhere if needed
- The refactored structure makes the logic more visible and explicit compared to the previous version with shared mutable state

**Q39**: Regarding analyzability, is the refactored code easier or harder to read, analyze, and understand?

**A39**:

- The refactored code is much easier to understand because removing shared mutable vectors eliminates uncertainty about what has already happened in the flow
- By expressing the flow explicitly through states, it becomes clear where the code is conceptually, which makes reasoning and re-entry much simpler

**Q40**: Regarding modifiability, did the refactoring make it easier or harder to change the code without introducing new errors?

**A40**:

- It is harder to introduce new errors, because the structure now prevents many classes of mistakes by construction
- The type-based flow forces changes to be made in the correct place instead of accidentally modifying shared state
- Removing shared mutable state eliminates a major source of accidental breakage
- As a result, developers can rely more on the existing logic and make changes with higher confidence

**Q41**: Can you say something about the testability of the code? Did it improve or worsen, and how does this affect the need for mocks or handling side effects?

**A41**:

- The expert cannot make a concrete assessment because there are no tests for this function and they are not sufficiently familiar with the codebase
- Despite that, the refactored version would be preferred for testing because it is easier to understand
- Improved analyzability helps identify what should be tested and what a sensible test setup looks like
- Since tests also serve as documentation, clearer code leads to clearer and more useful tests

**Q42**: If you look at the processor dependencies struct: previously the whole bundle of services was passed into the function, but now only the specific services needed are passed. Do you think this affects testability?

**A42**:

- Passing the whole dependency bundle is annoying because it forces setting up and mocking services that are not needed
- If the refactoring now passes only the required dependencies, this is clearly better and more convenient for tests
- Needing less setup and fewer mocks is generally a direct improvement for testability

**Q43**: Regarding the last criterion, faultlessness, do you want to add anything beyond what you already mentioned?

**A43**:

- There is nothing additional to add, since the same aspects discussed before apply here as well, because the same pattern leads to the same benefits

**Q44**: Do you have any general remarks on the refactoring?

**A44**:

- The coding style is very different from what the team is used to, which can feel overwhelming at first
- Some naming choices are criticized, for example using the term “Result,” which has a specific meaning in Rust but is used differently here
- These issues are considered minor nitpicks rather than fundamental problems
- Other aspects, such as existing API error handling, were already present before and are not caused by the refactoring

**General Remarks**

**Q45**: Do you have any final general remarks on the patterns, or anything you would like to add in conclusion?

**A45**:

- The expert evaluates the patterns very positively and sees strong value in them
- It would be interesting to involve business stakeholders and let them model one of the modules themselves to compare their understanding with the code
- The approach is seen as both more formal and closer to the business domain, but it requires learning and practice within the team

### A.6.4 Expert D

**Expert**: B.Sc. Fabian Diez
**Date**: January 9, 2026, 1:00 pm
**Location**: Online meeting
**Background**: Senior Software Developer, Otto GmbH & Co. KGaA, 3.5 years Rust experience

The following is a summary of the expert interview, which was created with the support of large language models. The summary has been confirmed to be correct in its content by the expert.

**1 - Introduction**

**Q1**: How long have you been working with Rust and in what field or on what products?

**A1**:

- Private experience for about 3.5 years and professional experience for 1.5 to 2 years
- Works in the Boxfish team at Otto Marketplace, specifically on generating return labels
- Replaced an old Java service with a new Rust service that uses Typst as a library for PDF rendering
- Chose Rust and Typst to meet the requirement for high-efficiency, real-time label generation for the frontend
- Implemented the service using AWS Lambda and the official AWS toolstack for Rust

**Q2**: What distinguishes Rust from Java or Kotlin regarding maintainability and faultlessness?

**A2**:

- Faultlessness is the biggest advantage of Rust compared to other languages
- While Kotlin improves on Java through nullability and its val/var distinction for mutability, Rust goes further with its sensible defaults
- In Rust, variables are immutable by default unless explicitly marked, whereas Java requires the extra effort of adding “final” to achieve safety
- Most real-world bugs the expert encountered over five years resulted from missing data or lack of verification, issues that strict type systems prevent
- Rust forces developers to explicitly acknowledge when they are bypassing safety features like using unwrap, making dangerous code sections easy to spot
- Unlike Java, which suffers from unchecked exceptions that can unexpectedly break control flow, Rust makes the control flow explicit and prevents hidden crashes

**Q3**: What do you consider the primary causes of technical debt in large Rust codebases, particularly when looking at the potential disadvantages of the language?

**A3**:

- The language is extremely feature-rich and complex, often offering too many different ways to achieve the same result
- Onboarding new team members is difficult because it takes a long time to master the patterns and specific syntax, which is a major disadvantage during team rotations
- In the domain of business software, the mandatory focus on memory management often feels like unnecessary overhead since they aren’t writing low-level software
- While Go might be a simpler fit for business logic due to its garbage collection, it lacks the unique safety features that still make Rust attractive
- Experienced developers from languages like Java cannot easily transfer their intuition for code structure and must essentially start fresh to learn idiomatic Rust, such as using the From trait or specific control flow macros

---

**2 - Problems in the current code**

**Q4**: What are the biggest pain points in the current codebase, and what bothers you about maintenance and making new changes?

**A4**:

- Advantages of the type system are not used consistently, such as passing raw strings instead of using Enums for defined cases
- This neglect of the type system leads to a direct loss of faultlessness
- There are many impure functions with side effects that mutate lists internally and take unnecessary dependencies
- The application flow is difficult to follow and requires deep diving through multiple functions to understand the extent of side effects

- The code is unwieldy and lacks modularity, making it nearly impossible to test parts in isolation or reuse existing logic

**Q5**: To summarize, everything is tightly coupled, making it hard to extract or reuse logic for other use cases. Let's move on to the next question: How difficult is it for new developers to understand how the domain is modeled in the code and why?

**A5**:

- It is extremely difficult because the domain logic is not self-documenting; instead, functions often perform 20 tasks in a row
- Knowledge about the different states and cases is implicit in the control flow rather than being explicitly defined in the type system
- The code appears to be a direct 1:1 translation from old Java to Rust without any attempt to rethink or remodel the architecture for a new language
- Developers porting code from Java often lack the necessary patterns or "handicraft" to find a fresh approach, leading them to simply replicate procedural functions

**Q6**: The same issue occurred during the migration of the Detailview Service from Kotlin. We followed the existing structure too closely instead of remodeling the business case from scratch.

**A6**:

- Agrees that failing to remodel during a rewrite is a common reality in the industry
- Finds it interesting that both the complex Detailview Service and the simpler Tag Handler suffered from the same root cause

**Q7**: Do you have specific examples of poorly maintainable code?

**A7**:

- The expert views functions with heavy side effects and mutable references as the primary drivers of unmaintainable code
- Deep nesting and high indirection levels make it difficult to follow the logic, often requiring developers to jump multiple layers deep to find trivial operations
- Use of helper functions that merely wrap simple tasks like logging or pushing to a list are bad, as they hide behavior without providing meaningful abstraction
- This lack of locality makes it dangerous to change code because the impact on the overall flow is hard to predict and isolate

**Q8**: We have similar problems where many services call further services and so on. This probably comes a bit from the Java mindset, but it is very interesting.

**A8**:

- In Java, this problem is rare because Dependency Injection makes almost everything indirect, which simplifies mocking
- Rust does not provide this by default unless you explicitly create Trait objects or Traits for every service struct
- The team is considering using Traits for database repositories to make them easier to mock, applying the Dependency Inversion Principle from SOLID
- It is much cleaner when a service depends on a Trait so you can replace the concrete implementation, like MongoDB, with something else

**Q9**: It is interesting that traditional software engineering principles don't really help once you are in Rust, right?

**A9**:

- These principles can be mapped to Rust, but they don't work out of the box like they do in Spring where Dependency Injection and Inversion are built-in
- Developers often try to transfer their gut feeling from Java 1:1 to Rust, which fails unless they take the extra step of building interfaces or extracting logic

- A better approach in Rust is to encapsulate the core domain logic in pure functions without side effects, making them easy to test without mocks
- While mocking is easy in Java, it can reduce the value of tests because they drift further away from the actual runtime behavior

**Q10**: We use six criteria for evaluation. What methods do you know to quantify these, and how relevant are they in your daily work?

**A10**:
- Code coverage is the only metric consistently used in their professional practice, particularly in the Java context
- Modularity metrics exist in academia and some tools, but the expert has not encountered or used them in a Rust professional environment
- Reusability can be measured by the amount of duplicated code; highly coupled functions with technical dependencies (e.g., MongoDB) are significantly less reusable in different contexts
- Analyzability is often associated with the Cognitive Complexity score, a concept known from tools like SonarQube in the Java world, though not currently used by his team in Rust
- Modifiability is best measured by the time required for a change and the total lines of code that need to be adjusted; high nesting and side effects make atomic changes much harder

**Q11**: Those would be more like Ops metrics, looking at how many commits change how many lines of code over time—process-based rather than static code analysis. Do you know the DORA metrics?

**A11**:
- The expert is not familiar with the DORA Software Delivery Performance Metrics
- These metrics play no role in the team's daily business or regular discussions
- It sounds like something from university that hasn't translated into their current work routine

**Q12**: I understand. That is already a good insight. Please continue with Testability.

**A12**:
- Test coverage is the primary metric used for this category
- It has been used extensively in past projects to measure how much of the codebase is exercised by tests

**Q13**: Do you think test coverage actually says something about testability? Coverage only shows how much you cover, not how testable the code is. For example, coverage could be low simply because someone was too lazy to write tests.

**A13**:
- There is an indirect connection because the team often skips testing certain parts of the code specifically because they are extremely annoying to test in Rust
- Unit tests are limited to isolated areas that are easy to test, while the rest is handled by integration tests
- This is a significant gap compared to Java, where basically every line and class is tested and everything untestable is simply mocked
- Rust's lack of reflection makes testing frameworks less powerful, and tools like Mockall require the extra overhead of writing Traits
- Testability depends heavily on code quality; the more isolated and separate the logic, the easier it is to test
- In reality, the team achieves about 60% coverage through unit tests and 20% through integration tests, leaving a 10% gap that is difficult to identify and close

**Q14**: Have you ever heard of cyclomatic complexity? It's defined via the control flow graph. Basically, you look at how many independent execution paths a function has. It's often called McCabe Complexity and was originally developed to determine how many paths you need to cover with tests, so it's actually a testability metric, though today it's often used for analyzability.

**A14**:
- Confirms that it sounds very similar to Cognitive Complexity since both factor in control flow and branching

- Admits he hadn't specifically considered it as a metric for testability before

**Q15**: That is quite interesting. Now, moving to the last criterion, Faultlessness: can you think of any quantifiable metrics or KPIs for that?

**A15**:

- You could count the number of unsafe operations, such as `unwrap` or `unsafe` code blocks in Rust, or the `!!` operator in Kotlin
- Mutation testing is a valuable method for verifying robustness; the expert mentions using tools like "Cargo Mutants" manually
- Mutation testing helps see how robust the code is by intentionally breaking parts of the logic to see if tests fail or if the code even compiles
- Ideally, the code should be structured so that invalid states or "broken pieces" are caught by the compiler itself
- Combining compiler safety with strong test coverage ensures a high degree of faultlessness

---

**Use Case 1 - Status Quo**

**Q16**: Let's move on to the actual use cases, starting with the Detailview Service. Which code smells did you notice in its current state, and are there areas that are particularly difficult to maintain or prone to errors?

**A16**:

- The primary issue is a massive, sequential function called `create_detailview_view_model` that handles too much logic at once
- Making changes is risky because an adjustment early in the function affects everything downstream, increasing the chance of unintended side effects
- Testing is difficult and requires an excessive number of mocks because the logic is not modular enough to be tested in isolation
- Analyzability suffers because you have to read through the entire sequence to understand what the function actually achieves
- Faultlessness is compromised because the code doesn't strictly enforce the correct order of operations; it relies on implicit flow rather than compiler-enforced states
- There's a risk of passing incorrect data (like swapping a Variation ID with a Benefit ID) to the wrong call; the code would still compile but the logic would be broken

---

**Use Case 1 - Refactoring**

**Q17**: Let's move on to the refactoring. To what extent has the modularity of the code worsened or improved through the refactoring?

**A17**:

- Modularity has significantly improved
- Previously, everything was trapped in one massive function; now, the logic is distributed across many small structs
- Changes are now contained within specific functions defined on those structs, preventing global side effects
- The refactoring established clear boundaries between the individual processing steps
- By breaking the sequential flow into distinct modules, the code is much more flexible and structured

**Q18**: To what extent has the reusability of the code worsened or improved through the refactoring?

**A18**:

- Reusability has improved because logic is now encapsulated in individual blocks that can be reused more easily than logic buried in a massive function
- It is now very clear what responsibility each step has, making it simpler to extract specific logic for new use cases

- However, the Type State Pattern introduces a significant amount of boilerplate code that is highly specific to this particular workflow
- While the business logic is now modular, the structural glue of the pattern itself is not easily transferable or reusable elsewhere because it is so tightly tailored to this flow
- Compared to the original giant function, where extracting even a small piece of logic was exhausting, this is still a step forward

**Q19**: Exactly. And since this refactoring only targeted this one specific spot, the states we extracted are very specialized. To make them truly reusable, you'd probably need to apply this pattern across multiple similar code areas. Otherwise, you're always tied to the specific struct or state as a starting point, which might contain fields you don't even need in another use case. This is just my opinion, please elaborate on whether you agree or disagree with this.

**A19**:

- It's difficult to decide where to draw the line for a state machine in a real-world application
- In practice, a hybrid approach might be best: keeping technical interactions, like DB calls, as standard services while modeling complex logic, validation, and processing sequences as state machines
- A major challenge is that many workflows overlap significantly but require slightly different data at different stages
- If a new flow is 90% identical but needs one extra piece of data passed through, the rigid nature of the Type State Pattern might force you to copy and adjust almost all the structs and transitions
- This high specificity can become a hurdle for reusability if the domain logic isn't perfectly aligned across different use cases

**Q20**: Is the code easier or harder to read and understand after the refactoring?

**A20**:

- The method names (e.g., `get_benefit_ids`, `get_valid_benefits`) are very tangible and describe the business process clearly
- Piping the calls sequentially looks clean and helps someone looking at the code for the first time understand the high-level flow in seconds
- Error handling is improved because errors occur within specific `impl` blocks, making it much easier to pinpoint exactly where something went wrong compared to a massive function
- The data flow is more transparent because each Type strictly defines which data is available and required for that specific step
- A minor disadvantage is jumping between structs; unlike the old version where you just scrolled down, you now have to navigate different parts of the code to see the full implementation
- Overall, this is well-balanced: you get a high-level overview quickly and can choose to dive into specific details only when needed

**Q21**: Other experts have noted that the line count doubled and it's mostly boilerplate. One argued that while it's more code, it's simple code, mostly structs, that isn't hard to understand. How do you position yourself on this? Does the boilerplate make it harder to understand, or is it negligible?

**A21**:

- Every line of code technically adds maintenance cost, but the argument that boilerplate is a burden is largely invalid here because the structs contain no complex logic
- The main trade-off isn't the boilerplate itself, but rather that the behavior is no longer local, it's split up, which can make the code feel stretched out while reading
- Since code is read much more often than it is written, the effort to write the boilerplate is a small price to pay for the clarity it provides
- Forcing the developer to sit down and explicitly define what data a method needs and what its possible outcomes are is actually a beneficial exercise
- The increase in line count is real, but as far as the structs are concerned, the expert sees them as a positive addition rather than a problem

**Q22**: Let's move on to Modifiability. Has the refactoring made it easier or more difficult to make changes without introducing new bugs?

**A22**:

- Modifiability is much better because changes are now isolated within specific modules, whereas changing the old super-function almost guaranteed breaking side effects
- The Type State Pattern forces the compiler to act as a safety net; if you change a data requirement, the code turns red everywhere that needs an update
- Clear definitions of inputs and outputs for each step prevent the accidental swapping of similar variables, like different IDs, which was a major risk in the procedural version

**Q23**: On one hand, adding a simple filter or aggregation takes longer because you might have to create a new state or adjust existing structs. On the other hand, it forces you to think deeply about where the logic truly belongs. Do you agree that this overhead is a disadvantage, or is the forced intentionality a benefit?

**A23**:

- Simple one-liners like mapping or filtering shouldn't necessarily require a whole new state or helper function; you could ideally just chain a `.map()` or `.filter()` within the pipeline to keep minor changes local
- For more complex logic, the overhead of adjusting the next step's requirements like expecting a filtered list instead of a raw one is a price worth paying for the structure it provides
- The perceived disadvantage of extra work is outweighed by the benefits, especially since the original code was an extreme example of an unmanageable super-function
- If the original code had already been split into small, isolated functions, the leap to the Type State Pattern might feel more burdensome, but in this context, the improvement in modifiability is clear.

**Q24**: To what extent has the testability of the code worsened or improved through the refactoring?

**A24**:

- Testability has significantly improved because you can now write unit tests for actual individual steps rather than having to test the entire massive function at once
- Each step only carries the specific dependencies it needs, which greatly reduces the number of mocks required per test compared to the original procedural flow

**Q25**: Has the code's faultlessness improved? Does the refactoring prevent invalid application states, and has the risk of bugs or incorrect function calls been minimized?

**A25**:

- This is the greatest advantage of the refactoring because the explicit nature of the pattern makes it very difficult to accidentally introduce bugs
- You have a guaranteed execution order and strict definitions of which data must be available at each step, making it nearly impossible to bypass the intended logic
- You could still technically put the wrong ID into a struct field
- Because each state transition consumes `self`, it is ensured that steps cannot be called twice or used out of sequence

---

**Use Case 2 - Status Quo**

**Q26**: Let's move on to the Tag Component Handler and look at the status quo. Which code smells did you notice? Are there parts that make the code hard to maintain or error-prone, and which criteria are negatively affected?

**A26**:

- The validation logic is written inline, which significantly reduces both modularity and reusability because the logic cannot be used elsewhere
- The presence of deeply nested control flow with statements like `let`, `match`, `if` and `else` makes it difficult for someone not deeply involved in the code to follow the logic

- The code includes workaround comments and scattered test-related snippets, which clutter the implementation and make it harder to understand what is actually happening
- A major faultlessness issue is the reliance on untyped strings; even if the data is validated once, the rest of the system has no type-level guarantee that it remains valid
- Because there is no type-level enforcement, future developers writing new handlers could bypass the validation entirely, as the rest of the system cannot make safe assumptions about the data's state

---

**Use Case 2 - Refactoring**

**Q27**: Let's move to the refactoring and start with Modularity. To what extent has the code's modularity worsened or improved?

**A27**:

- The modularity has improved because the validation logic has been moved into its own dedicated function or module, rather than being tangled inline
- While it is a improvement, the change feels less significant than the previous use case because the original starting point wasn't quite as monolithic as the first service

**Q28**: To what extent has the reusability of the code worsened or improved through the refactoring?

**A28**:

- Reusability went from straight-up impossible to completely reusable because the logic is no longer hardcoded inline
- If you were to create a second handler that needs to validate the same Dreson data, you can now simply call the existing validation module instead of having to copy-paste or refactor the original code

**Q29**: Has the refactoring made the code easier or harder to read, analyze, and understand?

**A29**:

- Removing the deeply nested `let` statements and replacing them with a functional chain using `.map()`, `.transpose()`, etc. makes the code much cleaner and easier for any experienced Rust developer to follow
- One point of confusion is the `ValidatedDreson.from_string` function; it takes a list of exceptions, which is confusing at first
- While the previous nested version was annoying, the special case logic was very visible, whereas here you might need IDE type hints to realize that a list of strings is being passed as an exception list rather than being the source for the Dreson itself

**Q30**: Do you have a suggestion for handling these exceptions better when using the Newtype pattern? Usually, you'd use `From` or `TryFrom` for a string, but those traits don't allow passing additional parameters like an exception list.

**A30**:

- One option is to implement `TryFrom` for a tuple, containing both the raw string and the exception list, though this isn't always the cleanest look
- A more explicit approach would be creating a specific `ValidationRequest` struct with two fields: the raw Dreson and the exceptions.
- While a dedicated struct is very clear and removes any guesswork, it introduces more boilerplate, which is a common trade-off between "ideal" pattern application and practical convenience

**Q31**: Has the refactoring made it easier or more difficult to make changes without introducing new errors?

**A31**:

- Modifiability has improved because the validation logic is now isolated in a single method; if the validation rules change, you only have to update that one specific spot
- In the original version, there was a risk of having to update multiple locations, which creates error potential where you might forget to sync a change across different parts of the code

**Q32**: Is the code easier or harder to test, and is there a change in the number of required mocks or freedom from side effects?

**A32**:

- Testability is significantly better because the validation logic is now isolated and can be verified independently without needing to call the entire handler
- Previously, you had to provide all of the handler's dependencies, like the Cache, Tag Service, and Variation Service, just to test a simple string validation, which is no longer necessary
- The special `new_unvalidated` helper method, restricted to test code only via the `#[cfg(test)]` macro, allows for a simple test setup
- This approach makes it easy to intentionally create invalid states for testing purposes without compromising type safety in the production code

**Q33**: Then we come to the final criterion: Faultlessness - has it improved or worsened?

**A33**:

- Faultlessness is significantly higher because validating once at the entry point guarantees a valid state for the entire rest of the system
- Catching errors as early as possible prevents them from leaking deep into the control flow, which is a common problem in languages like Java where a failure might only surface 20 functions deep during a repository call
- By mapping to a specific Newtype immediately, you either get a valid instance or the process stops right there, ensuring that no further logic or side effects are triggered unnecessarily
- This pattern provides a level of certainty that makes the system much more robust, as subsequent functions no longer need to worry about the validity of the data they receive

---

**Use Case 3 - Status Quo**

**Q34**: Then we move to the third use case: the Boxfish Lambda Handler. Looking at the status quo code, which code smells did you notice? Which parts are hard to maintain or error-prone, and which criteria are negatively affected?

**A34**:

- Analyzability is a major issue because the code suffers from deeply nested functions and a convoluted call stack that is hard to follow
- Testability is poor due to excessive side effects in almost every function, making it nearly impossible to verify logic in isolation
- Reusability is almost non-existent; even when building similar flows, the current structure forces you to start from scratch because the logic is so tightly coupled
- Faultlessness is compromised by the use of string data and mutable lists for errors, which provide no compiler-level guarantees that data has been validated
- When migrating the lambda from Java to Rust, there was a requirement to match the old Java implementation quirks from years ago to avoid breaking changes for consumers of the endpoint, so some legacy problems were carried over into the new system
- The control flow is extremely difficult to trace, and the system relies entirely on manual tests rather than the compiler to enforce correctness

---

**Use Case 3 - Refactoring**

**Q35**: Then let's move to the refactoring. To what extent has the modularity of the code worsened or improved?

**A35**:

- Modularity has significantly improved because you no longer have a single, long control flow where everything happens at once

- The structure is now broken down into individual structs that can be handled in isolation, which means you don't have to understand the entire system just to look at one specific part
- It is interesting to note that the benefits here are very similar to those seen in the Detailview Service, showing a consistent pattern of improvement across different use cases

**Q36**: And what about Reusability? Has it worsened or improved through the refactoring?

**A36**:

- The logic is much better contained and easier to reuse or adapt because it isn't buried in a monolithic block of code
- Since the new types are highly specialized for this specific use case, they carry a bit of boilerplate that can make them harder to reuse directly in a generic way

**Q37**: Has the refactoring made the code easier or harder to read, analyze, and understand?

**A37**:

- Making state transitions explicit through types is a major benefit because it provides a clear overview of the various edge cases that the system actually handles
- Specifically, the "Final Result" type with multiple distinct states is incredibly helpful for instantly understanding all the possible outcomes that can occur in a flow
- Even though these edge cases existed implicitly before, making them visible all at once makes them feel much more manageable and less daunting than when they were buried in logic

**Q38**: Could you summarize it by saying that the Type State Pattern essentially forces you to give an explicit name to states that were previously only implicit?

**A38**:

- Exactly. You are forced to name each state, write it down, and ensure the compiler enforces that you actually handle it
- Because Rust requires matches to be exhaustive, if a function can potentially produce a certain state, you are strictly required to deal with it, making it impossible to accidentally ignore an outcome

**Q39**: I feel that while Type State can be used in other languages, Rust features like ownership (consuming `self`) and exhaustive matching make it feel much more natural. It's almost "married" to the language.

**A39**:

- A notable downside is that this pattern in this case replaces traditional `Result` types, making error handling less ergonomic since you can't easily use the question mark operator (`?`) to bubble up errors or add context
- Team feedback suggests the code becomes "less typical Rust" because developers are used to scanning for the `?` to identify failure points; without it, the possibility of failure becomes more implicit
- There is a lack of distinction between "Technical Exceptions" (e.g., Database down) and "Business Exceptions" (e.g., Request has no items), where technical failures should probably still use `Result` for quick termination
- Combining all outcomes into a single state enum can make the logic clunky because you lose the automatic conversion and ergonomics provided by Rust's standard error-handling traits

**Q40**: That's an interesting point. It's likely due to how I personally applied the Type State Pattern rather than an inherent rule of the pattern itself. There's no requirement that every error must be an enum variant. Using `Result` types alongside it is definitely possible, and I should probably include that as a recommendation in my bachelor thesis.

**A40**:

- Using `Result` for unexpected errors (Technical Exceptions) allows you to use the question mark operator to bubble them up to the top level immediately
- If an unexpected error occurs, you simply exit the state machine since it represents undefined behavior, which keeps the rest of the application from having to manually handle resource failures like a database being down
- At the top level, you could simply map these technical errors to a 500 Internal Server Error, while the specific enum variants handle expected business outcomes like "Not Found" or success states

- While having an enum inside a `Result` might seem slightly complex at first, it would likely be cleaner in practice because it separates the unrecoverable failures from the expected flow transitions

**Q41**: I actually modeled all the use cases using BPMN. It felt like a good fit to capture exactly what the code does and define the various states.

**A41**:

- BPMN is a great common ground that matches how business partners already think on boards like Miro, making transitions, and conditions explicit
- Incorporating state machine logic during the planning phase of a story would be a great recommendation for your thesis; it clarifies exactly which state is being modified or added
- Long-term, it would be ideal to have a visual flow for handlers where changes aren't just defined in text, but as visual updates to a diagram that developers can map directly to the code
- This approach makes technical logic much more accessible to non-technical stakeholders and helps developer pairs know exactly where to start their implementation
- An idea for a tool would be a library similar to Swagger that automatically generates a BPMN diagram directly from Rust Type State code

**Q42**: I've actually already thought of that. My Future Research section already mentions the idea of automatically detecting the Type State pattern to generate a model that can be synchronized with business stakeholders.

**A42**:

- I'm on the same page, that would be a fantastic idea

**Q43**: In BPMN, you have these "Intermediate Error Events" that can be attached to any activity. If business stakeholders model the logic with XOR gateways for business decisions but use Intermediate Error Events for unexpected technical failures (like a database being down), it translates directly to the code. Seeing an Intermediate Error Event in the model tells the developer: "I need a Result type here."

**A43**:

- That would be great

**Q44**: Let's move through the remaining criteria. We have Modifiability left: has the refactoring made it easier or more difficult to make changes without introducing new errors?

**A44**:

- It is definitely easier because the code is now split into individual steps and blocks that can be modified without causing unintended ripple effects
- In the original version, some functions were used so deeply that changing one would almost certainly break five others; that risk is now largely eliminated

**Q45**: And what about Testability? How has it changed through the refactoring, specifically regarding the number of required mocks or freedom from side effects?

**A45**:

- Testability has improved because you can write tests for individual components like validation logic without any mock setup at all
- Previously, you had to inject a massive amount of dependencies just to test small details, which is a problem we've seen across all the examples today
- While passing a single `ProcessorDependencies` object is somewhat simpler than passing 20 individual variables, it becomes problematic when that object is dragged deep into the control flow alongside multiple mutable lists
- Having those dependencies at the top level is fine, but forcing them deep into every function creates a mess that offers only disadvantages for testing and maintenance

**Q46**: Then we come to the final criterion: Faultlessness. To what extent has the error-free nature of the code improved, especially regarding invalid states, the risk of bugs, and so on?

**A46**:

- This is one of the biggest advantages of the pattern: you essentially prevent invalid states from existing in the first place
- If errors do occur, they are caught immediately at the point of origin, which is a major benefit for system reliability
- The pattern also prevents calling functions in the wrong order, which was a significant risk before; overall, faultlessness has definitely increased

---

**General Remarks**

**Q47**: Do you have any general comments about the refactoring that you'd like to add?

**A47**:

- I noticed you used an `Either` type in this part of the code and I was wondering if there was a specific reason for that
- It made me wonder if it wouldn't have been better to use a custom enum where the two variants are named explicitly, rather than a generic Left/Right structure

**Q48**: Exactly. I based that on a research paper explaining the Type State pattern and wanted to try it out. But you're right, there are different ways to do it. `Either` is essentially similar to a `Result`, with a value on the left and a value on the right.

**A48**:

- Using `Either` actually hurts readability because it isn't immediately obvious what "Left" or "Right" represents without looking back at the context
- It contradicts the rest of the logic where everything else is an explicit struct or enum with clearly named variants
- While it might seem like a convenient shortcut to avoid boilerplate when you only have two cases, the benefit is lost because you lose that self-documenting quality
- I found myself jumping back and forth in the code to figure out what was happening, which was much more difficult than just having a named enum

**Q49**: I think the idea is that when you have a larger state machine where different paths eventually merge back together, `Either` helps because it's generic. You can construct it with any type without needing a specific enum, especially if you don't know yet how the flow might split.

**A49**:

- Even in that case, you could just define an input enum with variants A and B to keep it explicit
- The `Either` approach feels very Haskell-like or functional on automata theory, but it's less practical for the reader
- Since it was only used in one place, it wouldn't have hurt to just use an enum

**Q50**: That's a fantastic insight! It serves as a practical evaluation of what that research paper suggested. It's a very interesting finding that this generic approach is actually harder to read in a Rust context.

**A50**:

- Unless you use `Either` everywhere and get used to it, it's better to leave it out entirely
- Being able to see exactly what is happening at a single glance is one of the primary advantages of this entire pattern, so generic types tend to undermine that strength

**Q51**: I'm looking through the Rust Design Patterns documentation to see if there is a convention like "prefer enums over generics"… I can't find the specific quote, but it feels like that would be a solid idiom for business code. Since Rust has ADTs where enum variants can hold different data, you can replace generic structures entirely as long as you know your states.

**A51**:

- I just realized this might be a crucial point for maintainability. If the business requirements change and you need a third entry point instead of two, a generic `Either` type scales very poorly

- Generic structures like `Either` or binary trees are great for abstract data structures or parsers where you truly only have two mathematical directions (0 or 1)
- In business logic, requirements are always flexible and likely to expand; using explicit enums makes the code much more adaptable to these changes

**Q52**: If you're writing a library where you don't know the end-user's specific use cases, generics are essential. But in business logic, the context is known. It's a very interesting distinction to make based on the type of software you are developing.

**A52**:

- Mhm. (agreement)

**Q53**: Do you have any other comments on the refactoring? Also, to follow up on my earlier question: when you showed this to the team, did they have any specific feedback? I imagine they might have said it looks unfamiliar since the pattern is new to them.

**A53**:

- The feedback was actually very positive; the main critique (from our Tech Lead) was the point about technical exceptions and the lack of `Result`, which we've already agreed is a valid point that can be integrated
- The team was already somewhat dissatisfied with the current implementation, viewing it as a "messy rewrite," so they were looking for a cleaner skeleton or template to use with Cargo Lambda
- There is a concern about consistency: if the pattern isn't followed strictly, for example, if a team member is away and comes back without fully grasping the logic, the value of the pattern disappears quickly as the implementation starts to tilt
- We might look for a middle ground where we take the best learnings, like the simple rule of using more explicit enums; they are essentially free and clarify exactly what states or results a method can return even without a full state machine

**Q54**: I definitely pushed the pattern to its academic limit for this project, but in practice, you can pick the best parts. What really helped me was shifting my mindset regarding enums. In Java, enums are just a fixed list of constants. In Rust, they are Sum Types, part of ADTs. While Structs are Product Types (all fields must exist), Enums allow for "one of these" cases where each variant can hold different data. Thinking in terms of Sum Types might help the team grasp how to model these "either-or" states.

**A54**:

- Exactly; it's very similar to Sealed Interfaces in Kotlin or the newer Sealed Classes/Interfaces in Java
- This represents a specific level of abstraction where, in traditional Java, you would have immediately reached for Object-Orientation and inheritance to solve the problem
- Shifting to this functional Sum Type approach provides a cleaner way to handle state without the overhead of complex class hierarchies

*The expert describes an alternative implementation of the Type State pattern using generic Typestates on a single struct, which differs from using Enums and separate Structs:*

- Instead of having 20 separate structs or one large enum, you define a single struct (e.g., `KafkaVault<Mode>`) that has a generic type parameter representing its current state
- The states themselves (like `Encryption`, `Decryption`, or `Uninitialized`) can be implemented as simple, empty marker structs without any data, or regular stucts that carry additional data
- You can use `impl<Mode> KafkaVault<Mode>` to write functions that are available regardless of the state (e.g., accessing shared IDs or configurations)
- You use `impl KafkaVault<Encryption>` to define functions that are only valid when the vault is in the encryption state
- This approach prevents the user from calling a function in the wrong state (like trying to `decrypt` an `Uninitialized` vault) because the compiler will literally not find that method on the current type

- In the Enum-based Type State pattern, if every state needs a specific `ID`, you have to pass it through or duplicate it in every enum variant. With the generic struct, common fields live in the parent struct and don't need to be moved during transitions
- Like the other pattern, state transitions are handled by functions that consume `self` (claim ownership) and return a new version of the struct with a different generic type (e.g., `KafkaVault<Uninitialized>` -> `KafkaVault<Initialized>`).
- This keeps the state machine "contained" within one logical unit (the struct) rather than scattering it across many independent types, making it feel more like a cohesive class while remaining functionally pure

## A.7 Static code analysis

### A.7.1 Use case 1

| Function | CyC | CoC | Halstead Difficulty | MI |
|---|---|---|---|---|
| `create_detailview_view_model` | 17 | 8 | 32.22 | 50.20 |
| `get_valid_benefits` | 3 | 0 | 14.67 | 108.83 |
| `get_detailview_auto_activated` | 6 | 4 | 25.48 | 77.83 |

**Table 7:** Static code metrics collected for the status quo of use case 1. The table shows the collected metrics for each major function that was included in the refactoring.

| Function | CyC | CoC | Halstead Difficulty | MI |
|---|---|---|---|---|
| `create_detailview_view_model` | 3 | 0 | 15.11 | 83.67 |
| `get_benefit_ids` | 1 | 0 | 10.67 | 111.87 |
| `get_valid_benefits` (top-level) | 3 | 0 | 14.86 | 106.03 |
| `BenefitIdsFetched::get_valid_benefits` | 2 | 0 | 15.94 | 98.65 |
| `ValidBenefitsFiltered::determine_detailview_benefit` | 8 | 3 | 24.76 | 78.44 |
| `DetailviewDetermined::check_for_up_contracts` | 3 | 1 | 17.14 | 89.49 |
| `UpContractTaskSpawned::determine_activation_status` | 4 | 4 | 22.50 | 80.06 |
| `ActivationStatusDetermined::create_adjust_link` | 4 | 1 | 16.00 | 88.40 |
| `AdjustLinkCreated::determine_up_membership` | 3 | 1 | 17.18 | 92.68 |
| `UpMembershipDetermined::build` | 1 | 0 | 12.83 | 107.82 |
| `get_detailview_auto_activated` | 6 | 4 | 25.48 | 77.83 |

**Table 8:** Static code metrics collected for the refactoring of use case 1. The table shows the collected metrics for each major function that was refactored.

### A.7.2 Use case 2

| Function | CyC | CoC | Halstead Difficulty | MI |
|---|---|---|---|---|
| `handle_request` | 16 | 17 | 31.07 | 52.61 |

**Table 9:** Static code metrics collected for the status quo of use case 2. The table shows the collected metrics for each major function that was included in the refactoring.

| Function | CyC | CoC | Halstead Difficulty | MI |
|---|---|---|---|---|
| handle_request | 16 | 8 | 27.93 | 54.89 |
| ValidatedDreson::from_string | 3 | 3 | 10.83 | 112.10 |
| ValidatedDreson::new_unvalidated | 1 | 0 | 2.67 | 136.18 |
| ValidatedDreson::inner | 1 | 0 | 4.67 | 133.39 |
| ValidatedDreson::contains | 1 | 0 | 6.00 | 131.30 |
| ValidatedDreson::eq | 1 | 0 | 6.75 | 131.44 |

**Table 10:** Static code metrics collected for the refactoring of use case 2. The table shows the collected metrics for each major function that was refactored.

### A.7.3 Use case 3

| Function | CyC | CoC | Halstead Difficulty | MI |
|---|---|---|---|---|
| handle | 5 | 2 | 19.38 | 70.16 |
| fetch_item_return_status_for_valid_items | 4 | 2 | 19.45 | 79.24 |
| fetch_item_return_status_list | 2 | 0 | 12.28 | 86.47 |
| build_internal_server_errors | 2 | 1 | 8.25 | 105.74 |
| add_missing_position_item_errors | 3 | 1 | 12.50 | 85.61 |
| extract_position_item_ids_from_request | 2 | 0 | 10.83 | 111.62 |
| process_position_items | 5 | 6 | 18.64 | 80.39 |
| add_internal_server_error | 1 | 0 | 6.88 | 110.14 |
| build_final_response | 8 | 8 | 19.83 | 78.13 |
| build_unprocessable_entity_error_response | 2 | 0 | 10.23 | 97.54 |
| AnnouncedPositionItemsRequest::validate | 1 | 0 | 6.29 | 114.88 |
| AnnouncedPositionItems::new | 1 | 0 | 5.83 | 120.77 |
| AnnouncedPositionItems::validate | 1 | 0 | 10.50 | 100.10 |
| AnnouncedPositionItemsRequest::try_from | 6 | 1 | 8.80 | 109.86 |
| validate_announced_items_request | 5 | 3 | 20.06 | 82.57 |
| validate_return_item | 4 | 3 | 16.00 | 85.31 |
| validate_announced_position_items | 3 | 4 | 14.50 | 84.17 |

**Table 11:** Static code metrics collected for the status quo of use case 3. The table shows the collected metrics for each major function that was included in the refactoring.

| Function | CyC | CoC | Halstead Difficulty | MI |
|---|---|---|---|---|
| `handle` | 4 | 1 | 14.17 | 88.59 |
| `handle_valid_request` | 3 | 1 | 18.00 | 86.22 |
| `AfterValidationResult::fetch_return_statuses` | 4 | 2 | 24.00 | 70.20 |
| `fetch_item_return_status_list` | 2 | 0 | 12.50 | 86.64 |
| `build_internal_server_errors` | 2 | 0 | 7.70 | 109.25 |
| `build_missing_position_item_errors` | 4 | 1 | 14.50 | 81.86 |
| `ItemsWithReturnStatuses::process` | 6 | 5 | 23.43 | 79.73 |
| `ValidItemReturnStatus::process` | 3 | 1 | 20.77 | 87.85 |
| `internal_server_error` | 1 | 0 | 6.14 | 114.95 |
| `AnnouncedPositionItem::from` (impl @ ~303) | 1 | 0 | 6.00 | 122.69 |
| `AnnouncedPositionItem::from` (impl @ ~312) | 1 | 0 | 6.00 | 122.69 |
| `FinalResponse::from` | 9 | 7 | 23.40 | 72.97 |
| `FinalResponse::build_response` | 8 | 1 | 17.00 | 77.10 |
| `AnnouncedPositionItemsRequest::try_from` | 6 | 1 | 11.90 | 99.08 |
| `AnnouncedPositionItem::validate` | 1 | 0 | 10.50 | 100.10 |
| `validate_announced_items_request` | 4 | 2 | 14.73 | 91.09 |
| `ValidAnnouncedPositionItemsRequest::validate_announced_items` | 4 | 3 | 15.00 | 95.91 |
| `validate_announced_position_item` | 2 | 2 | 13.93 | 88.55 |
| `validate_item_return_status` | 4 | 3 | 16.73 | 89.16 |

**Table 12:** Static code metrics collected for the refactoring of use case 3. The table shows the collected metrics for each major function that was refactored.

## A.8 Time behaviour measurement

### A.8.1 Use Case 1

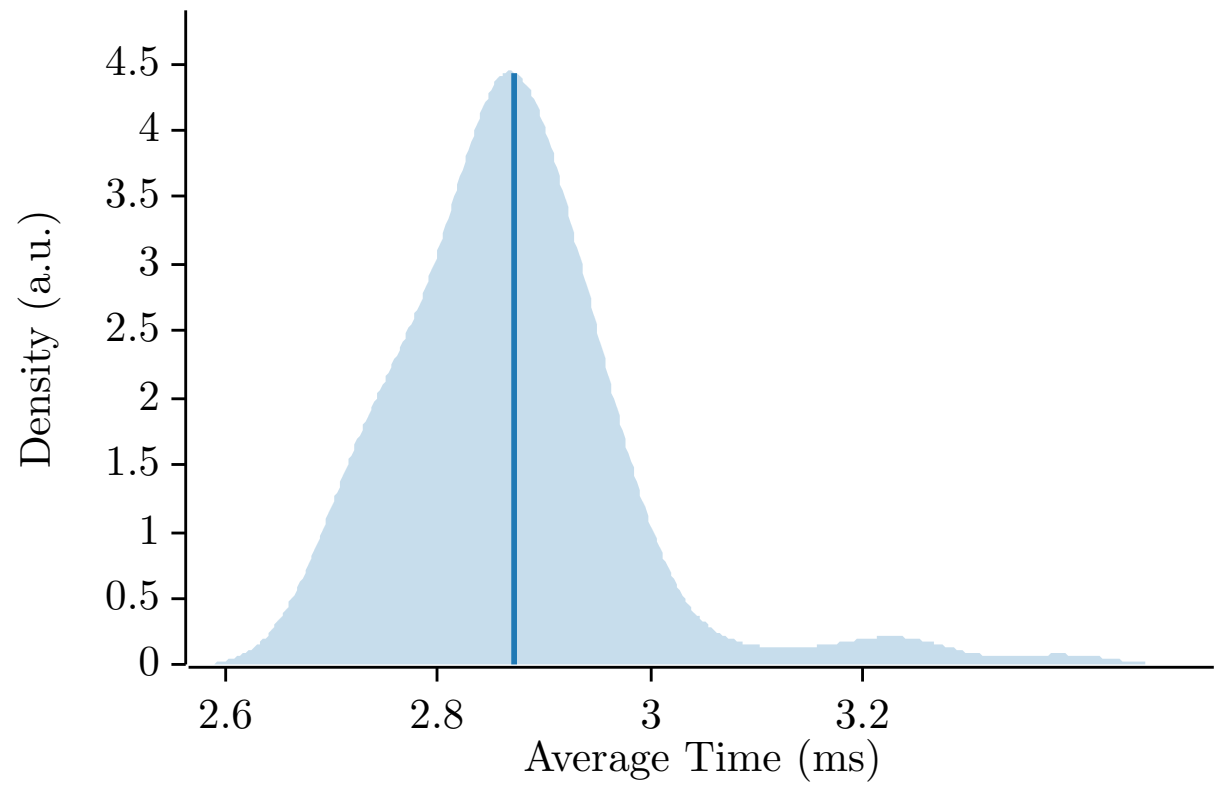

**Figure 8:** Time density of use case 1 benchmark. The x-axis shows the time in milliseconds, while the y-axis shows the density. A distribution of execution time can be seen, including a marker for the average time. The distribution is concentrated around the average time, with a few outliers around 3.2ms.

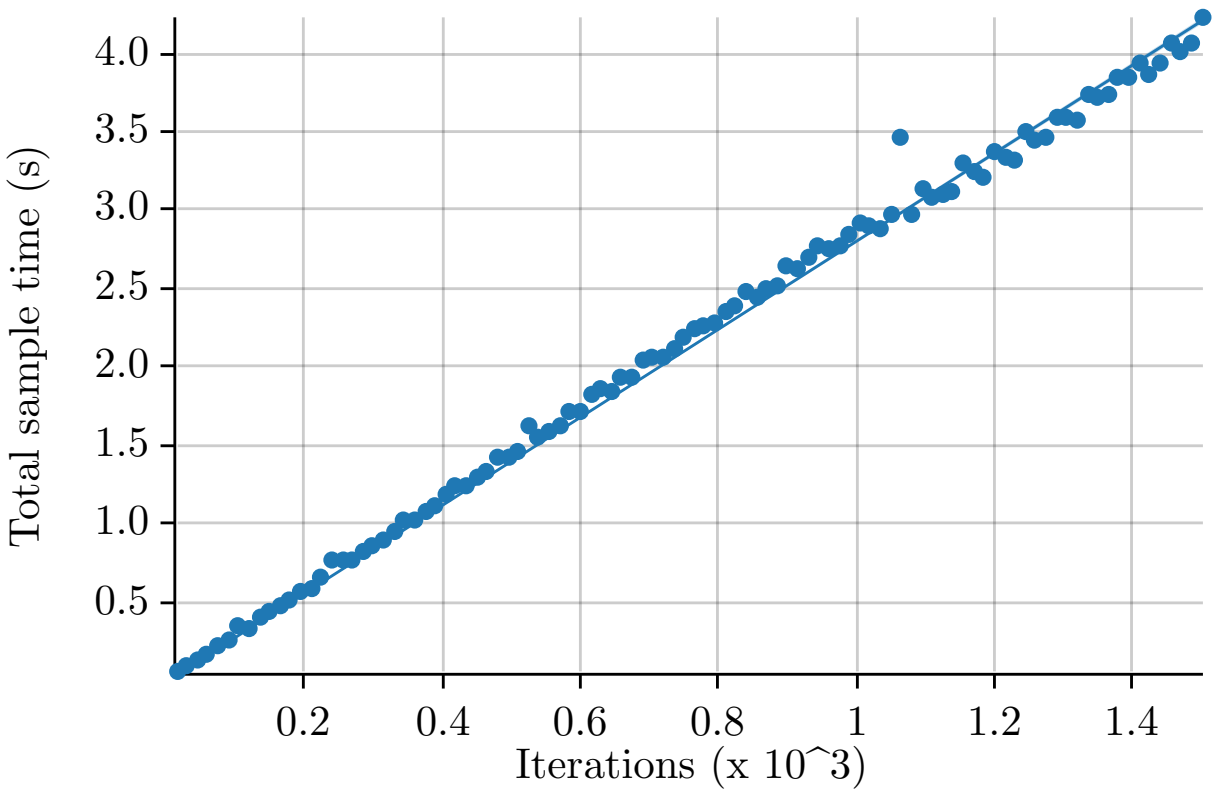

**Figure 9:** Total sample time per iteration of use case 1 benchmark. The x-axis shows the iteration number, while the y-axis shows sample time in seconds. A linear regression line can be seen that fits the data points. The data points become slightly more spread towards the end, and there is one outlier at around x = 1.05, y = 3.5.

| Metric | Lower bound | Estimate | Upper bound |
|---|---|---|---|
| Slope | 2.7942 ms | 2.8134 ms | 2.8373 ms |
| $R^2$ | 0.8690985 | 0.8740415 | 0.8663407 |
| Mean | 2.8502 ms | 2.8714 ms | 2.8944 ms |
| Std. Dev. | 80.703 µs | 111.68 µs | 140.10 µs |
| Median | 2.8537 ms | 2.8639 ms | 2.8698 ms |
| MAD | 50.362 µs | 75.196 µs | 104.83 µs |
| Change in time | −2.8563% | −1.8605% | −0.7604% |

**Table 13:** Statistical estimations and confidence intervals for use case 1. The lower bound, point estimate and upper bound of each measure are displayed. The change in time was reported as statistically significant, but within the noise threshold by the tool Criterion. The $R^2$ value indicates that the regression model fits well, but is slightly lower due to the spread out data points towards the end.

### A.8.2 Use Case 2

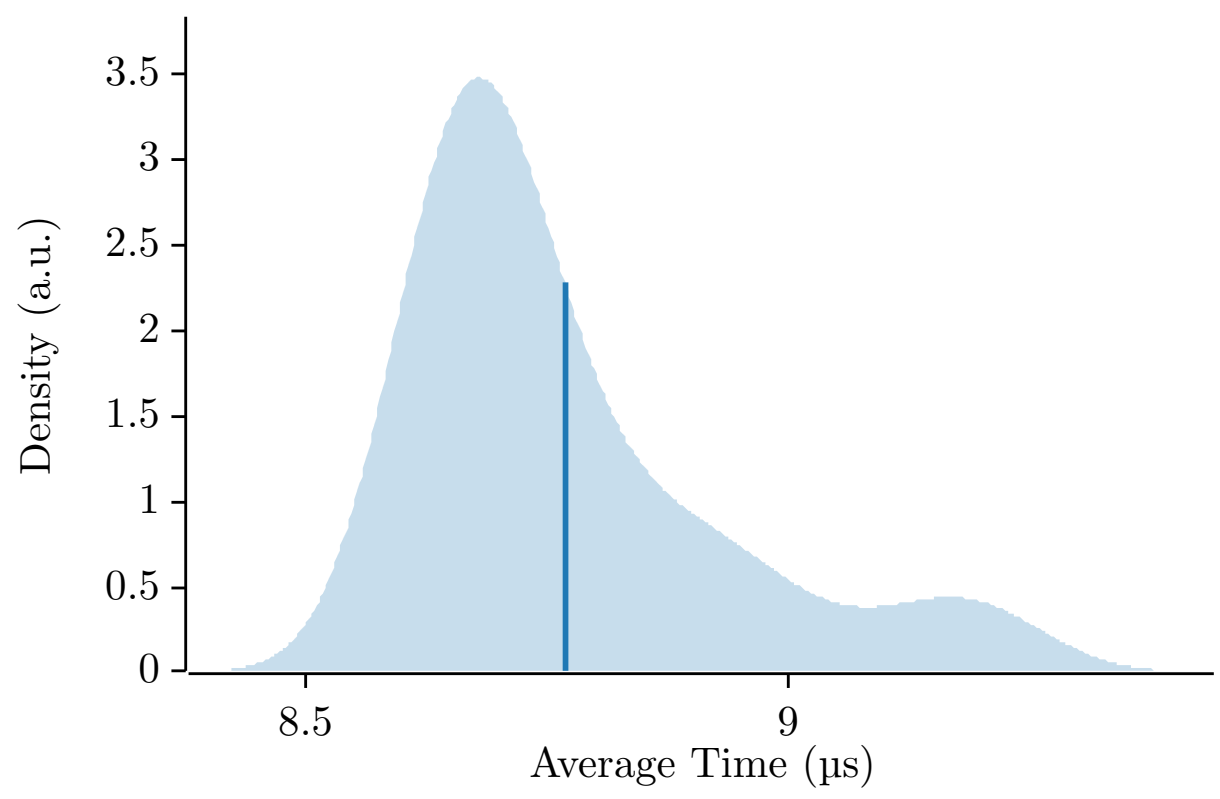

**Figure 10:** Time density of use case 2 benchmark. The x-axis shows the time in milliseconds, while the y-axis shows the density. A distribution of execution time can be seen, including a marker for the average time. The distribution is concentrated a bit to the left of the average time, and the distribution is stretched out to the right at a lower level.

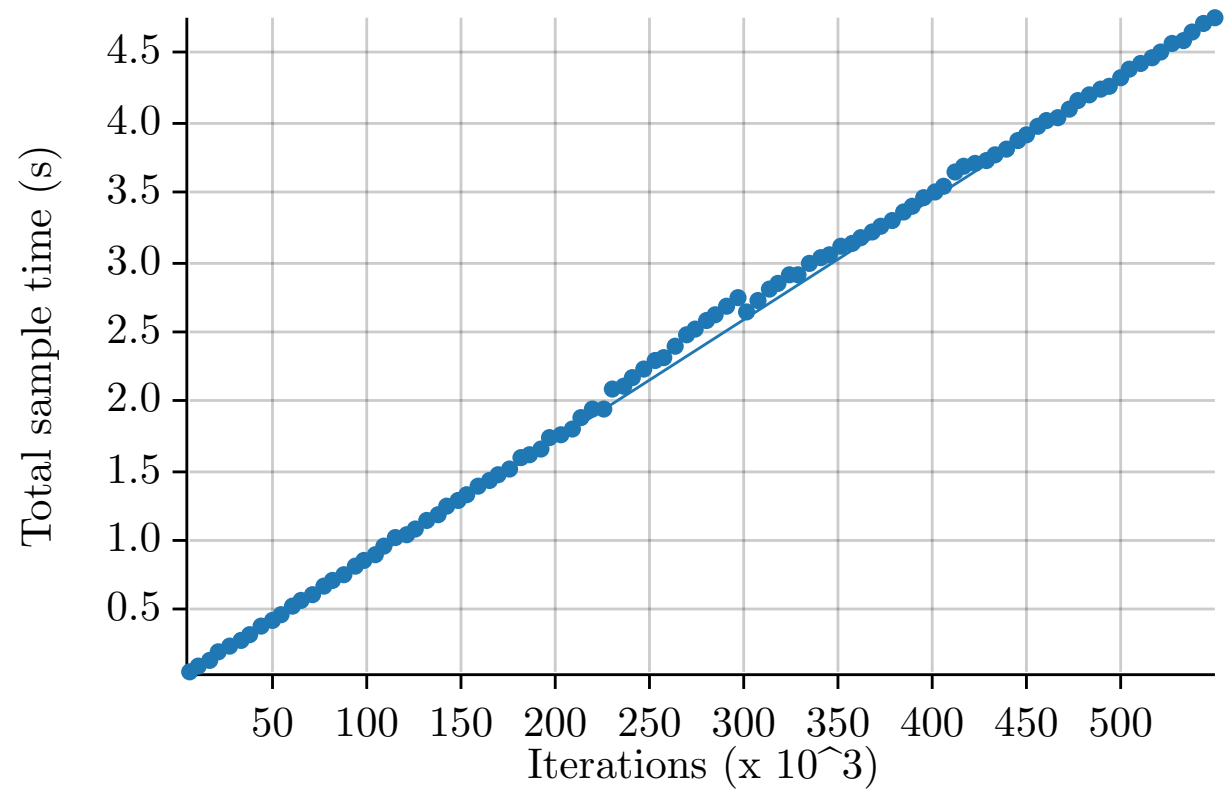

**Figure 11:** Total sample time per iteration of use case 2 benchmark. The x-axis shows the iteration number, while the y-axis shows sample time in seconds. A linear regression line can be seen that fits the data points. A few data points are slightly higher than the regression line between x = 230 and x = 300.

| **Metric** | **Lower bound** | **Estimate** | **Upper bound** |
|---|---|---|---|
| Slope | 8.7232 µs | 8.7503 µs | 8.7831 µs |
| $R^2$ | 0.9623165 | 0.9636155 | 0.9617049 |
| Mean | 8.7387 µs | 8.7691 µs | 8.8016 µs |
| Std. Dev. | 130.17 ns | 161.20 ns | 186.87 ns |
| Median | 8.6884 µs | 8.7025 µs | 8.7360 µs |
| MAD | 65.199 ns | 87.682 ns | 130.32 ns |
| Change in time | +0.9048% | +1.2524% | +1.6533% |

**Table 14:** Statistical estimations and confidence intervals for use case 2. The lower bound, point estimate and upper bound of each measure are displayed. The change in time was reported as statistically significant, but within the noise threshold by the tool Criterion. The $R^2$ value indicates that the regression model fits well.

### A.8.3 Use Case 3

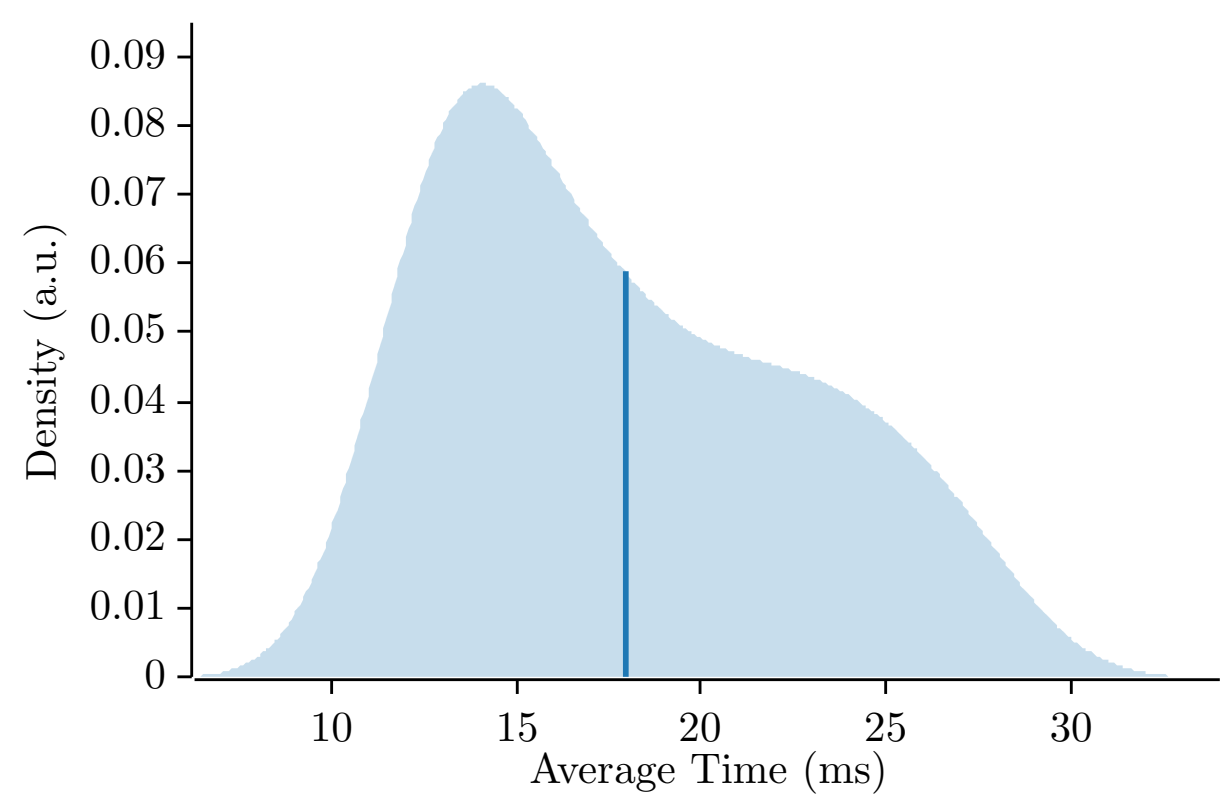

**Figure 12:** Time density of use case 3 benchmark. The x-axis shows the time in milliseconds, while the y-axis shows the density. A distribution of execution time can be seen, including a marker for the average time. The distribution is centered around the average time, but its peak lies left of the average.

**Figure 13:** Total sample time per iteration of use case 3 benchmark. The x-axis shows the iteration number, while the y-axis shows sample time in seconds. A linear regression line can be seen that doesn't fit the data points well. The data points follow a nonlinear, steady increase, possibly indicating a superlinear growth.

| Metric | Lower bound | Estimate | Upper bound |
|---|---|---|---|
| Slope | 21.059 ms | 22.096 ms | 22.978 ms |
| $R^2$ | 0.3526551 | 0.3677720 | 0.3566953 |
| Mean | 17.015 ms | 17.926 ms | 18.858 ms |
| Std. Dev. | 4.2030 ms | 4.7230 ms | 5.1472 ms |
| Median | 15.490 ms | 16.900 ms | 18.423 ms |
| MAD | 3.8419 ms | 5.3993 ms | 6.7121 ms |
| Change in time | −10.668% | −3.6920% | +3.8755% |

**Table 15:** Statistical estimations and confidence intervals for use case 3. The lower bound, point estimate and upper bound of each measure are displayed. The change in time was reported as statistically insignificant by the tool Criterion, with the lower and upper bound being relatively far apart, possibly due to the low $R^2$ value. It indicates that the linear regression model doesn't fit the data points well, which is due to the nonlinear progression. This is also a possible reason for the high standard deviation of around 4.7ms, when compared to the mean of around 17.9ms.